%% file: main_iclr2027.tex
\documentclass{article} 
\usepackage{iclr2027_conference,times}

\input{preamble}

\usepackage{hyperref}
\usepackage{url,bbm}
\usepackage{tikz}
\usetikzlibrary{arrows.meta,positioning,calc,decorations.markings}

\allowdisplaybreaks

\title{Simulating Eutopia: Revisiting Long-term Fairness with Outcomes, Performativity, and Dynamics}

\iclrfinalcopy 

\author{%
  Vedant Palit\thanks{Corresponding author: \texttt{vedantpalit@kgpian.iitkgp.ac.in}} \\
  Indian Institute of Technology
  Kharagpur\\
  Kharagpur, India 
  \And
  Udvas Das\thanks{Corresponding author: \texttt{udvas.das@inria.fr}}, Brahim Driss, Debabrota Basu \\
  Univ. Lille, Inria, CNRS,\\
  Centrale Lille, UMR 9189 – CRIStAL\\ 
  F-59000 Lille, France 
}

\begin{document}

\maketitle

\setcounter{parttocdepth}{2}
\doparttoc 
\faketableofcontents 

\input{iclr_2027/Chapters/abstract}

\input{iclr_2027/Chapters/introduction_short}
\input{iclr_2027/Chapters/Problem_formulation}
\input{iclr_2027/Chapters/final_results}

\input{iclr_2027/Chapters/Algorithm_Design}
\input{iclr_2027/Chapters/Theory_final}
\input{iclr_2027/Chapters/Experiments}

\input{iclr_2027/Chapters/Discussion_FutureWork}
\bibliography{ref}
\bibliographystyle{iclr2027_conference}

\appendix
\newpage
\part{Appendix}
\parttoc\newpage
\input{iclr_2027/Appendix/Notation_table}\clearpage
\input{iclr_2027/Chapters/Related_works}\clearpage

\input{iclr_2027/Appendix/Theory}\newpage
\input{iclr_2027/Appendix/detailed_experimentation}

\end{document}

%% file: preamble.tex
\usepackage[colorlinks=true,urlcolor=blue,linkcolor=red,citecolor=teal,backref=page]{hyperref}
\usepackage{makecell}
\usepackage{float,xcolor,todonotes,longtable}
\usepackage{mathtools}
\usepackage{amsmath, amssymb, natbib, graphicx, url,tabularx}
\usepackage[capitalize,noabbrev]{cleveref}

\usepackage{wrapfig}
\usepackage{booktabs}

\usepackage{algorithm}
\usepackage{algorithmic}

\usepackage{dsfont,enumitem}
\usepackage{footnote}
\usepackage{caption}
\usepackage{subcaption}
\usepackage{mwe}
\usepackage{cancel}
\usepackage{multirow}

\makesavenoteenv{tabular}
\makesavenoteenv{table}
\makesavenoteenv{tabularx}

\newcommand{\bx}{\mathbf{x}}
\newcommand{\by}{\mathbf{y}}

\newcommand \defn {\mathrel{\triangleq}}

 \DeclareMathOperator*{\argmin}{arg\,min}

\newcommand \expect {\mathop{\mbox{\ensuremath{\mathbb{E}}}}\nolimits}
\newcommand \prob {\mathop{\mbox{\ensuremath{\mathbb{P}}}}\nolimits}

\newcommand{\reals}{\mathbb{R}}

\usepackage{amsthm}

\newtheorem{proposition}{Proposition}
\newtheorem{definition}{Definition}

\newtheorem{remark}{Remark}

\makeatletter
\newtheorem*{rep@theorem}{\rep@title}
\newcommand{\newreptheorem}[2]{%
	\newenvironment{rep#1}[1]{%
		\def\rep@title{\textbf{#2} \ref{##1}}%
		\begin{rep@theorem}}%
		{\end{rep@theorem}}}
\makeatother
\newreptheorem{theorem}{Theorem}
\newreptheorem{lemma}{Lemma}
\newreptheorem{proposition}{Proposition}
\newreptheorem{assumption}{Assumption}
\newreptheorem{corollary}{Corollary}

\allowdisplaybreaks%

\newif\ifdoublecol
\doublecolfalse

\usepackage[nohints]{minitoc}

\usepackage{mathrsfs}

\usepackage{tikz}
\usetikzlibrary{automata}
\usetikzlibrary{bayesnet, arrows, calc, shapes, backgrounds,arrows,decorations.pathmorphing,fit,positioning}
\tikzset{
   container/.style = {rectangle, rounded corners, draw=yellow, dashed,
fit=#1, inner sep=6mm, node contents={}},
circle-label/.style = {circle, draw}
        }
\tikzset{box/.style={draw, diamond, thick, text centered, minimum height=0.5cm, minimum width=1cm, text width=0.9cm}}
\tikzset{line/.style={draw, thick, -latex'}}
\allowdisplaybreaks
\usepackage{pgfplots}

\def\bA{\mathbf{A}}

\def\bS{\mathbf{S}}

\def\bX{\mathbf{X}}

\def\bx{\mathbf{x}}
\def\by{\mathbf{y}}

\def\bpi{{\boldsymbol\pi}}

\def\cA{\mathcal{A}}

\def\cH{\mathcal{H}}

\def\cL{\mathcal{L}}
\def\cM{\mathcal{M}}

\def\cS{\mathcal{S}}

\newcommand{\transition}{\mathbf{P}}
\newcommand{\reward}{r}

\newcommand{\framework}{{\color{cyan}\ensuremath{\mathsf{Eutopia}}}}

\newcommand{\papply}{\ensuremath{{p_{\rm apply}}}}

\usepackage[most]{tcolorbox}
  \newtcolorbox{contribution}[1]{%
    colback=cyan!4, colframe=cyan!50!black, boxrule=0.6pt, arc=2pt,
    left=6pt, right=6pt, top=4pt, bottom=4pt, boxsep=2pt,
    title={#1}, fonttitle=\bfseries\small, fontupper=\small}

%% file: iclr_2027/Chapters/abstract.tex
\begin{abstract}
As AI-driven Decision Makers (ADMs) influence our socioeconomic reality, their roles in both enhancing efficiency and amplifying the social biases have drawn significant attention. In this paper, we revisit the nuances of \textit{long-term fairness} achievable by an ADM-- specifically, through the lens of a loan approver inducing a population-level wealth dynamics. 
The literature generally (a) considers passive environments, i.e. the decisions of an ADM does not change the population's behaviour, and (b) measures bias in terms of disparity in the instantaneous decisions rather than downstream equity. Modern ADMs challenge both the notions. 
To address these caveats, we first formalise the wealth dynamics induced by a loan approving ADM interacting with a multi-demographic population as a Performative Markov Decision Process.
Then, we mitigate the absence of such a performative test-bed by developing \framework{}: a lending-process simulator enabled with a novel performative data generator to learn long-term fair strategies. 
With \framework, we test reinforcement learning algorithms with different fairness-aware utilities dependent on approval decisions and downstream wealth.
Results show that performative dynamics-aware learning with fairness-aware utility that incorporates the downstream outcomes induce better long-term equity and inclusivity.
\end{abstract}

%% file: iclr_2027/Chapters/introduction_short.tex
\section{Introduction}
\begin{figure}[t]
    \centering
    \includegraphics[width=0.5\linewidth]{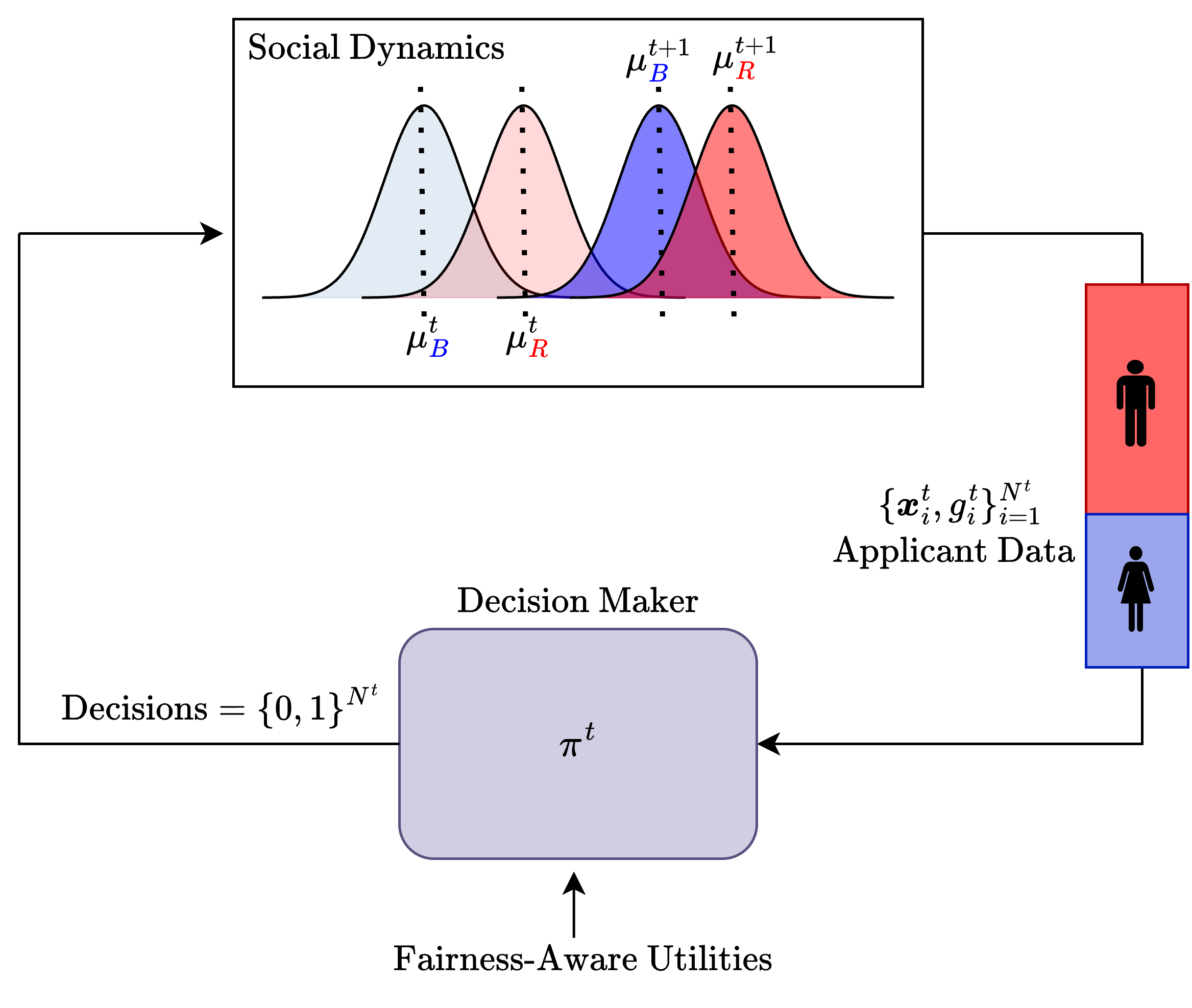}
   \caption{Sequential Credit Lending Problem}\label{fig:seq_credit_lending}
\end{figure}

Modern AI-driven Decision Makers (ADMs) are not passive. We live in a symbiosis with these digital companions. While they bring efficiency to our high and low-stake daily decisions, they also change our socio-politico-economic behaviours over time.
ADMs bring efficiency in applications ranging from credit scoring and loan underwriting~\citep{fuster2022predictably} to hiring~\citep{raghavan2020mitigating}, and criminal justice~\citep{mehrabi2021survey}. 
In tandem, they influence how we conduct job or loan applications, how we do medical diagnosis, how we even write papers and conduct research. 
This reactive effect induced by an algorithm is called \textit{performativity}~\citep{perdomo2020performative,gois2024performative} and ADM's ability to influence human behaviour is measured by its \textit{performative power}~\citep{hardt2022performative}.
This phenomenon invokes the question:
\vspace*{-0.5em}\begin{center}
    \textit{What are the long-term effects of performative interactions between ADMs and society,\\ 
    in terms of equity?} 
\end{center}\vspace*{-.3em}
\noindent\textbf{Fairness in Short and Long-term.} \emph{Efficiency} and \emph{equity} are not the same.  
Since ADMs learn from historical data mirroring the existing social inequalities, an unexamined pursuit of predictive accuracy can reproduce, even amplify the structural disadvantages faced by marginalised groups~\citep{obermeyer2019dissecting,angwin2022machine}.  
This has propelled the field of \textit{algorithmic fairness}.
Researchers developed different metrics to quantify bias~\citep{corbett2023measure}, auditors to measure bias induced by ADMs~\citep{ajarra2025active}, and algorithm designs to mitigate bias of predictive systems~\citep{barocas2023fairness}.
The initial works evaluated fairness at a \emph{single decision point}, treating the population of applicants as a fixed, exogenous distribution~\citep{d2020fairness}.
But decision making in real-world is a continual, interactive process.
For example,  a lending decision is not a passive observation, rather an intervention.
Approving a loan changes borrower's wealth (Figure~\ref{fig:seq_credit_lending}) that affects their future creditworthiness and also the dynamics of future applicants.
\citet{liu2018delayed,hashimoto2018fairness} show that learning to maximise \textit{short-term fairness} (e.g. demographic parity in one-shot lending) can \emph{reduce} long-run welfare of disadvantaged groups by denying them the very opportunities.
Hence, \textit{long-term fairness is studied as sequential problem over multiple interactions}.

\setlength{\textfloatsep}{10pt}
This has led to Reinforcement Learning (RL) driven study of long-term fairness in ADMs.
\cite{jabbari2017fairness} formalise fairness in \emph{Markov Decision Process} (MDP) setting-- requiring no action to make any group worse off relative to a baseline. Subsequent works developed general fairness constraints, and computational-statistical techniques to ensure better efficiency-equity trade-offs in long-term~\citep{hu2022achieving,deng2022reinforcement,ju2023achieving,wang2024online}.
But an overarching assumption persists that the \emph{transition dynamics of the population are exogenous}. The human population is assumed to evolve according to fixed, policy-independent rules, which is not true for modern ADMs.

 
\begin{contribution}{Contribution 1: A Performative Formalism}
We formalise sequential credit lending as a \emph{Performative MDP} (Section~\ref{sec:formulation pemdp}), i.e., the deployed policy shapes both wealth dynamics and future applications. It extends the MDP-centric long-term fairness literature. 
\end{contribution}
 
In this model, the state consists of features from the cohort of arriving applicants, the action is a vector of accept/reject decisions, and the reward is ADM's profit (Eq.~\eqref{eq:reward_def}). Approval increases an applicant's wealth if the loan is repaid (Eq.~\eqref{eq:wealth-dynamics}), and a group's application rate responds to the decisions it has observed throughout. The ADM optimises discounted cumulative reward over interactions, which forces it to reason about the long-term steady state. Additionally, we show precisely why this algorithmic feedback loop cannot be absorbed simply by enlarged state-augmentation (Remark~\ref{remark:state-aug}). 

A formalism defines the algorithmic feedback loop but to use it for real-life sequential decisions, one has to realise the shift in population dynamics under a deployed policy. No standard empirical resource allows this. \textit{Static credit} datasets are snapshots taken under a single, fixed acceptance rule. They contain no repeated interaction, or variation in the deployed policy, which are exactly the ingredients needed to identify performative responses~\citep{perdomo2020performative}. 
In contrast, \textit{live deployment} captures the dynamics but allows little space to experiment. Measuring long-term harm to a protected group means deploying rules one expects to fail, which is neither ethically defensible nor compatible with fair-lending regulations. Simulation is the established answer to this
bind~\citep{d2020fairness}, and it is missing for the performative sequential lending problem.
 
\begin{contribution}{Contribution 2: $\mathsf{Eutopia}$: a Performative Lending Simulator with Deomgraphics}
We build \framework{} (Section~\ref{sec:eutopia_main}), which realises the Performative MDP as a closed loop of
\emph{Decision}, \emph{Evaluation}, and \emph{Social Dynamics} blocks fed by a data
generator. To our knowledge, it is the first simulator for long-term lending in a performative
environment.
\end{contribution}

In \textit{Decision block}, an ADM deploys a policy trained on a chosen fairness-aware utility to guide decisions for the arriving cohort.
\textit{Evaluation block} realises the consequences of the decisions taken. Loans are repaid or defaulted, as the ADM obtains its profit. 
\textit{Social Dynamics block} closes the loop as wealth evolves, and a Hawkes process decides who applies next. The data generator grounds all of this on real data~\citep{adult_2}, and supplies the observations as per its response to the policy. 
\textit{We deploy \framework{} to compare standard RL against performative RL learners across outcome and decision-level fairness measures with three ways of turning them into a constraint.}
 
\begin{contribution}{Contribution 3: Long-term fairness is achievable with outcome parity and not decision-level parity}
Long-term equity requires deploying a performativity-aware learner trained with a dynamic fairness-aware utility measured on outcomes. 
\end{contribution}
 
Four empirical findings drive this conclusion (Section~\ref{sec:experimental_analysis}). \emph{(a) Performative learning matters}: Performative RL agent beats its standard RL counterparts on all the metrics in every scenario, while the RL agent collapses to approving almost everyone it receives.
\emph{(b) Decision-level parity renders delayed impact of fairness}: An ADM trained for Equality of Opportunity meets its
criterion early. After which its constraint stops binding and it behaves like a profit-driven utilitarian agent-- reproducing the \emph{delayed impact of fairness}~\citep{liu2018delayed} inside a performative environment. \emph{(c) FL under Equality of Outcome renders best equity}: Fairness Lagrangian under
Equality of Outcome closes the wealth gap while retaining two-thirds of the maximal profit. \emph{(d) Equality of Opportunity is selective, Outcome is inclusive}: Under Equality of Opportunity, $15$
to $26\%$ of each group never receives a loan and women gradually stop applying, whereas under
Equality of Outcome nearly everyone is served.

%% file: iclr_2027/Chapters/Problem_formulation.tex
\section{Sequential Lending as an Average Reward Performative MDP}\label{sec:formulation pemdp}

First, we formalise the sequential credit lending problem (Figure~\ref{fig:seq_credit_lending}) that induces a social wealth dynamics. We consider a population of $N$ applicants-- each with a feature vector $\bx_i \in \mathbb{R}^d$, and a demographic group $g_i \in \{\text{red}, \text{blue}\}$. We use the first feature $\bx_i[1]$ to denote an applicant's wealth.\footnote{\textit{Notations.} $X,x$ denote a random variable and its realisation, respectively. $y,\by$ denote a scalar and a vector, respectively. Whenever we write $z^t_i$, it corresponds to applicant $i$. Otherwise, $z^t$ denotes $\{z_i^t\}_{i=1}^{N^t}$.  } 
Each applicant requests for a loan $l_i \in \mathbb{R}_{>0}$ and yields a binary repayment outcome $Y_i \in \{0,1\}$, where $Y_i = 1$ denotes successful repayment of loan. Importantly, $Y_i$ is observed \textit{only for approved applicants}. The ADM never learns what would have happened if a denied applicant would have received a loan, known as \textit{selective label bias}~\citep{lakkaraju2017selective}.

\textbf{Interaction Dynamics.} At each time-step $t\in\mathbb{N}$, a set of $N^t \in [N]$ applicants arrive with their corresponding probability of loan application $\{{p_{\mathrm{apply}}}_i^t\}_{i=1}^{N^t}$. Thus, the \textbf{state} is defined as the full context of all arriving applicants as $   \bS^t = \bigl\{\bx_i,\, l_i,\, g_i\bigr\}_{i=1}^{N^t}$. 
Upon observing the state, the ADM takes a batch of binary \textbf{actions}  $\bA^t \in \{\perp, \not\perp\}^{N^t}$, where $\perp$ and $\not\perp$ denote approval and denial. The expected \textbf{transition} of applicant $i$'s wealth under a deployed policy $\bpi$ is expressed as: 
\begin{align}\label{eq:wealth-dynamics}
    \expect\left[  \bX_i^{t+1}[1] \right] =& \underbrace{(\bx_i^{t}[1]+\kappa_i) \cdot {p_{\rm apply}}_i^t \cdot\prob_{\bpi}\left(\bA_{i}^t =\perp\right) \cdot\prob(Y_i^t=1)}_{\text{successful repayment}}\notag\\
    &+ \underbrace{\bx_i^{t}[1]\cdot{p_{\rm apply}}_i^t \cdot \prob_{\bpi}\left(\bA_{i}^t =\perp\right)\cdot \prob(Y_i^t=0)}_{\text{default on repayment}}
    + \underbrace{\bx_i^{t}[1]\cdot{p_{\rm apply}}_i^t \cdot \prob_{\bpi}\left(\bA_{i}^t =\not\perp\right)}_{\text{loan denied}}\,.
\end{align}

Equation~\eqref{eq:wealth-dynamics} encapsulates three disjoint events: (a) the expected wealth gain $(\kappa_i >0)$ upon successful repayment that depends on $l_i$; (b) no expected wealth gain if applicant $i$ fails to repay the loan amount $l_i$; and finally, (c) no expected wealth gain if the loan is denied.    

\textbf{ADM's Decision Making as a PeMDP.} Since the standard MDP formalism~\citep{sutton1998reinforcement} of RL remains oblivious towards the \textit{algorithmic feedback loop} causing \textit{endogenous} shifts in applicants' behaviour based on the ADM's decisions in this wealth process, we model long-term fair lending as a \textit{Performative Markov Decision Process} (PeMDP)~\citep{mandal2023performative}.
Given a policy set $\Pi: \cS \rightarrow \Delta(|\cA|)$ mapping states to (distribution over) actions, the PeMDP is defined as the set of MDPs $\cM(\Pi) \triangleq \{ \cM(\bpi) \mid \bpi \in \Pi\}$, where each MDP is a tuple $ \cM(\bpi) \triangleq (\cS, \cA, {\transition_{\bpi}}, {\reward_{\bpi}})$. 
${\transition_{\bpi}}$ and $ {\reward_{\bpi}}$ denote the transitions (Eq.~\eqref{eq:wealth-dynamics}) and rewards (Eq.~\eqref{eq:reward_def}) corresponding to a deployed policy $\bpi$. Specifically, the ADM's per-step expected reward at interest rate $\tau_{\mathrm{interest}} \in (0,1]$ is
\begin{align}\label{eq:reward_def}
     \expect_{\bpi}\left[r_{\bpi}^t\right] = \sum_{i=1}^{N_t}
    \Bigl(\tau_{\mathrm{interest}}\prob(Y_i^t = 1) - \prob(Y_i^t = 0)\Bigr)\cdot
    \prob_{\bpi}(\bA_i^t = \perp)\cdot l_i^t \cdot \papply_i^t \,.
\end{align} 
The ADM adopts a infinite-horizon discounted objective and uses a Reinforcement Learning (RL) algorithm to maximise the value $V_{\bpi} \defn   \expect_{\bpi}\left[\sum_{t=0}^{\infty} \zeta^t r^t_{\bpi}\right]$.
Optimising $V_{\bpi}$ needs reasoning about the long-term steady-state behaviour, and not just immediate returns.

A \textit{non-performative} MDP setting assumes $\papply^t$ to be stationary i.e., the application behaviour remains unaffected by the ADM's historical behaviour. 
In contrast, a \textit{performative} PeMDP setting models the application rate based on the observed approval patterns: $\papply^{t+1} = f_\psi\left(\papply^t,\bS^t, \prob_{\bpi}(\bA^t = \perp \mid \bS^t)\right)$, where $f_\psi$ is a parametric response function, and $\prob_{\bpi}(\bA^t = \perp \mid \bS^t)$ is the approval probability in the current state $\bS^t$ under the deployed policy $\bpi$. Intuitively, this captures the effect that if a demographic group is consistently rejected, they will gradually stop applying, making it harder for the ADM to learn a fair policy for that group. This feedback, where the policy shapes application behaviour, and in turn, constrains what future policies can learn, \textit{is absent from the existing MDP-based long-term fairness works}~\citep{jabbari2017fairness, hu2022achieving}. 

\begin{remark}[Why PeMDP and not Augmented MDP?]\label{remark:state-aug}
An alternative formalism is to absorb the feedback loop into an augmented state $\bar \bS^t \triangleq (\bS^t, \papply^t)$.
Since $\papply^t$ is not explicitly observed, it becomes a POMDP~\citep{smallwood1973optimal}, and standard discounted reward RL algorithms do not work.
In addition, under sustained denial, if a certain group's application rate collapses, no data from that group would be generated. Yet the dynamics continues to depend on $\bpi$'s impact on the applicants. Thus, two policies agreeing on the visited region of $\cS$ may still induce different wealth trajectories while being indistinguishable from any collected trajectory. Repairing this by state augmentation
would require encoding $\bpi$ itself in the state, which is not a well-posed RL problem.
\end{remark}

\section{\framework: Simulating Long-term Impact of Algorithmic Decisions}\label{sec:eutopia_main}
In this section, we develop the fair lending simulator \framework{} that numerically deploys the sequential lending PeMDP of Section~\ref{sec:formulation pemdp}. \framework{} realises $\cM(\bpi^t)$ and its long-term evolution as a closed loop of three blocks (Figure~\ref{fig:eutopia_pipeline}): (1) \textit{Decision block} that allows an ADM to learn a fairness-aware policy $\bpi^t$ using an RL algorithm of choice and yields the lending decisions $\bA^t$; (2) \textit{Evaluation block} that realises the consequences of credit lending actions; and (3) \textit{Social Dynamics block} that allows the ADM's actions to reshape the population dynamics that the ADM encounters in the next step. 
\begin{wrapfigure}{r}{0.4\textwidth}
\vspace*{-1.6em}\begin{center}
\resizebox{0.8\linewidth}{!}{\begin{tikzpicture}[
    font=\sffamily\small,
    >={Stealth[length=2mm, width=1.6mm]},
    box/.style={draw=#1!80!black, fill=#1!18, line width=0.7pt,
    rounded corners=2.5pt, align=center,
    minimum width=2.4cm, minimum height=0.62cm, inner sep=3pt},
    flow/.style={->, line width=0.7pt, draw=black!70, rounded corners=2pt},
    state/.style={flow, draw=cData!85!black},
    feedback/.style={flow, dashed, draw=cEval!85!black},
    score/.style={flow, dashed, draw=cAgent!85!black},
    lbl/.style={font=\sffamily\small, inner sep=2pt},
    block/.style={draw=black!45, densely dotted, line width=0.8pt,
                  rounded corners=4pt, inner sep=0.15cm},
    blocklbl/.style={font=\sffamily\scriptsize\bfseries, text=black!55,
                     inner sep=1.5pt, anchor=south west},
    midarrow/.style={decoration={markings,
        mark=at position 0.5 with {\arrow{Stealth[length=2mm, width=1.6mm]}}},
        postaction={decorate}},
]
\definecolor{cData}{HTML}{3B7BBF}   
\definecolor{cDyn}{HTML}{4E9A4A}    
\definecolor{cEval}{HTML}{D08A1E}   
\definecolor{cAgent}{HTML}{8A5CB8}  
 
\node[box=cData, minimum height=0.7cm] (loanly) at (2.0,  0.00) {(c) Data Generator};
\node[box=cDyn]   (app)    at (2.0, -1.30) {(b) Application};
\node[box=cDyn]   (wealth) at (2.0, -2.60) {(a) Wealth};
\node[box=cEval]  (eval)   at (2.0, -4.00) {Evaluation};
\node[box=cAgent] (adm)    at (2.0, -5.55) {ADM$(\bpi^t)$};
 
\node[lbl, align=center] (rl) at (0.55, -6.50) {(a) RL\\Algorithm};
\node[lbl, align=center] (fu) at (3.65, -6.50) {(b) Fairness-\\aware Utility};
 
\begin{scope}[on background layer]
  \node[block, fit=(loanly)(app)(wealth)] (bSoc)  {};
  \node[block, fit=(eval)]                (bEval) {};
  \node[block, fit=(adm)(rl)(fu)]         (bDec)  {};
\end{scope}
\node[blocklbl] at (bSoc.north west)  {III. Social Dynamics};
\node[blocklbl] at (bEval.north west) {\hspace*{-2em}II. Evaluation};
\node[blocklbl] at (bDec.north west)  {\hspace{1.5em}I. Decision};
 
\draw[flow] (adm.north)    -- node[lbl, right]
      {$\bA^{t}\!\in\!\{\perp,\not\perp\}^{N^t}$} (eval.south);
\draw[flow] (eval.north)   -- node[lbl, right] {$\{\kappa_i\}_{i=1}^{N^t}$}     (wealth.south);
\draw[flow] (wealth.north) -- node[lbl, right] {$\bx^{t+1}[1]$}     (app.south);
\draw[flow] (app.north)    -- node[lbl, right] {$N^{t+1}$}        (loanly.south);
 
\draw[state] (2.9, 0 |- loanly.north) -- (2.9, 1.05) -- (4.95, 1.05)
      -- (4.95, -5.55) -- (adm.east);
\node[lbl, above, text=cData!85!black] at (3.92, 1.05)
      {$\big(\bx^{t}, g^{t}, l^{t}\big)$};
 
\draw[feedback, midarrow] (loanly.east) -- (4.5, 0) -- (4.5, -4.0)
      -- node[lbl, above, text=cEval!85!black] {$Y^{t}$} (eval.east);
 
\draw[state] (eval.west) -- (0, -4.00)
      -- node[lbl, right, text=cAgent!85!black] {} 
      (0.0, -5.55)
      -- (adm.west);
\draw[flow] (rl.east) -| (1.65, -5.55 |- adm.south);
\draw[flow] (fu.west) -| (2.35, -5.55 |- adm.south);
\end{tikzpicture}}
\caption{\framework~: Interaction pipeline.}\label{fig:eutopia_pipeline}
\end{center}\vspace*{-4em}
\end{wrapfigure}

\textbf{Interaction protocol.} \framework{} yields a single continual trajectory to accumulate the long-term effects of algorithmic and performative feedback loop. Thus, cumulative counters such as wealth and application histories are never reset. At every step $t$:

\textit{I. Decisions:} The Data Generator generates the arriving applicant cohort $\bS^t = \{\bx^t,g^t,l^t\}=\{(\bx_i^t, g_i, l_i)\}_{i=1}^{N^t}$. The ADM observes $\bS^t$ provided by the Data Generator, and deploys a history-aware policy $\bpi^t$ rendering decisions $\bA^t \in \{\perp, \not\perp\}^{N^t}$.
   
\textit{II. Evaluation:} As repayment $\{Y_i^t\}$ of each approved applicant is observed, the ADM accumulates reward $r_{\bpi}^t$ according to Equation~\eqref{eq:reward_def}, and each repaying borrower gains $\kappa_i$ in wealth.

\textit{III. Social dynamics:} The observed wealth gains $\{\kappa_i\}_{i=1}^{N^t}$ update population wealth to $\{\bx^{t+1}_i[1]\}$ (Eq.~\eqref{eq:wealth-dynamics}). Shifted wealth, and the approval history dictate each individual's probability of next application ${p_{\textrm{apply}}}_i^{t+1}$ and the next cohort size $N^{t+1}$. Finally, the Data Generator yields the observation space $\bS^{t+1}$ for the new cohort of applicants.

\textit{Binding the loop: Learning to lend fairly.} After $T/\Delta t$ steps, referred as an episode, the chosen RL algorithm updates $\bpi^t$ such that it optimises a chosen fairness-aware utility under the shifted social dynamics.  

Steps 2--3 realise the transition kernel $\mathbf P_{\bpi}$. Holding the application probability constant removes its dependence on $\bpi$ and recovers the non-performative MDP. Thus, \framework{} remains compatible for direct comparison. We refer to Appendix~\ref{sec:pseudocodes} for the detailed pseudo-codes. In the next three subsections, we present each block of \framework~and its components, in details.

\vspace*{-.5em}\subsection{Decision Block: How the ADM Acts} \label{sec:decision}\vspace*{-.5em}
 
The Decision Block comprises the ADM, i.e. (a) the RL algorithm used to train decision-making policy $\bpi^t$, and (b) the fairness-aware utility used to train.


\textbf{(a) RL Algorithm: Learning to decide.}  The RL algorithm takes a cohort of applicants' data, and its history of loan application and repayment as an input and computes an approval policy $\bpi^t$ across applicants. The decisions are drawn as $\bA^t \sim \bpi^t (\cdot\mid \bS^t, \cH^{t-1})$, where $\cH^{t-1}$ consists of the application and repayment history of the cohort till time $t-1$. We choose to work with Beta policies to have scores for and against our yielded decisions, and keep them differentiable for further algorithm design. After each episode, we update the policy to maximise an average fairness-aware utility (Equation~\eqref{eq:utility}) estimated with the historical trajectory.

\textbf{(b) Fairness-aware utility.} The decision behaviour of a learning agent is guided by the utility function it aims to maximise. Thus, we make this choice explicit along two axes. The first axis is the \textit{fairness perspective} that leads to the question:
\vspace*{-0.3em}\begin{center}
    \textit{Should the ADM seek parity in the decisions it takes, or in the actual outcomes after deployment?}\vspace*{-0.3em}
\end{center}
This invokes two perspectives on fairness (Table~\ref{tab:fairness_rewards}): (a) \textit{Equality of Outcome} asks whether the two groups end up equally wealthy, captured by the group's mean wealth till time $t$, i.e., $\mu_g^t$, and (b) \textit{Equality of Opportunity}~\citep{hardt2016equality} asks whether equally creditworthy applicants are approved at equal rates regardless of the group, captured by group-specific true positive rate till time $t$, i.e., $\mathrm{TPR}_g^t$. These two fairness perspectives need not agree. A policy can equalise approval rates among creditworthy applicants while still producing sharply diverging wealth trajectories. We study both perspectives separately to understand their individual impacts in long-term.

The second axis is the \textit{fairness constraint} that turns the two groups' wealth/opportunity statistics into the quantity $C^t$ (Table~\ref{tab:fairness_rewards}): Social Welfare (SW) aims to maximise the aggregated wealth/opportunity~\citep{harsanyi1955cardinal}, Rawlsian Max-Min (RMM) considers maximising wealth/opportunity of the worse-off protected group~\citep{41832fbd-4a82-3c18-ae99-e02c0f77c64f}, and the Fairness Lagrangian (FL)~\citep{10.5555/3692070.3694332} penalises discrepancies between the two protected groups.
\setlength{\textfloatsep}{6pt}
\begin{table}[t!]
\centering
\caption{The constraint $C^t$ instantiated under each fairness perspective.} \label{tab:fairness_rewards}\vspace*{-.5em}
    \begin{tabular}{c c c}
    \toprule
     Fairness Constraint & Equality of Outcome & Equality of Opportunity\\
     \midrule
     Social Welfare (SW) & $\mu_{\rm blue}^t + \mu_{\rm red}^t$ & ${\rm TPR}_{\rm blue}^t + {\rm TPR}_{\rm red}^t$\\
     Rawlsian Max-Min (RMM) & $ \min \{ \mu_{\rm blue}^t,\mu_{\rm red}^t\}$ & $\min\{{\rm TPR}_{\rm blue}^t,{\rm TPR}_{\rm red}^t\}$ \\
     Fairness Lagrangian (FL) & $ \mid \mu_{\rm blue}^t - \mu_{\rm red}^t\mid$ & $  \mid {\rm TPR}_{\rm blue}^t - {\rm TPR}_{\rm red}^t\mid$ \\
    \bottomrule
    \end{tabular}
\end{table}

The ADM aims to maximise discounted cumulative reward subject to the constraint $C^t$, and thus, deploys an RL algorithm that episodically optimises the dual objective $\max_{\bpi\in\Pi}\cL(\bpi, \gamma)$, where
\begin{align}\label{eq:utility}
  \cL(\bpi, \gamma) \defn \frac{1}{T}\sum_{t=1}^{T}
  \Big( \expect_\bpi\big[r_{\bpi}^t\big] + s\gamma C^t \Big),
  \qquad
  s = \begin{cases} +1 & \text{SW, RMM},\\ -1 & \text{FL}, \end{cases}
  \quad \gamma \ge 0 .
\end{align}
The Lagrangian multiplier $\gamma$ is learned using dual ascent. It rises while the constraint is violated and decays once it is satisfied (Algo.~\ref{alg:dual}). Setting $\gamma \equiv 0$ recovers the solely profit-driven \textsc{MaxUtil} agent \citep{d2020fairness}. For each fairness constraint and perspective, we deploy and compare both RL (Policy Gradient and Soft Actor-Critic) and performative (PePG) RL algorithms (Section~\ref{sec:q1}).

\subsection{Evaluation Block: What the Decisions Realise}
\label{sec:evaluation}
Evaluation block turns decisions into outcomes. For each approved applicant $i$, repayment is drawn as $Y_i^t \sim \mathrm{Bernoulli}(1 - d_i)$. $d_i$ is the applicant's default probability obtained from the data generator, which the ADM never observes. Denied applicants generate no label emulating the selective-label bias~\citep{lakkaraju2017selective} in a controlled setting. The ADM receives the reward according to Equation~\eqref{eq:reward_def}: $\tau_{\mathrm{interest}} l_i$ on repayment, $-l_i$ on default, and $0$ on denial. The applicant gains wealth $\kappa_i = \tau_{\mathrm{gain}} l_i Y_i^t$, where $\tau_{\mathrm{gain}} > 0$ is the return from investing the loan net of interest. 

Thus, a repaid loan grows ADM profit and applicant's wealth at once, while a default costs the ADM and leaves the applicant no better off. This is the tension our RL agent must navigate while learning any of the utility functions (Equation~\eqref{eq:utility}). The realisations are sent to the Dynamics block that forms the group statistics $\mu_g^t$ and $\mathrm{TPR}_g^t$ from which $C^{t+1}$ is computed, and to the Decision block that keeps track of the cumulative counts of loans and defaults by applicants as $\cH^t$. 
\subsection{Social Dynamics Block: How the Population Responds}\label{sec:dynamics}

Social Dynamics block closes the feedback loop of \framework. Equipped with realisations from the Evaluation block, it updates wealth, decides on who applies next, and realises both with real data.
 
\textbf{(a) Wealth evolution.} Each borrower's gain is added to their wealth, $\bx_i^{t+1}[1] = \bx_i^t[1] + \kappa_i$, depending on whether $i$ applies, is approved and repays recovers Equation~\eqref{eq:wealth-dynamics}. For simplicity, we keep the financial type $(d_i,l_i,\kappa_i)$ fixed for each applicant in experiments but \framework{} provides the flexibility to vary them over time.
 
\textbf{(b) Application dynamics.} At each time step $t\in\mathbb{N}$, loan applications arrive in cohorts of $N^t$. In real-life lending systems, an approval signals access to credit and encourages further applications, while sustained denial discourages them. This phenomenon is known as ``Matthew effect''~\citep{Merton1968The}. Thus, we model each group's application intensity as a Hawkes process~\citep{hawkes}:
\begin{align}\label{eq:hawkes}
  \lambda_g^t = f(\mu_g^t) + \alpha_g \sum\nolimits_{t_j \in \mathcal H_g^t}
  e^{-\beta_g (t - t_j)} .
\end{align}
The base rate $f(\mu_g^t)$ increases with the group's mean wealth relative to the population that means people from wealthier groups apply more. The second term adds an excitation for each past approval $t_j \in \cH_g^t$ decaying at rate $\beta_g$. We set $\alpha_g = 0.3$ and $\beta_g = 2.0$ for both groups. Thus, any asymmetry among groups that emerges over time is caused by the policy, not built
into the application arrivals. 
Individual $i$ then applies with probability $\papply_i^t \propto \left(\frac{{\lambda^t_{g}}}{{N_{g}}}\right)_i$. 
This deployment realises $f_\psi$ of Section~\ref{sec:formulation pemdp}, i.e., the policy influences future applications through the approval history $\mathcal H_g$ and through wealth.
Thus, a group that is consistently denied loses both, applies less, and generates less of the data a ADM would need to serve it fairly. Conversely, loan approvals encourages people from a protected group to apply more in future, accumulate more wealth, hence preserving the Matthew effect in a sequential lending system in a performative environment.

\textbf{(c) Data Generator: realising the state.} For the ADM to act, the state $\bS^t$ must be populated with applicants. The Data Generator supplies them. Unlike a static dataset, it produces both an initial population and the parameters governing its response to the policy. It builds on the UCI Adult Income dataset~\citep{adult_2}. Hence, the initial population carries real demographic and economic heterogeneity.
For each individual $i$, it generates a wealth $\bx_i[1]$; a protected group $g_i$ (gender-- red for male and blue for female as per the dataset's documented wealth disparity); a fixed financial type $(l_i, d_i, \kappa_i)$, i.e. loan amount and \textit{fixed} default probability, and gain on repayment respectively. At each step $t\in\mathbb{N}$, Data Generator passes the arriving applicants' data $\{\bx_i^t, g_i, l_i, \kappa_i\}_{i=1}^{N^t}$ to the ADM except $\{d_i\}_{i=1}^{N^t}$ that Evaluation block uses to tally repayments $\{Y_i^t\}_{i=1}^{N^t}$ following loan approvals. With this 12-dimensional state space yielded by the data generator, the ADM's policy combines the applicants' features and repayment history $\cH^t$ for further decision making. Appendix~\ref{app:loanly} and~\ref{app:eutopia} provide further details on the data generator and \framework.

%% file: iclr_2027/Chapters/final_results.tex
\section{Design Choices Matter for Achieving Long-term Equity}\label{sec:experimental_analysis}

We now use \framework{} to study which learner and which utility can actually deliver long-term equity. We first fix the setting, and then answer three questions.
 
\textbf{Experimental setting: which learners, and why.} We deploy two RL and a performative RL algorithm to compare the effect of performative learning in long-term fairness-utility trade-off.
 
\textit{The RL algorithm: PG.} As RL algorithms, we test vanilla Policy Gradient (PG)~\citep{Williams1992} and Soft Actor-Critic (SAC)~\citep{haarnoja2018soft}. In the main text, we describe PG as it emerges as the better baseline. SAC learns a fixed score cutoff in every scenario. Its Lagrangian multiplier sits at the floor for most of deployment, and four out of six scenarios end at the same point (Appendix~\ref{app:sac}). As SAC ignores its fairness-aware utility, it cannot tell us how these utilities shape long-term outcomes. 
 
\emph{The performative RL algorithm: PePG.} Most of the performative RL algorithms~\citep{mandal2023performative,
rank2024performative} aim to converge to \emph{performative stability} i.e., a policy that is optimal for the environment it induces. A stable policy can be far from the optimal one~\citep{perdomo2020performative,
miller2021outside}. ADM aims to deploy the policy that maximises $V_{\bpi}$ over the whole family $\cM(\Pi)$, i.e. achieving \textit{performative optimality}. Performative Policy Gradient (PePG)~\citep{basu2025performativepolicygradientoptimality} exactly solves this. It differentiates through the population's response to the policy. It is the performative version of PG, thus making the comparison fair and accurate. \textit{We call them RL and PeRL agents for the rest of the paper.} Note that there exists no performative variant for SAC. 
 
\textit{Training details.} Both agents use a Beta policy over the approval probability, a two-layer MLP (width $128$) and AdamW optimiser~\citep{loshchilov2017decoupled} (learning rate $10^{-3}$). Each group has $N_M = N_F = 12{,}000$ individuals. An episode spans $100$ time units in $200$ decision steps. Each agent is trained for $1{,}000$ episodes, and then deployed for $3{,}000$ episodes in \framework. The multiplier $\gamma^t$ is set via dual ascent (step $0.1$). Every experiment is run and averaged across $10$ seeds. Further training details and hyperparameters are described in Appendix~\ref{appendix:hyperparameters}.
 
\textbf{Performance metrics.} We track three macro metrics while simulating the long-term interactions between the RL and PeRL agents and \framework~ across the two fairness perspectives and three fairness constraints (Table~\ref{tab:fairness_rewards}).
\textit{(a) Wealth gap}: $\Delta_\mu^t \triangleq |\mu_M^t - \mu_F^t|$. $\Delta_\mu^t=0$ indicates outcome parity. \textit{(b) Inequality ratio}:
$\rho^t \triangleq (\mu_M^t - \mu_M^0)/(\mu_F^t - \mu_F^0)$. $\rho^t=1$ means both groups grow at the same rate, and $\rho^t > 1$ means the male group grows faster. \textit{(c) Cumulative profit}: $\mathrm{CP}^t \triangleq \sum_{\tau \le t} r_{\bpi}^\tau$. This is the ADM's utilitarian gain, while the other two metrics capture equity. 

Specifically, we ask three empirical questions:
\begin{tcolorbox}[top=2pt,bottom=2pt,left=2pt,right=2pt]
\textbf{Q1.} Does being reactive to the performative shifts in the environment yields better equitable outcomes for the applicants in long-term?\\
\textbf{Q2.} Which fairness-aware utility function (fairness perspective + constraint) should be chosen by a practitioner to ensure fairness-utility trade-off in long-term? \\
\textbf{Q3.} How much different fairness-aware utilities can ensure intra-group fairness in long-term?
\end{tcolorbox}

\subsection{Q1: Performative RL for Better Long-term Fairness-Utility Trade-off}\label{sec:q1}

To address Q1, we report the best-performing agents per metric achieved by the end of deployment in Table~\ref{tab:app_rl_vs_perl_main}. Trajectory-level results and detailed discussions are in Appendix~\ref{sec:passive_vs_performative}.

\textit{Performative RL leads to better fairness-utility trade-off.} PeRL performs significantly better in terms of higher cumulative profit, smaller wealth gap, and an inequality ratio closer to one than RL (Table~\ref{tab:app_rl_vs_perl_main}). Under Equality of Opportunity, PeRL earns $6$--$7\times$ RL's profit while cutting the wealth gap by a factor of $3$--$15$, and under FL with Equality of Outcome, it reduces the wealth gap $16\times$ relative to RL, with an inequality ratio $\rho = 1.045$ against RL's $0.19$.

\textit{Outcome- and performativity-oblivious RL converges to approve most applicants.} Except the case of FL under Equality of Outcome, RL's end-of-deployment performance metrics are surprisingly indistinguishable from one another. These are, to within seed noise, the metrics demonstrated by Always Approve rule (Appendix~\ref{app:heuristic}). Under RMM and SW with either perspective, and under FL with Equality of Opportunity, RL eventually settles on approving nearly every applicant of both groups. A lender who approves everyone treats the groups identically by construction. This leads to lower approval rate disparity, while the Matthew effect drives the wealth gap and $\rho$ to similar values as a heuristic policy. Only in case of FL under Equality of Outcome, it recalibrates. As a result, the Female group ends up growing five times faster than the red one ($\rho = 0.19$). 



\input{iclr_2027/Chapters/Best_Tables_New}
\vspace*{-.5em}\subsection{Q2: A Practitioner's Guide to Designing a Fairness-aware Utility}\label{sec:q2}
\input{iclr_2027/Chapters/learning_fairness}

\vspace*{-.5em}\subsection{Q3: Equality of Outcome Matters for Intra-group Fairness}\label{sec:q3}

\input{iclr_2027/Chapters/Inclusivity}

%% file: iclr_2027/Chapters/Best_Tables_New.tex
\setlength{\textfloatsep}{8pt}
\begin{table}[t!]
\centering
\caption{End-of-deployment Performance of RL and PeRL agents across fairness-aware utilities (mean $\pm$ std over $10$ seeds after $3000$ episodes). For each metric, the better agent is in \textbf{bold}. $^{\star\star}$ denotes the winners for each metric.}\label{tab:app_rl_vs_perl_main}\vspace{-.5em}
\resizebox{\textwidth}{!}{%
\begin{tabular}{llcccccc}
\toprule
\multirow{2}{*}{\textbf{Perspective}} & \multirow{2}{*}{\textbf{Constraint}}
    & \multicolumn{2}{c}{\textbf{Cumulative Profit ($\times10^6$)}}
    & \multicolumn{2}{c}{\textbf{$|$Wealth Gap$|$}}
    & \multicolumn{2}{c}{\textbf{Inequality Ratio $\rho$}} \\
\cmidrule(lr){3-4}\cmidrule(lr){5-6}\cmidrule(lr){7-8}
    & & RL & PeRL & RL & PeRL & RL & PeRL \\
\midrule
 & FL & 0.374 $\pm$ 0.385 & $\mathbf{4.909 \pm 0.579}$ & 745.0 $\pm$ 1288.3 & $\mathbf{46.5 \pm 82.1}^{\star\star}$ & 0.190 $\pm$ 0.132 & $\mathbf{1.045 \pm 0.032}^{\star\star}$ \\
Outcome & RMM & 1.096 $\pm$ 0.096 & $\mathbf{1.267 \pm 0.058}$ & 3190.0 $\pm$ 68.9 & $\mathbf{3081.9 \pm 26.5}$ & 10.198 $\pm$ 0.032 & $\mathbf{0.101 \pm 0.000}$ \\
 & SW & 1.103 $\pm$ 0.071 & $\mathbf{4.721 \pm 1.346}$ & 3169.8 $\pm$ 82.5 & $\mathbf{1956.0 \pm 1012.6}$ & 10.294 $\pm$ 0.082 & $\mathbf{0.138 \pm 0.048}$ \\
\midrule
 & FL & 1.131 $\pm$ 0.062 & $\mathbf{7.443 \pm 0.410}$ & 3154.8 $\pm$ 103.9 & $\mathbf{1086.8 \pm 332.2}$ & 10.328 $\pm$ 0.045 & $\mathbf{3.981 \pm 0.568}$ \\
Opportunity & RMM & 1.105 $\pm$ 0.100 & $\mathbf{6.711 \pm 1.229}$ & 3177.3 $\pm$ 55.4 & $\mathbf{207.9 \pm 1328.7}$ & 10.231 $\pm$ 0.043 & $\mathbf{1.229 \pm 0.643}$ \\
 & SW & 1.118 $\pm$ 0.099 & $\mathbf{6.757 \pm 1.137}$ & 3115.9 $\pm$ 136.1 & $\mathbf{535.2 \pm 1187.0}$ & 10.304 $\pm$ 0.079 & $\mathbf{1.776 \pm 0.878}$ \\
\midrule
\multicolumn{2}{l}{MaxUtil} & 1.092 $\pm$ 0.101 & $\mathbf{7.495 \pm 0.194^{\star\star}}$ & 3137.6 $\pm$ 125.1 & $\mathbf{1181.2 \pm 110.0}$ & 10.369 $\pm$ 0.105 & $\mathbf{4.672 \pm 0.195}$ \\
\midrule
 \multicolumn{2}{l}{\textbf{Winner}} & \multicolumn{2}{c}{\textbf{PeRL$+$Opportunity$+$FL}} & \multicolumn{2}{c}{\textbf{PeRL$+$Outcome$+$FL}} & \multicolumn{2}{c}{\textbf{PeRL$+$Outcome$+$FL}} \\
\bottomrule
\end{tabular}}
\end{table}

%% file: iclr_2027/Chapters/learning_fairness.tex
Section~\ref{sec:q1} establishes the fact that deploying performative agents yields better long-run equitable outcomes in society. Now, using performative RL, we investigate which combination of \textit{fairness perspective and constraint} (Table~\ref{tab:fairness_rewards}) should be learnt for obtaining the best fairness-utility trade-off. 

\textit{FL under Equality of Outcome renders best long-term equity.}
This is the only utility function that closes the wealth gap. It leads to a gap of $46.5$ and inequality ratio $\rho = 1.05$, which are the best values across the grid in Table~\ref{tab:app_rl_vs_perl_main}. Moving from evaluating FL under Opportunity vs. Outcome shrinks the wealth gap $23\times$ and brings down $\rho$ from $3.98$ to $1.05$. It still yields two thirds of the \textsc{MaxUtil}'s profit ($4.91$ against $7.50$ Million). Thus, the utility trade-off is modest. 

FL is the only constraint that penalises the gap between the protected groups rather than optimising an aggregate of their statistics. SW, on the other hand maximises a sum, oblivious to how a fixed total is distributed across groups~\citep{harsanyi1955cardinal}. RMM is flat in the better-off group's statistic~\citep{41832fbd-4a82-3c18-ae99-e02c0f77c64f}, so the learning signal is one-sided and vanishes wherever the $\argmin$ is stable. The resulting behaviour is visible in terms of $\rho$ being far away from $1$, and huge wealth gaps.

\textit{Equality of Opportunity reproduces the delayed impact of fairness.} Equality of Opportunity agents fail in long-term exactly as the literature predicts. FL under Opportunity ends with a gap of $\sim 1{,}087$ and $\rho = 3.98$. This is close to \textsc{MaxUtil} ($1{,}181.2$, $\rho = 4.67$), i.e., an agent that ignores fairness altogether. The reason is that TPR parity is a decision-level criterion. As soon as the agent meets it early: approval-rate disparity falls to $0.01$ (Appendix~\ref{app:perl_metric_by_metric}), and the constraint no more binds. But equal approval rates are not equal outcomes. As the male group starts wealthier, they apply more. Over deployment, they file four times more applications and receives four times
the loans per person (Appendix~\ref{app:balanced}). Its wealth pulls ahead. This is known as the \textit{``delayed impact of fairness''}-- a criterion satisfied at the decision-level harms long-term equity~\citep{liu2018delayed}. In the non-performative case, \citet{d2020fairness} observed that an Equal-Opportunity (EO) agent ensuring instantaneous decision-level parity causes long-term harm in outcomes of the disadvantageous group, while \textsc{MaxUtil} agent does not. \framework~reproduces the results of EO agent for end-of-deployment macro metric and trajectory-level micro metrics  (Section~\ref{sec:q3}), while demonstrating \textsc{MaxUtil} agent to be equally harmful in long-term. \textit{FL under Outcome escapes it by constraining the outcome disparity itself as the constraint cannot be satisfied while the harm accumulates.}

\begin{wrapfigure}{r}{0.5\textwidth}\vspace*{-1.5em}
    \centering
    \resizebox{!}{0.12\textwidth}{
    \begin{minipage}{0.45\textwidth}
        \centering
        \includegraphics[width=\textwidth]{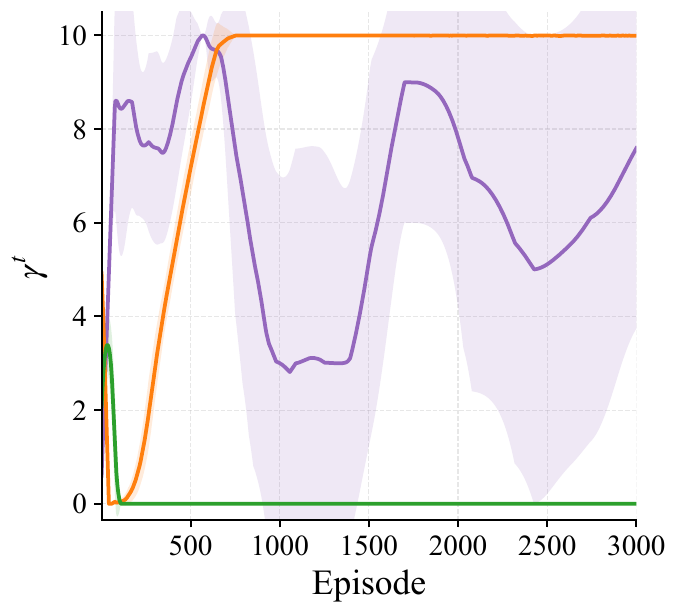}\vspace{-.5em}
        \subcaption{Equality of Outcome}
    \end{minipage}\hfill
    \begin{minipage}{0.45\textwidth}
        \centering
        \includegraphics[width=\textwidth]{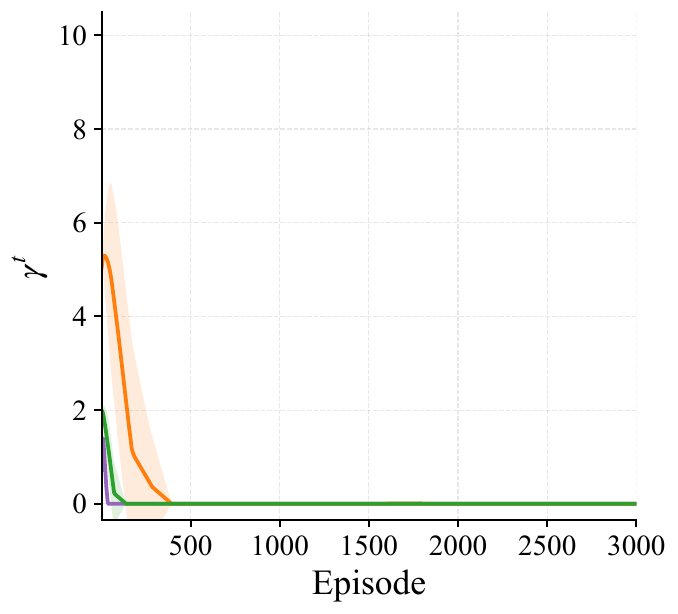}\vspace{-.5em}
        \subcaption{Equality of Opportunity}
    \end{minipage}}\\[0.5em]
    \includegraphics[width=0.3\textwidth]{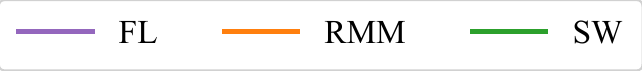}
    \caption{Evolution of the multiplier $\gamma^t$ over deployment for PeRL agent.}\label{fig:lambda_overlay_main}\vspace*{-3.5em}
\end{wrapfigure}

\textit{Only FL under Equality of Outcome keeps the fairness-utility trade-off alive.} 
Figure~\ref{fig:lambda_overlay_main} tracks the multiplier $\gamma^t$ quantifying the balance between profit and fairness. For four out of six utility functions, it decays to floor as $t$ grows. At this point, the fairness constraint no longer binds and the agent maximises profit alone-- essentially becoming a utilitarian \textsc{MaxUtil} agent. Its end-of-deployment performance is then guided by the bias acquired early and further amplified by the Matthew effect. FL and RMM under Outcome are the exceptions. While RMM's Lagrangian multiplier converges around $500$ episodes, $\gamma^t$ for FL keeps evolving throughout. This persistent balancing between profit and fairness is precisely what a fairness-aware utility is meant to reflect. While the TPR gap is met once approval rates match, the wealth gap works differently. The Matthew effect keeps pushing it to open, and thus, the outcome-based constraint must keep binding like FL under Outcome.

Another concern might surface from Table~\ref{tab:app_rl_vs_perl_main}: for RMM and SW, every macro metric is better under Opportunity than Outcome unlike FL. A practitioner choosing utilities by macro metrics alone might prefer them. Here, we suggest looking at how credit is distributed inside the groups and over time.


%% file: iclr_2027/Chapters/Inclusivity.tex
\begin{figure}[t!]
    \centering\vspace*{-1em}
    \begin{minipage}{0.32\textwidth}
        \centering
        \includegraphics[width=0.6\textwidth]{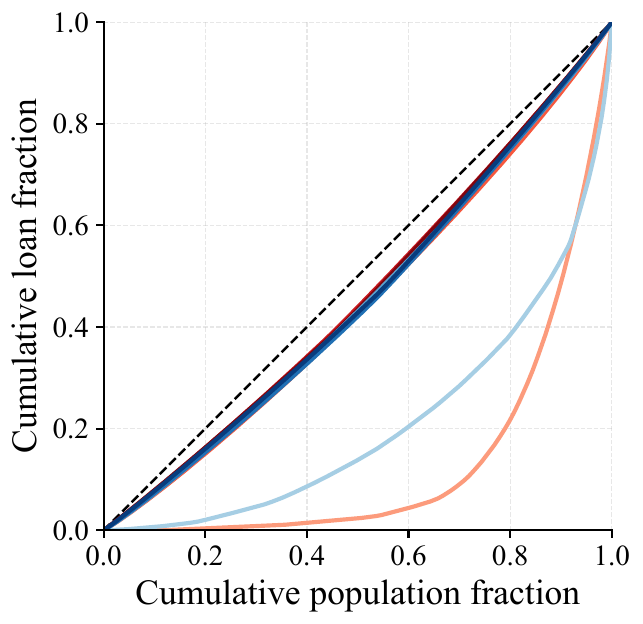}\vspace{-.5em}
        \subcaption{FL, Outcome}\label{fig:lorenz_out_fl_main}
    \end{minipage}\hfill
    \begin{minipage}{0.32\textwidth}
        \centering
        \includegraphics[width=0.6\textwidth]{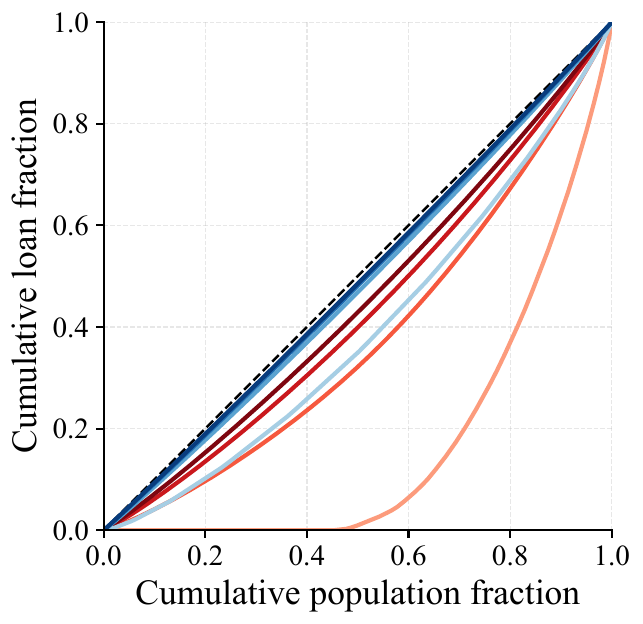}\vspace{-.5em}
        \subcaption{RMM, Outcome}\label{fig:lorenz_out_rmm_main}
    \end{minipage}\hfill
    \begin{minipage}{0.32\textwidth}
        \centering
        \includegraphics[width=0.6\textwidth]{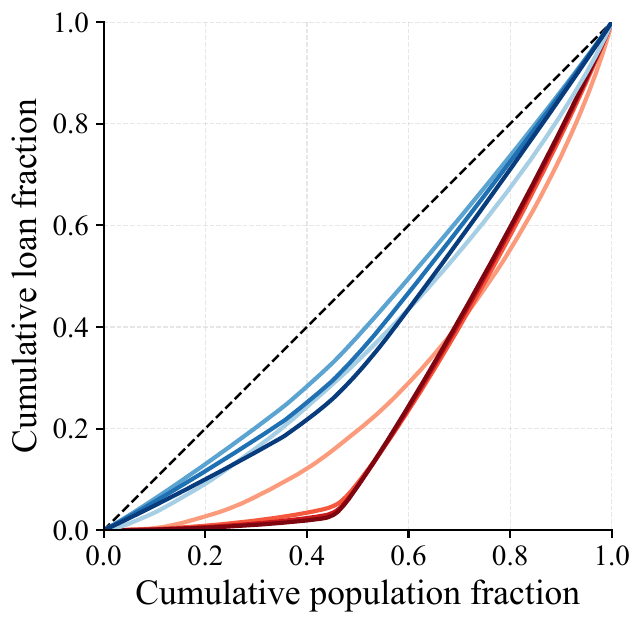}\vspace{-.5em}
        \subcaption{SW, Outcome}\label{fig:lorenz_out_sw_main}
    \end{minipage}\\
    \begin{minipage}{0.32\textwidth}
        \centering
        \includegraphics[width=0.6\textwidth]{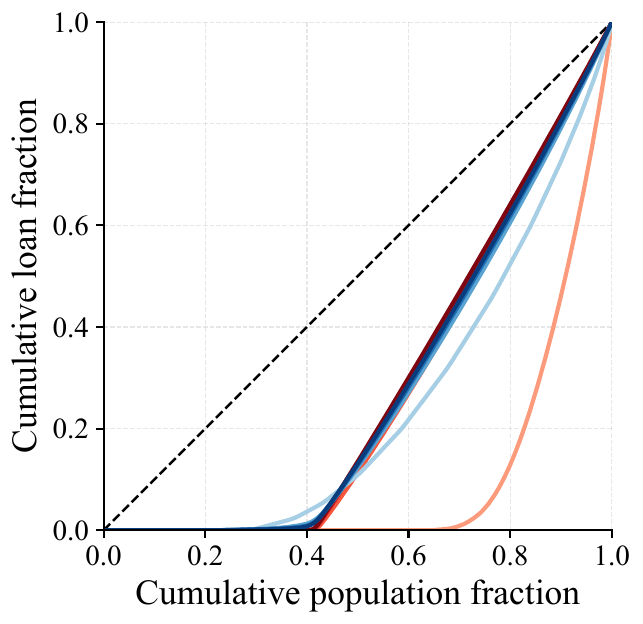}\vspace{-.5em}
        \subcaption{FL, Opportunity}\label{fig:lorenz_opp_fl_main}
    \end{minipage}\hfill
    \begin{minipage}{0.32\textwidth}
        \centering
        \includegraphics[width=0.6\textwidth]{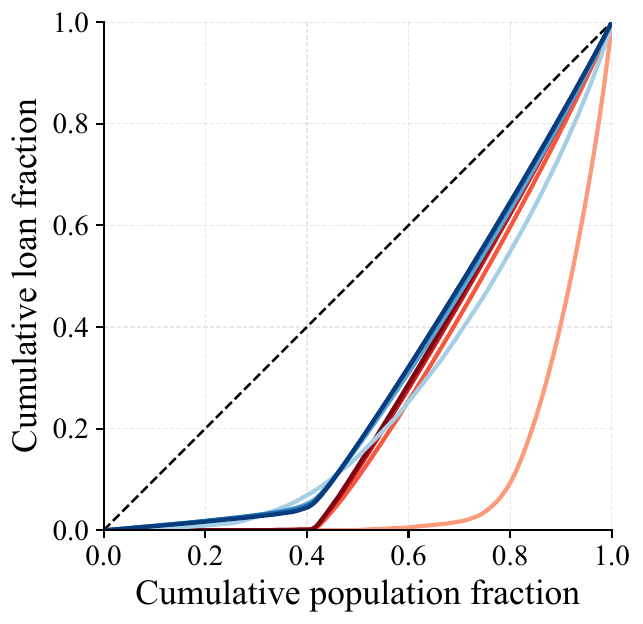}\vspace{-.5em}
        \subcaption{RMM, Opportunity}\label{fig:lorenz_opp_rmm_main}
    \end{minipage}\hfill
    \begin{minipage}{0.32\textwidth}
        \centering
        \includegraphics[width=0.6\textwidth]{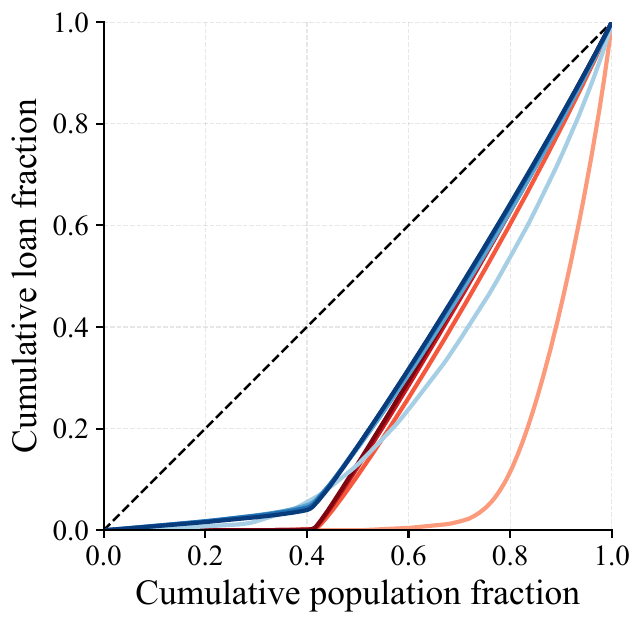}\vspace{-.5em}
        \subcaption{SW, Opportunity}\label{fig:lorenz_opp_sw_main}
    \end{minipage}\\[0.5em]
    \includegraphics[width=0.2\textwidth]{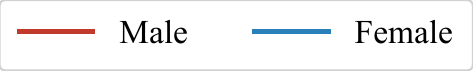}\hspace{1em}
    \includegraphics[width=0.6\textwidth]{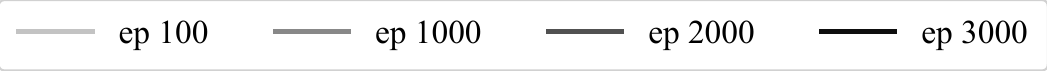}\vspace{-.4em}
    \caption{\textbf{Lorenz curves of cumulative loans for PeRL agent across utilities and over time.}}\label{fig:lorenz_main}\vspace*{-.4em} 
\end{figure}

Metrics such as wealth gap and inequality ratio measure fairness between groups, but they are oblivious towards how credit access is distributed within groups. A policy can achieve near-perfect group-level parity while systematically concentrating loans among the same high-wealth individuals in each group. To surface this, we conduct a micro-analysis. We report \textit{Lorenz curves} to track cumulative loan concentration within groups over deployment (Figure~\ref{fig:lorenz_main}), where the diagonal represents that the whole population in a group gets loan at least once. In Appendix~\ref{app:micro}, we also report the \textit{reach rate} plots for tracking the fraction of unique recipients per episode relative to the full group size, and cumulative loan distributions that reveal the within-group skewness of loans received in short-term vs. the full deployment horizon.

\textit{Equality of Outcome is inclusive; Equality of Opportunity is selective.} Under Equality of Outcome, almost everyone in each group gets an opportunity. Almost no individual goes without a chance of getting loan under FL or RMM, and only $6\%$ of men miss under SW. Under Opportunity, $15$ to $26\%$ of each group never receive a loan rendering the flat segments in corresponding Lorenz curves. 
The reason is what each perspective pays for. An Opportunity-driven utility pays for approving the creditworthy and is oblivious to the rest. An Outcome-driven utility pays for wealth created across the group. Hence, leaving part of it unserved is costly. 

\textit{Under Opportunity, the disadvantaged group stops applying.} The female reach rate peaks early, at $\sim 5-7\%$, then declines to $1.8\%$ under FL, $3.5\%$ under SW and $5\%$ under RMM. The male reach rate rises throughout, to $9$-$12\%$. TPR parity is therefore maintained on a shrinking pool of applicants. This is representation disparity~\citep{hashimoto2018fairness}, and it is the micro-level trace of the decaying Lagrangian multiplier (Sec.~\ref{sec:q2}). A decision-level criterion constrains who
is approved among those who apply. It does not consider harms on the ones yet to apply~\citep{zhang2019group}.

Macro metrics place RMM under Opportunity second based on end-of-deployment outcomes ($207.9$, $\rho = 1.23$). Micro-analysis shows that it leaves a quarter of men and a sixth of women unserved, and lets women's access decay.\textit{ Only FL under Outcome is equitable between groups, within groups, and over time.} An aggregated end-of-deployment evaluation cannot surface this effect~\citep{d2020fairness}. Thus, we advocate for FL under Equality of Outcome as \textit{the} fairness-aware utility, since it is the one utility which pushes the ADM to be fair even when its instantaneous criterion is met, and hence, finally refutes the delayed impact of fairness~\citep{liu2018delayed}.

%% file: iclr_2027/Chapters/Experiments.tex




%% file: iclr_2027/Chapters/Discussion_FutureWork.tex
\vspace*{-.5em}\section{Discussions, Limitations, and Open Questions}\label{sec:discussion}\vspace*{-0.5em}
We revisit the long-term fairness achieved by an Algorithmic Decision Maker (ADM) interacting with a population. We built the first performative fair lending simulator \framework{} to observe: (a) a performative RL agent beats standard RL counterparts on profit, wealth gap and inequality ratio under all six fairness-aware utilities; (b) to ensure long-term equity, fairness must be measured based on outcomes rather than decisions; (c) while Equality of Opportunity based fairness objectives leave a fifth to a quarter of each group without a single loan and let the disadvantaged group's access decay throughout deployment, FL under Outcome is the only long-term utility ensuring equity between groups, within groups, and along the trajectory. It is, therefore, \textit{the} utility design that escapes delayed impact of fairness in long-term~\citep{liu2018delayed}.

Anticipating outcomes of long-term interactions is hard even in the simplest setups. \citet{liu2018delayed} show that a one-step feedback model already suffices for a static criterion to harm the group it protects. \citet{d2020fairness} show that single-step analyses fail to predict long-term behaviour even in a two-group lending model with a fixed applicant pool. Each additional feedback channel broadens the possibilities between what is modelled and what can be claimed. We, therefore, position \framework{} as a \textit{concrete test-bed for simulating long-term interaction between a ADM and a reactive population}, and not a universal recipe for ensuring long-term
fairness. Rather, its purpose is to provide a comparative guideline for algorithms aiming for long-term equity: specifically, which objectives and learners sustain/degrade under the algorithmic feedback loop, in what way, and at what cost.
In addition, \framework{} serves as a test-bed for performative RL algorithms.
Our findings also call for further theoretical and algorithmic investigations of sequential decision making with outcome-based fairness measures.

%% file: iclr_2027/Appendix/Notation_table.tex
\section{Notations}

\begin{table}[h]
\centering
\small
\caption{Notation used throughout the paper.}
\label{tab:notation}
\renewcommand{\arraystretch}{1.15}
\begin{tabularx}{\linewidth}{@{}l>{\raggedright\arraybackslash}X@{}}
\toprule
\textbf{Symbol} & \textbf{Description} \\
\midrule
\multicolumn{2}{@{}l}{\textit{Population and applicants}}\\
$N$, $N_g$ & Total population size, and size of protected group $g$ \\
$N^t$ & Number of applicants arriving at step $t$ (cohort size) \\
$g_i \in \{\mathrm{red},\mathrm{blue}\}$ & Protected group of applicant $i$ (male, female) \\
$\bx_i \in \reals^d$, $\bx_i[1]$ & Feature vector of applicant $i$, and its first coordinate, the wealth \\
$l_i \in \reals_{>0}$ & Loan amount requested by applicant $i$, fixed across its applications \\
$d_i \in (0,1)$ & Default probability of applicant $i$; latent, never observed by the ADM \\
$\kappa_i = \tau_{\rm gain} l_i$ & Wealth gained by applicant $i$ upon repayment \\
$c_i \in \{0,1\}$ & Creditworthiness label; used only to evaluate $\mathrm{TPR}$, never observed \\
$Y^t_i \in \{0,1\}$ & Repayment outcome, $Y_i^t=1$ on repayment; observed only if approved \\
\midrule
\multicolumn{2}{@{}l}{\textit{Interaction and decisions}}\\
$t$, $T$, $\Delta t$ & Decision step, episode horizon in time units, and time increment \\
$\bS^t = \{\bx^t_i, g_i, l_i\}_{i=1}^{N^t}$ & State: the context of all applicants arriving at step $t$ \\
$\bA^t \in \{\perp,\not\perp\}^{N^t}$ & Action vector; $\perp$ denotes approval and $\not\perp$ denial \\
$p_i \in [0,1]$ & Approval probability output by the policy for applicant $i$ \\
$\pi_t$, $\Pi$ & Deployed policy at step $t$, and the policy set \\
$\papply_i^t$ & Probability that applicant $i$ applies at step $t$ \\
$f_\psi$ & Performative response function governing $\papply_i^{t}$ \\
$\cM(\pi) = (\cS,\cA,\mathbf{P}^{\pi},r^{\pi})$ & MDP induced by the deployed policy $\pi$ \\
$\cM(\Pi) = \{\cM(\pi)\mid \pi\in\Pi\}$ & The PeMDP, i.e.\ the family of induced MDPs \\
$\tau_{\rm interest}$, $\tau_{\rm inv}$ & Lender's interest rate and applicant's return on investing the loan \\
$\tau_{\rm gain} = \tau_{\rm inv}-\tau_{\rm interest}$ & Net gain rate of a repaid loan, strictly positive \\
$r^t_{\pi}$, $V_{\pi}, \zeta$ & Lender's per-step reward, and its average-reward objective, discount factor \\
\midrule
\multicolumn{2}{@{}l}{\textit{Social dynamics}}\\
$\mu^t_g$, $\bar{\mu}^t$ & Mean wealth of group $g$ at step $t$, and population mean wealth \\
$\lambda^t_g$ & Hawkes application intensity of group $g$ \\
$f(\mu^t_g)$ & Base arrival rate, increasing in the group's relative wealth (Matthew effect) \\
$\alpha_g$, $\beta_g$ & Excitation strength and decay rate of the Hawkes process \\
$\cH_g$ & History of past approvals of group $g$ \\
$\left(\frac{{\lambda^t_{g}}}{{N_{g}}}\right)_i$ & group-specific parametric response function for an individual $i$ in protected group $g$\\  
\midrule
\multicolumn{2}{@{}l}{\textit{Fairness-aware utility}}\\
$\mathrm{TPR}^t_g$ & True positive rate: creditworthy applicants of group $g$ approved by step $t$ \\
$C^t$ & Constraint statistic (Table~\ref{tab:fairness_rewards}) \\
SW, RMM, FL & Social Welfare, Rawlsian Max-Min, and Fairness Lagrangian constraints \\
$\gamma^t \ge 0$, $s = \pm 1$ & Lagrange multiplier, learned by dual ascent, and its sign per constraint \\
$\cL(\pi,\gamma)$ & Fairness-aware utility optimised by the ADM \\
\textsc{MaxUtil} & Profit-only agent, recovered by $\gamma^t \equiv 0$ \\
\midrule
\multicolumn{2}{@{}l}{\textit{Evaluation metrics}}\\
$\Delta^t_{\mu} = |\mu^t_M - \mu^t_F|$ & Wealth gap; $0$ indicates parity of outcomes \\
$\rho^t = \frac{\mu^t_M-\mu^0_M}{\mu^t_F-\mu^0_F}$ & Inequality ratio; $\rho^t=1$ means equal growth rates \\
$\mathrm{CP}^t = \sum_{\tau\le t} r^{\tau}_{\pi}$ & Cumulative profit of the lender \\
Reach rate & Fraction of a group receiving at least one loan in an episode \\
Gini, Lorenz curve & Within-group concentration of cumulative loans \\

\bottomrule
\end{tabularx}
\end{table}

%% file: iclr_2027/Chapters/Related_works.tex
\section{Detailed Related Works}

The study of fairness in machine learning has evolved from static, snapshot-based metrics toward a dynamic understanding of how algorithms reshape the populations they serve.

\textbf{Fairness in modern Machine Learning.}
Early foundational work established formal definitions of algorithmic fairness, including individual fairness~\citep{dwork2012fairness} and group fairness criteria such as equalized odds and calibration \citep{hardt2016equality}. A key theoretical insight from this era is that many natural fairness criteria are mutually exclusive except in degenerate cases~\citep{chouldechova2017fair, kleinberg2016inherent}. More critically, these definitions treat fairness as a property of a single decision point, ignoring feedback dynamics inherent in repeated deployment. Short-term fairness metrics in lending can reduce the long-term welfare of disadvantaged groups by denying them opportunities to build credit history~\citep{liu2018delayed} . Similarly, instantaneous fairness constraints create feedback loops: a policy trained on historically biased data may perpetuate exclusion, as the population distribution at time $t+1$ depends on decisions made at time $t$~\citep{wyllie2024fairness} . These findings motivate reasoning about fairness across extended temporal horizons.

\textbf{Fairness in Sequential Decision making.}
To address temporal dependencies, researchers recast fair decision-making as a sequential problem. Initial work in the bandit setting \citep{joseph2016fairness, nanda2021fairness} introduced fairness constraints that account for shifting outcome distributions. The natural generalization is to formalise the problem as a Markov Decision Processes (MDPs), where the agent must balance immediate fairness requirements against long-term consequences encoded in state dynamics. Subsequent work developed tractable algorithms: Lyapunov-based methods for high-probability constraint satisfaction \citep{hu2022achieving}, primal-dual approaches for online fair policy learning \citep{deng2022reinforcement, wang2024online}, and sample-efficient methods for unknown dynamics~\citep{ju2023achieving}. A common assumption across this literature, however, is that transition dynamics are exogenous-- the environment evolves according to fixed, policy-independent rules. This assumption breaks down when the population adapts its behaviour in response to the deployed policy~\citep{miller2020strategic,10.1145/3593013.3594006}, or the underlying state dynamics of the MDP depends on the deployed policy directly~\citep{mandal2023performative}.

\textbf{Fairness in Performative Environment and Long-term Equity.}
The intersection of long-term fairness and performative prediction \citep{perdomo2020performative} represents the current frontier. Performative prediction recognises that deploying a model changes the distribution it was trained on: a loan approval model affects who applies for loans in future, and a crime recidivism predictor influences defendant behaviour~\citep{dressel2018accuracy}. Furthermore, a foundational analysis of different policy objectives demonstrates the steady-state distribution of welfare across demographic groups~\citep{wu2025policy}. It is important to note that the choice of objective has profound implications for long-run inequality. However, pre-existing frameworks do not account for the \textit{performative} shift in transition dynamics that arises when populations respond strategically to deployed policies. Recent theoretical advances have begun to address this gap by establishing conditions for performative stability in discrete MDPs~\citep{PRL,MPRL}, as well as by developing policy gradient methods with convergence guarantees for performative optimality~\citep{basu2025performativepolicygradientoptimality}.

%% file: iclr_2027/Appendix/Theory.tex
\clearpage

\section{Reactivity of a Learner}
\label{app:reactivity}

We propose a structural classification of reactivity for a learning algorithm operating in a performative environment. Standard analyses of long-term fairness compare policies by their cumulative profit, wealth gap, or inequality ratio, but none of these isolate how a learner lets information from the environment propagate into its decision rule. Reactivity is the missing axis: it asks whether the learner's responsiveness coefficient is itself updated as the environment shifts, or remains pinned to its training-time value.

\paragraph{Model.}
We model the learner-environment interaction as a coupled linear system with exogenous input $Z_t$ and i.i.d.\ zero-mean Gaussian noises $\xi_t$ (environment) and $\nu_t$ (learner):
\begin{align}
X_{t+1} &= m\, X_t + q\, U_t + \chi\, Z_t + \xi_t, \label{eq:react-x} \\
U_{t+1} &= a\, U_t + b\, X_t + \nu_t, \label{eq:react-y}
\end{align}
with $|m|, |a| < 1$ for boundedness. $X_t$ denotes the environment state and $U_t$ the learner's decision variable. The coefficients $(m, q, \chi)$ of the environment and $(a, b)$ of the learner are specific to this section.

\paragraph{Trajectory unrolling.}
Substituting \eqref{eq:react-x} into \eqref{eq:react-y} recursively and gathering terms by source yields a three-part decomposition,
\begin{equation}
\label{eq:react-unroll}
U_{t+1} \;=\; a\, U_t
\;+\; q \underbrace{\sum_{s=0}^{t-1} m^{t-1-s} b_s U_s}_{\text{(Past)}}
\;+\; \underbrace{\sum_{s=0}^{t-1} m^{t-1-s} b_s \chi_s Z_s \;+\; b_0\, m^t X_0}_{\text{(Data)}}
\;+\; \underbrace{\sum_{s=0}^{t-1} m^{t-1-s} b_s \xi_s + \nu_t}_{\text{(Noise)}}.
\end{equation}
The (Past) term sums contributions from earlier learner states; the (Data) term is the running sum of inputs $\{Z_s\}_{s<t}$ weighted by the per-step responsiveness $b_s$, plus the propagated initial condition $b_0 m^t X_0$; and finally the (Noise) term collects the noise from both the environment and the learner.\textit{ Reactivity is governed by the (Data) term, and specifically by where $b_s$ enters the temporal sum.}

\paragraph{Local sensitivities.}
The two channels through which information moves are the partial derivatives
$\chi_t \;\triangleq\; \frac{\partial X_t}{\partial Z_t}, \quad b_t \;\triangleq\; \frac{\partial U_t}{\partial X_t}$,
both allowed to be time-varying in \eqref{eq:react-x}--\eqref{eq:react-y}. The qualitative behaviour of the learner is determined by whether $b_t$ remains pinned to its initial value $b_0$ or evolves with the dynamics.

\begin{definition}[Reactivity classes]
\label{def:reactivity}
Under \eqref{eq:react-x}--\eqref{eq:react-y}, a learner is (i) \textbf{non-reactive} if $b_t \equiv 0$; (ii) \textbf{passive} if $b_t = \Phi(X_t)$ for some $t$-independent function $\Phi$; (iii) \textbf{reactive} if $b_t = \Phi_t(X_t)$ with $\Phi_t$ depending genuinely on $t$.
\end{definition}

\begin{figure}[t!]
  \centering
  \begin{subfigure}{0.32\linewidth}
    \includegraphics[width=\linewidth]{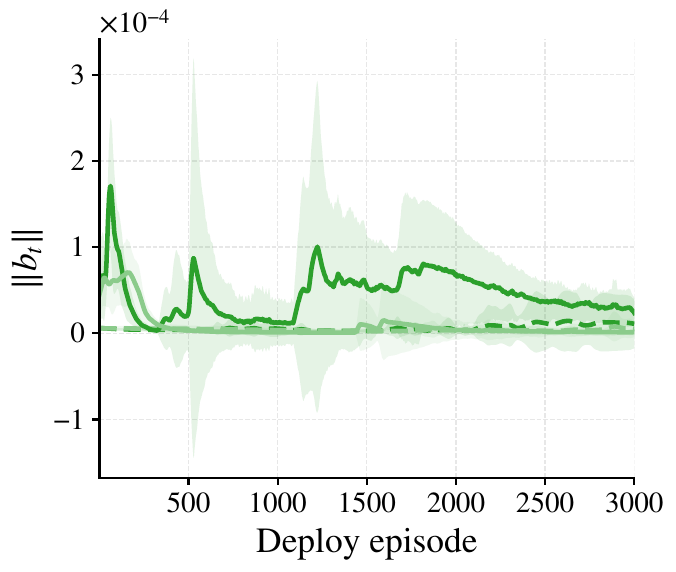}
    \caption{Social Welfare (SW)}
  \end{subfigure}\hfill
  \begin{subfigure}{0.32\linewidth}
    \includegraphics[width=\linewidth]{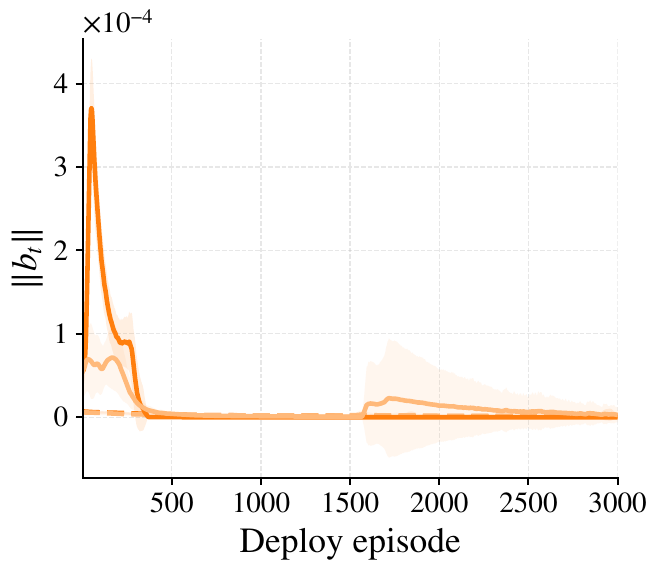}
    \caption{Rawlsian Max-Min (RMM)}
  \end{subfigure}\hfill
  \begin{subfigure}{0.32\linewidth}
    \includegraphics[width=\linewidth]{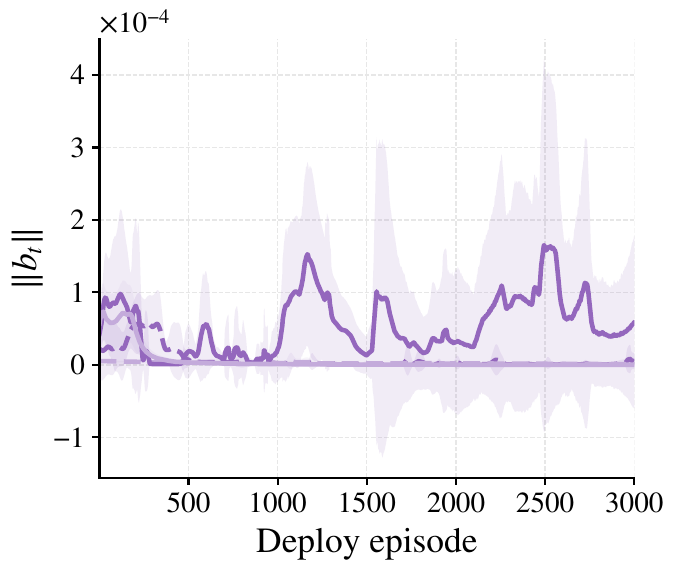}
    \caption{Fairness Lagrangian (FL)}
  \end{subfigure}\\[0.4em]
  \includegraphics[width=0.75\linewidth]{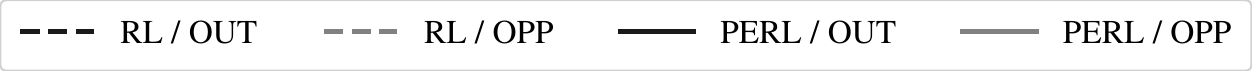}
  \caption{\textbf{Episodic reactivity.} $\|b_t\|$ over $E=3{,}000$ deploy episodes, one panel per fairness constraint, RL (dashed) versus PeRL (solid) under Equality of Outcome (OUT, dark) and Equality of Opportunity (OPP, light). Mean over $10$ seeds, $25$-episode moving average, bands $\pm$ one standard deviation across seeds.}
  \label{fig:reactivity-per-episode}
\end{figure}

\begin{figure}[t!]
  \centering
  \begin{subfigure}{0.32\linewidth}
    \includegraphics[width=\linewidth]{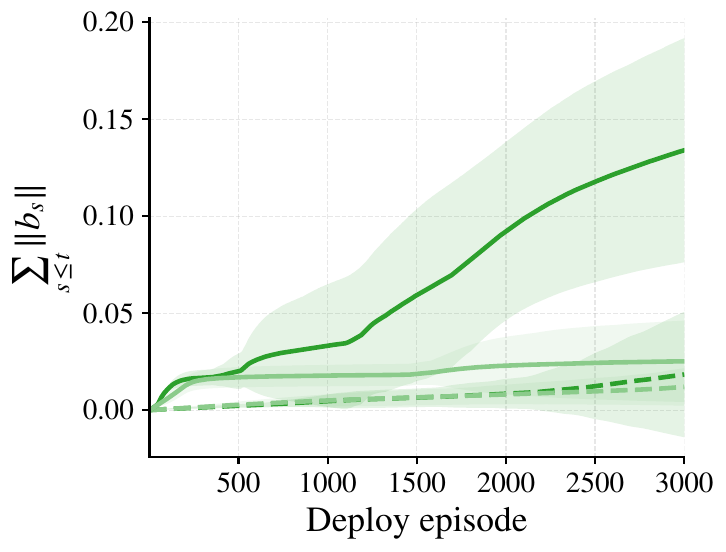}
    \caption{Social Welfare (SW)}
  \end{subfigure}\hfill
  \begin{subfigure}{0.32\linewidth}
    \includegraphics[width=\linewidth]{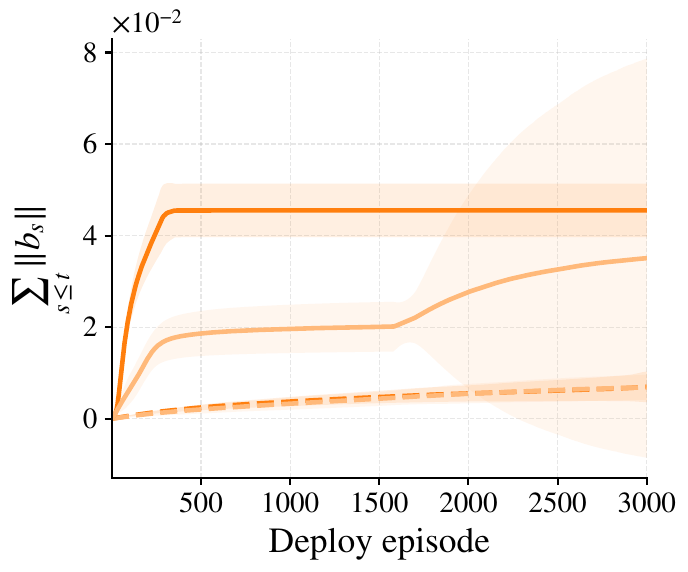}
    \caption{Rawlsian Max-Min (RMM)}
  \end{subfigure}\hfill
  \begin{subfigure}{0.32\linewidth}
    \includegraphics[width=\linewidth]{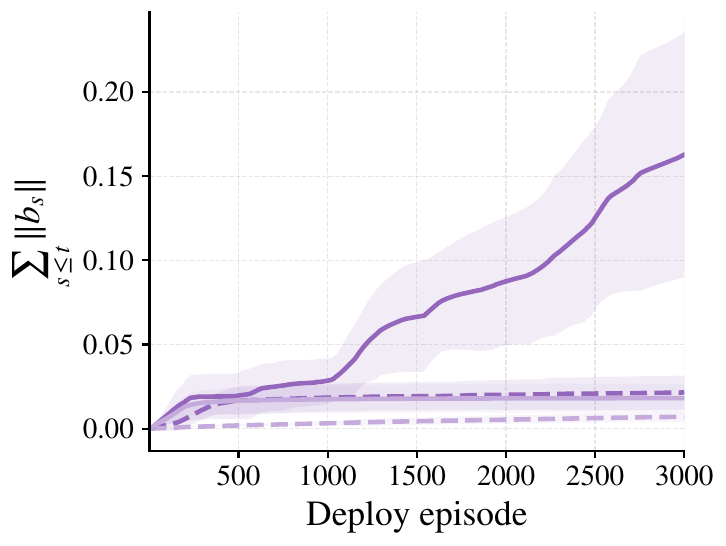}
    \caption{Fairness Lagrangian (FL)}
  \end{subfigure}\\[0.4em]
  \includegraphics[width=0.75\linewidth]{iclr_2027/iclr_2027_plots/reactivity_run11/reactivity_legend.pdf}
  \caption{\textbf{Cumulative reactivity.} Running sum $\sum_{s\le t}\|b_s\|$ over $E=3{,}000$ deploy episodes; same panels, curves and bands as Figure~\ref{fig:reactivity-per-episode}, without smoothing.}
  \label{fig:reactivity-cumulative}
\end{figure}

\paragraph{Empirical evolution.}
For each agent we set $b_t = \nabla_\mu \pi_t$, the gradient of the policy's expected approval probability (the Beta mean) with respect to the two group-mean wealths $(\mu^{\text{red}}, \mu^{\text{blue}})$, estimated by auto-grad on the batch of states seen during deploy episode $t$ (same procedure for both agents). We record $b_t$ for the six cells of Table~\ref{tab:fairness_rewards} over the $E = 3{,}000$ deploy episodes and $10$ seeds of the main experiments. Figures~\ref{fig:reactivity-per-episode} and~\ref{fig:reactivity-cumulative} show the per-episode magnitude $\|b_t\|$ ($25$-episode moving average) and its running sum $\sum_{s \le t} \|b_s\|$, one panel per fairness constraint and one curve per (agent, perspective). Three observations follow:

\textit{1. PeRL is more reactive than RL in every scenario (Figure~\ref{fig:reactivity-cumulative}):} $6$--$7.5\times$ more under Equality of Outcome and $2$--$5\times$ more under Equality of Opportunity. RL's running sum grows slowly and linearly in every cell: its sensitivity to the group wealths stays where training left it, the \emph{passive} class of Definition~\ref{def:reactivity}. PeRL, whose gradient carries the dependence of the dynamics on the policy, is the \emph{reactive} learner.

\textit{2. PeRL reacts, but fairness perspective matters (Figure~\ref{fig:reactivity-per-episode}).} Under Equality of Opportunity every constraint shows one burst within the first $200$ episodes and is flat afterwards: true-positive rates can be equalised once and the constraints stops binding. FL under Equality of Outcome the constraint acts on a wealth gap that the Matthew dynamics keep re-opening, and PeRL keeps reacting, and shows the largest peak of the study near episode $2{,}500$.

\textit{3. Reactivity can vanish exactly.} For RMM under equality of Outcome every PeRL seed reaches $\|b_t\| = 0$ by episode ${\approx}\,1{,}000$: the policy has converged to the approve-everyone corner, where the guards of Appendix~\ref{appendix:hyperparameters} clip the input gradient, and has moved from \emph{reactive} to \emph{non-reactive} in the terms of Definition~\ref{def:reactivity}. As approve-everyone is the welfare-optimal fixed point of that scenario (Section~\ref{app:heuristic}), this is the end of learning rather than a failure of it; but a flat cumulative curve must be read together with where the policy ends up.

\clearpage
\newpage
\section{Pseudo-codes}
\label{sec:pseudocodes}

Algorithm~\ref{alg:loan-sim} gives \framework's interaction protocol, shared by every agent and utility in the paper; Equation~\eqref{eq:app_reward} the per-applicant reward; and Algorithm~\ref{alg:dual}, with Table~\ref{tab:dual-thresholds}, the dual-ascent update of the multiplier $\gamma^t$ for the six cells of Table~\ref{tab:fairness_rewards}. Notation follows Sections~\ref{sec:formulation pemdp} and~\ref{sec:eutopia_main}; $\pi_\theta$ is the deployed policy $\pi^t$, with parameters $\theta$ updated once per episode. Apart from wealth $\bx_i[1]$, every per-individual quantity ($c_i$, $g_i$, $l_i$, $d_i$, $\kappa_i$) is fixed by the Data Generator (Appendix~\ref{app:loanly}) and never changes. The ADM's creditworthiness estimate $\hat c_i^t$ uses scorer coefficients $(\phi_g, \phi_X, b)$ fit once on the initial population, applied to $\tilde\bx_i^t[1]$, the individual's current wealth standardised by the initial population's mean and standard deviation; since neither is refit, $\hat c_i^t$ rises with wealth over deployment. Algorithm~\ref{alg:loan-sim} is the deployment protocol, which PeRL also follows during pre-training; RL's pre-training environment instead resets wealth, the Hawkes history and the counters at the start of every episode (Appendix~\ref{appendix:hyperparameters}).

\begin{algorithm}[H]
\caption{\framework{} interaction protocol}
\label{alg:loan-sim}
\begin{algorithmic}[1]
\STATE \textbf{Input:} groups $g \in \{\mathrm{red}, \mathrm{blue}\}$ of size $N_g$; per-individual $(\bx_i[1], c_i, g_i, l_i, d_i, \kappa_i)$ from the Data Generator; scorer coefficients $(\phi_g, \phi_X, b)$; policy $\pi_\theta$; multiplier $\gamma^t$; horizon $T$, timestep $\Delta t$; Hawkes parameters $\alpha_g, \beta_g$; Matthew coupling $K$; application sensitivity $s_g$; reference population $N_{\mathrm{ref}}$
\STATE Initialise $\bx_i[1]$ from the Data Generator; $t \leftarrow 0$; $\mathcal{H}_g \leftarrow \emptyset$; $\text{loan}_g, \text{def}_g \leftarrow 0$
\FOR{each of the $E$ episodes}
    \STATE $(\mathrm{tp}_g, \mathrm{fn}_g) \leftarrow 0$;\quad $\mu_g^{\mathrm{start}} \leftarrow \mu_g^t$ \hfill \textit{(all other state carries over)}
    \FOR{$T/\Delta t$ timesteps}
        \STATE $\mu_g^t \leftarrow \frac{1}{N_g}\sum_{i \in g} \bx_i^t[1]$;\quad $\bar\mu^t \leftarrow \tfrac{1}{2}(\mu_{\mathrm{red}}^t + \mu_{\mathrm{blue}}^t)$;\quad $\bar d_g^t \leftarrow \text{def}_g / \max(\text{loan}_g, 1)$
        \STATE $\hat c_i^t \leftarrow \sigma\bigl(\phi_g \mathbf{1}[g_i = \mathrm{red}] + \phi_X \tilde\bx_i^t[1] + b\bigr)$ for every individual $i$ \hfill \textit{(creditworthiness estimate)}
        \STATE $f(\mu_g^t) \leftarrow \dfrac{N_g}{N_{\mathrm{ref}}}\,\max\!\Bigl(0.5,\ 2\bigl(1 + K\,(\mu_g^t/\bar\mu^t - 1)\bigr)\Bigr)$
        \STATE $\lambda_g^t \leftarrow f(\mu_g^t) + \alpha_g \sum_{t_j \in \mathcal{H}_g} e^{-\beta_g (t - t_j)}$ \hfill \textit{(Equation~\eqref{eq:hawkes})}
        \STATE $\papply_i^t \leftarrow \operatorname{clip}\!\Bigl(\dfrac{\lambda_{g_i}^t\,\Delta t}{N_{g_i}}\bigl(1 + s_{g_i}\,\hat c_i^t\bigr),\ 0,\ 1\Bigr)$ for every individual $i$
        \STATE $\mathcal{N}^t \leftarrow \{\, i : u_i < \papply_i^t,\ u_i \sim \mathcal{U}[0,1] \,\}$ \hfill \textit{(the cohort; $|\mathcal{N}^t| = N^t$)}
        \FOR{each applicant $i \in \mathcal{N}^t$}
            \STATE $o_i^t \leftarrow \bigl(\bx_i^t[1]/100,\ g_i,\ \mu_{\mathrm{red}}^t/100,\ \mu_{\mathrm{blue}}^t/100,\ \lambda_{\mathrm{red}}^t,\ \lambda_{\mathrm{blue}}^t,\ \bar d_{\mathrm{red}}^t,\ \bar d_{\mathrm{blue}}^t,\ \hat c_i^t,\ \phi_g,\ \phi_X,\ l_i\bigr)$
            \STATE $p_i \sim \pi_\theta(\cdot \mid o_i^t)$;\quad $\bA_i^t \leftarrow \perp$ with probability $p_i$, else $\not\perp$
        \ENDFOR
        \STATE $r_i^t \leftarrow$ Equation~\eqref{eq:app_reward} for each $i \in \mathcal{N}^t$, from the state before this cohort's decisions
        \FOR{each applicant $i \in \mathcal{N}^t$ with $\bA_i^t = \perp$}
            \STATE $Y_i^t \sim \mathrm{Bernoulli}(1 - d_i)$;\quad $\kappa_i^t \leftarrow \kappa_i Y_i^t$;\quad $\bx_i[1] \leftarrow \bx_i[1] + \kappa_i^t$
            \STATE Append $t$ to $\mathcal{H}_{g_i}$;\quad $\text{loan}_{g_i} \mathrel{+}= 1$;\quad $\text{def}_{g_i} \mathrel{+}= 1 - Y_i^t$
        \ENDFOR
        \STATE $\mathrm{tp}_g \mathrel{+}= \#\{i \in \mathcal{N}^t \cap g : c_i = 1,\ \bA_i^t = \perp\}$;\quad $\mathrm{fn}_g \mathrel{+}= \#\{i \in \mathcal{N}^t \cap g : c_i = 1,\ \bA_i^t = \not\perp\}$
        \STATE $t \leftarrow t + \Delta t$
    \ENDFOR
    \STATE $\mu_g^{\mathrm{end}} \leftarrow \mu_g^t$;\quad update $\theta$ to increase the discounted return $\sum_{t} \zeta_{\mathrm{discount}}^{\,t} \sum_{i \in \mathcal{N}^t} r_i^t$ over the episode \hfill \textit{(Appendix~\ref{appendix:hyperparameters})}
    \STATE Update $\gamma^t$ from $\mu_g^{\mathrm{start}}, \mu_g^{\mathrm{end}}, \mathrm{tp}_g, \mathrm{fn}_g$ \hfill \textit{(Algorithm~\ref{alg:dual})}
\ENDFOR
\end{algorithmic}
\end{algorithm}

\paragraph{Reward.}
Each applicant's reward is its expected profit (Equation~\eqref{eq:reward_def}) plus the fairness term of its cell of Table~\ref{tab:fairness_rewards}:
\begin{equation}
\label{eq:app_reward}
r_i^t \;=\; \underbrace{p_i\bigl[(1-d_i)\,\tau_{\mathrm{interest}} - d_i\bigr]\, l_i}_{\text{expected profit}}
\;+\; s\,\gamma^t \times
\begin{cases}
C^t / N^t & \text{Equality of Opportunity},\\[2pt]
u_i\; p_i\,(1-d_i)\,\kappa_i/\bar\kappa & \text{Equality of Outcome},
\end{cases}
\end{equation}
with $s$ as in Equation~\eqref{eq:utility}, $C^t$ the Opportunity constraint of Table~\ref{tab:fairness_rewards} evaluated on the true positive rates before the cohort's decisions, $\bar\kappa$ the population-mean gain on a repaid loan, and $u_i = 1$ (SW), $\mathbf{1}[\mu_{g_i} < \mu_{\bar g_i}]$ (RMM), $\operatorname{sign}(\mu_{g_i} - \mu_{\bar g_i})$ (FL), where $\bar g_i$ is the other group.
Under Equality of Opportunity the constraint enters as its level, split evenly over the $N^t$ applicants of the step, so that each step contributes one copy of $C^t$. Under Equality of Outcome the level is a function of the group means, which one step's approvals move by about $10^{-5}$ relative, too little to carry a per-applicant training signal. Each applicant is instead credited with its expected contribution to the change of the constraint: $p_i(1-d_i)\kappa_i/\bar\kappa$ is the wealth the approval is expected to create, and $u_i$ assigns it to the constraint, so that the credits of an episode are, in expectation, proportional to the episode's change in $C^t$. Dividing by $\bar\kappa$ puts $\gamma^t$ on the profit scale: it reads as the profit the ADM gives up per average loan's worth of borrower wealth.

\begin{table}[t]
\centering
\caption{Constraint value $C$, measured at the end of each episode, and threshold $\vartheta$ used by Algorithm~\ref{alg:dual} for each cell of Table~\ref{tab:fairness_rewards}. SW and RMM require $C \ge \vartheta$, FL requires $C \le \vartheta$. $\Delta W_g = N_g(\mu_g^{\mathrm{end}} - \mu_g^{\mathrm{start}})$ is the wealth created in group $g$ during the episode.}
\label{tab:dual-thresholds}
\begin{tabular}{l l c l c}
\toprule
 & \multicolumn{2}{c}{Equality of Outcome} & \multicolumn{2}{c}{Equality of Opportunity} \\
\cmidrule(lr){2-3} \cmidrule(lr){4-5}
Constraint & $C$ & $\vartheta$ & $C$ & $\vartheta$ \\
\midrule
SW  & $(\Delta W_{\mathrm{red}} + \Delta W_{\mathrm{blue}}) / \bigl(\bar\kappa\,(N_{\mathrm{red}} + N_{\mathrm{blue}})\bigr)$ & $0.035$ & $\mathrm{TPR}_{\mathrm{red}} + \mathrm{TPR}_{\mathrm{blue}}$ & $1.5$ \\
RMM & $\Delta W_{\mathrm{poorer}} / (\bar\kappa\,N_{\mathrm{poorer}})$ & $0.025$ & $\min(\mathrm{TPR}_{\mathrm{red}}, \mathrm{TPR}_{\mathrm{blue}})$ & $0.75$ \\
FL  & $|\mu_{\mathrm{red}} - \mu_{\mathrm{blue}}| / \bar\mu$ & $0.12$ & $|\mathrm{TPR}_{\mathrm{red}} - \mathrm{TPR}_{\mathrm{blue}}|$ & $0.10$ \\
\bottomrule
\end{tabular}
\end{table}

\begin{algorithm}[H]
\caption{Dual ascent on $\gamma^t$, after every episode}
\label{alg:dual}
\begin{algorithmic}[1]
\REQUIRE current $\gamma^t$; the episode's $C$ and threshold $\vartheta$ (Table~\ref{tab:dual-thresholds}); step $\eta = 0.1$; floor $\epsilon = 10^{-4}$; cap $\gamma_{\max} = \max(10, 2\gamma^0)$
\STATE $v \leftarrow s\,(\vartheta - C)/|\vartheta|$ \hfill \textit{($s$ as in Equation~\eqref{eq:utility}; $v > 0$ iff the constraint is violated)}
\STATE $\gamma^t \leftarrow \operatorname{clip}\bigl(\gamma^t + \eta\, v,\ \epsilon,\ \gamma_{\max}\bigr)$
\end{algorithmic}
\end{algorithm}

\paragraph{Dual ascent.}
After every episode, Algorithm~\ref{alg:dual} raises $\gamma^t$ while the constraint is violated and lowers it once it is satisfied. Under Equality of Outcome the constraint is measured per episode in normalised form: SW and RMM by the wealth the episode created, per capita and in units of $\bar\kappa$, the same quantity the per-applicant credits of Equation~\eqref{eq:app_reward} add up; FL by the wealth gap relative to mean wealth. Dividing the violation by $|\vartheta|$ gives the step $\eta$ the same meaning in every cell. The thresholds are set against a coin-flip ADM (approval probability $0.5$) in the deployment environment, so that every constraint binds against profit without being out of reach. $\gamma^t$ starts at $\gamma^0 = 2$ (SW), $5$ (RMM) and $0.5$ (FL) and is reset to $\gamma^0$ at the start of deployment, so that both agents enter deployment with the same multiplier. Formally, Algorithm~\ref{alg:dual} is projected gradient descent in $\gamma$ on the Lagrangian of the constraint $sC \ge s\vartheta$; the threshold $\vartheta$ does not change the policy's maximiser in Equation~\eqref{eq:utility}, which is why the main text omits it.

\clearpage


%% file: iclr_2027/Appendix/detailed_experimentation.tex
\section{Experimental Setting}
\label{app:reproduce}
\subsection{Data Generator}
\label{app:loanly}

The Data Generator of Section~\ref{sec:dynamics} builds the initial population from the UCI Adult Income dataset~\citep{adult_2} and fixes the per-individual parameters that govern its response to lending.

\paragraph{Wealth and creditworthiness.}
A logistic regression predicts $\mathbf{1}[\text{income}_i > \$50\text{K}]$ from age, years of education, weekly working hours, and log capital gains and losses; gender is not among the features. Its output $sc_i \in [0,1]$ sets both the initial wealth, $\bx_i[1] = 10 + 190\,sc_i \in [10, 200]$ (in \$K), and the creditworthiness label $c_i = \mathbf{1}[sc_i \ge 0.5]$, which the agent never observes and which is used only to compute true positive rates. Individuals are split by gender into the red (male) and blue (female) groups, with $N_M = N_F = 12{,}000$.

\paragraph{Financial type.}
Each individual's financial type $(l_i, d_i, \kappa_i)$ is fixed (Section~\ref{sec:dynamics}). The loan amount is $l_i \sim \mathcal{N}(30, 10^2)$, clipped to $[10, 100]$ (\$K), and the gain on repayment is $\kappa_i = \tau_{\mathrm{gain}}\, l_i = 0.17\, l_i$, with $\tau_{\mathrm{inv}} = 0.35$ and $\tau_{\mathrm{interest}} = 0.18$. The default probability is set by rank: within each group, individuals are ranked by initial wealth and assigned $d_i$ linearly from $\delta_{\max}$ (lowest) to $\delta_{\min}$ (highest), with $[\delta_{\min}, \delta_{\max}] = [0.05, 0.25]$ in deployment. Ranking keeps the mean default probability at $(\delta_{\min} + \delta_{\max})/2$ whatever the shape of the wealth distribution, and ties risk to an observable feature; an independent random $d_i$ would give a policy nothing to learn. Repayment is drawn afresh on every approval, $Y_i^t \sim \mathrm{Bernoulli}(1 - d_i)$.

\paragraph{ADM's scorer.}
The scorer behind $\hat c_i^t$ in Algorithm~\ref{alg:loan-sim} is a second logistic regression, fit once on the Adult records to predict $\mathbf{1}[\text{income}_i > \$50\text{K}]$ from the group indicator and standardised wealth. Unlike $sc_i$, it uses group membership, and neither its coefficients $(\phi_g, \phi_X, b)$ nor its standardisation are refit during deployment.

\subsection{\framework{}: Details on our proposed Simulator}
\label{app:eutopia}

Algorithm~\ref{alg:loan-sim} gives the protocol; this subsection gives its constants and the reasoning behind its choices.

\paragraph{Time.}
An episode has horizon $T = 100$ simulated time units at timestep $\Delta t = 0.5$, i.e.\ $200$ decision steps. Simulated time, wealth, the Hawkes history and the loan and default counters carry over from one episode to the next, so a run is a single continual trajectory. Each agent is pre-trained for $1{,}000$ episodes and then deployed for $3{,}000$, continuing to learn throughout deployment; Appendix~\ref{appendix:hyperparameters} provides details for the pre-training environments.

\paragraph{Applications.}
Arrivals follow the Hawkes process of Equation~\eqref{eq:hawkes}: each approval raises its group's intensity by $\alpha_g = 0.3$, decaying at rate $\beta_g = 2.0$ (under $0.1\%$ of its initial value after $3.5$ time units), while denials and defaults add nothing. The base rate is
\[
f(\mu_g^t) = \frac{N_g}{N_{\mathrm{ref}}}\,\max\!\Bigl(0.5,\ 2\bigl(1 + K\,(\mu_g^t/\bar\mu^t - 1)\bigr)\Bigr), \qquad K = 2,\quad N_{\mathrm{ref}} = 3{,}000 .
\]
It depends on a group's wealth \emph{relative} to the population mean $\bar\mu^t$: a base rate proportional to absolute wealth would make wealth grow exponentially, whereas the relative form lets the ratio of group wealths diverge, which is the inequality under study, while absolute wealth stays bounded in rate. The factor $N_g/N_{\mathrm{ref}}$ makes arrivals per capita. Without it the cohort size would not grow with $N_g$, each approval would move $\mu_g$ by $\kappa_i/N_g$, and the performative drift would vanish as the population grows; with it, a larger population reduces noise without weakening the dynamics. $N_{\mathrm{ref}}$ is the population size at which the constants were calibrated. An individual then applies with the probability of Algorithm~\ref{alg:loan-sim}, which scales the group rate by $1 + s_g\,\hat c_i^t$ with $s_g = 0.5$ for both groups, so that individuals the ADM is more likely to approve apply more often.

\paragraph{Actions.}
The policy outputs an approval probability $p_i \in [0,1]$ and the decision is drawn as $\mathrm{Bernoulli}(p_i)$, so the policy's output is continuous and differentiable while the realised decision stays binary. The $12$-dimensional observation is $o_i^t$ of Algorithm~\ref{alg:loan-sim}.

\subsection{Algorithmic Details}

\subsubsection{Rule-based Policies}
\label{app:heuristic}
We evaluate seven rule-based policies, summarised in Table~\ref{tab:heuristics}, as fixed reference points for the learned agents. None of them learns during deployment. Six have no parameters; the seventh, Pattern Prediction, uses the ADM's scorer, whose coefficients are fit once before deployment (Appendix~\ref{app:loanly}).

\begin{table}[h]
\centering
\caption{Rule-based reference policies. $p_i$ is the approval probability for applicant $i$, $d_i$ its default probability and $\hat c_i^t$ the ADM's creditworthiness estimate (Algorithm~\ref{alg:loan-sim}).}
\label{tab:heuristics}
\begin{tabular}{lll}
\toprule
Policy & Symbol & Decision rule \\
\midrule
Oracle & $\mathbb{O}$ & $p_i = \mathbf{1}\bigl[d_i < \tau_{\mathrm{interest}}/(1+\tau_{\mathrm{interest}})\bigr]$ \\
Always Approve & $\mathbb{A}_a$ & $p_i = 1$ \\
Always Reject & $\mathbb{A}_r$ & $p_i = 0$ \\
Uniform Acceptance & $\mathbb{A}_U$ & $p_i \sim \mathcal{U}(0,1)$ \\
Rich getting Richer & $\mathbb{A}_{\textcolor{red}{R}\uparrow}$ & $p_i = 1$ if $g_i = \mathrm{red}$, else $p_i \sim \mathcal{U}(0,1)$ \\
Reverse Rich getting Richer & $\mathbb{A}_{\textcolor{blue}{B}\uparrow}$ & $p_i = 1$ if $g_i = \mathrm{blue}$, else $p_i \sim \mathcal{U}(0,1)$ \\
Pattern Prediction & $\mathbb{A}_H$ & $p_i = \sigma(\hat c_i^t$) \\
\bottomrule
\end{tabular}
\end{table}

\paragraph{Oracle ($\mathbb{O}$).}
The Oracle observes each applicant's true default probability $d_i$, which no learned agent sees, and approves exactly when the loan is profitable in expectation: $(1-d_i)\,\tau_{\mathrm{interest}}\,l_i - d_i\,l_i > 0$, i.e.\ $d_i < \tau_{\mathrm{interest}}/(1+\tau_{\mathrm{interest}}) \approx 0.153$. It is the per-decision profit optimum under perfect knowledge of risk. Because this break-even point lies inside the deployment range $[0.05, 0.25]$, the Oracle is selective: it approves the $51\%$ least risky members of each group.

\paragraph{Pattern Prediction ($\mathbb{A}_H$).}
Pattern Prediction approves with the ADM's own creditworthiness estimate, $p_i = \sigma(\hat c_i^t)$, read directly from the observation. The scorer is never refit (Appendix~\ref{app:loanly}), so the rule is a static learned heuristic: its output moves with the applicant's current wealth and so responds to the population's drift, but it never learns from it.

\subsubsection{Learning Agents}
\label{appendix:hyperparameters}
RL and PeRL share the policy network, optimiser, discount, entropy bonus, gradient clipping and dual ascent (Table~\ref{tab:agent_hyperparams}), and differ in how the policy gradient is estimated. Both are deployed under identical conditions.

\paragraph{Policy.}
$\bpi_\theta(\cdot \mid o_i^t)$ is a Beta distribution over the approval probability $p_i$, whose two concentration parameters are the two heads of an MLP with two hidden layers of width $128$ and ReLU activations; its input is the observation $o_i^t$ of Algorithm~\ref{alg:loan-sim}. The concentrations are bounded above and below to keep the policy away from degenerate approve-all and reject-all solutions. Both agents use AdamW with learning rate $10^{-3}$, weight decay $0.1$ on the heads, gradient clipping at unit norm, and an entropy bonus of $10^{-3}$.

\paragraph{RL.}
The RL agent is REINFORCE~\citep{Williams1992}. Per-applicant rewards (Equation~\eqref{eq:app_reward}) are summed over each timestep's cohort and normalised online by their running mean and standard deviation. The return-to-go is discounted per timestep with $\zeta_{\mathrm{discount}}$, and a baseline, an exponential moving average of past returns-to-go at the same timestep index, is subtracted; every applicant in a cohort receives its cohort's advantage. RL treats the transition as fixed and follows the score-function gradient.

\paragraph{PeRL.}
The PeRL agent is Performative Policy Gradient~\citep{basu2025performativepolicygradientoptimality}, which differentiates through the population's response to the policy. Each episode it runs, from the same starting state as the simulator, a differentiable mean-field shadow rollout of the group means $\mu_g$ and Hawkes intensities $\lambda_g$: at every step, a batch of individuals drawn from the current population passes through the policy with a reparameterised Beta sample, and their expected outcomes update $\mu_g$, $\lambda_g$ and the reward of Equation~\eqref{eq:app_reward} with the current $\gamma^t$. Backpropagating through the unrolled episode credits each decision with its effect on every later state and reward, which is the performative term absent from the RL gradient. The shadow rollout only supplies the gradient; the simulator itself advances the population and produces every reported metric, exactly as for RL.

\begin{table}[h]
\centering
\caption{Hyperparameters, shared by RL and PeRL agents.}
\label{tab:agent_hyperparams}
\small
\begin{tabular}{lll}
\toprule
Hyperparameter & Value & Note \\
\midrule
Hidden layers & $2 \times 128$ & ReLU \\
Policy & Beta & Over the approval probability $p_i$ \\
Optimiser & AdamW & Weight decay $0.1$ on the heads \\
Learning rate & $10^{-3}$ & \\
Discount $\zeta_{\mathrm{discount}}$ & $0.99$ & Per timestep \\
Entropy coefficient & $10^{-3}$ & \\
Gradient clipping & $1.0$ & Global norm \\
Baseline decay (RL) & $0.9$ & Per timestep index \\
Dual-ascent step $\eta$ & $0.1$ & Algorithm~\ref{alg:dual} \\
$\gamma^0$ & $2$ / $5$ / $0.5$ & SW / RMM / FL; reset at deployment \\
$\gamma^t$ floor, cap & $10^{-4}$, $\max(10, 2\gamma^0)$ & Algorithm~\ref{alg:dual} \\
Episodes & $1{,}000$ + $3{,}000$ & Pre-training + deployment \\
Seeds & $10$ & Per experiment \\
\bottomrule
\end{tabular}
\end{table}

\newpage

\section{Experimental Results}

\subsection{Diagnosing \framework{} with Heuristic Policies}
\label{app:heuristic_results}
\begin{figure*}[h]
    \centering
    \begin{minipage}{0.47\textwidth}
        \centering
        \includegraphics[width=0.9\textwidth]{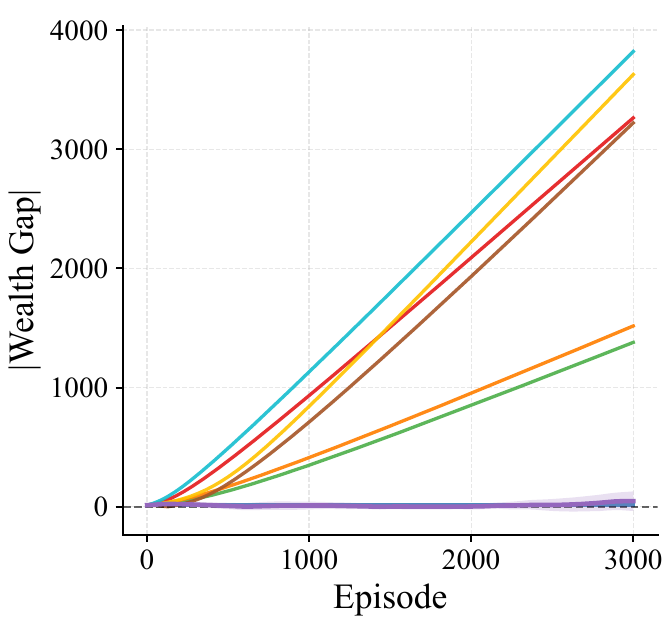}\vspace{-.5em}
        \subcaption{}\label{fig:heuristics1}
    \end{minipage}\hfill
    \begin{minipage}{0.47\textwidth}
        \centering
        \includegraphics[width=0.9\textwidth]{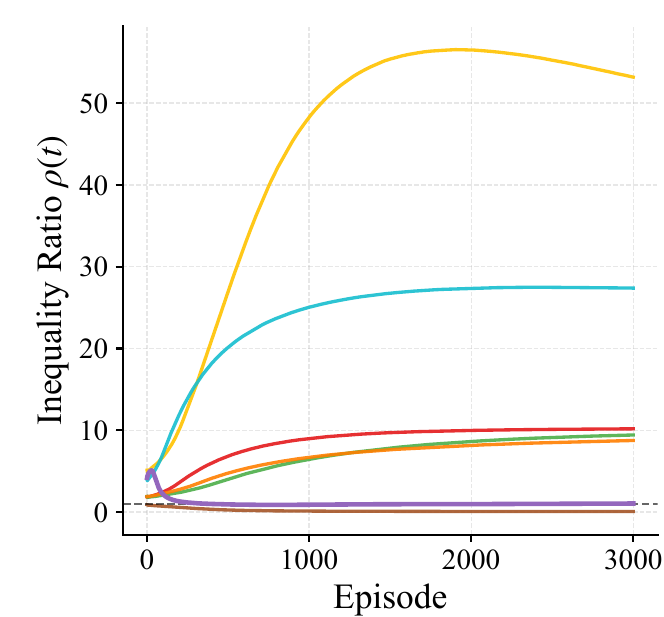}\vspace{-.5em}
        \subcaption{}\label{fig:heuristics2}
    \end{minipage}\\
    \begin{minipage}{0.47\textwidth}
        \centering
        \includegraphics[width=0.9\textwidth]{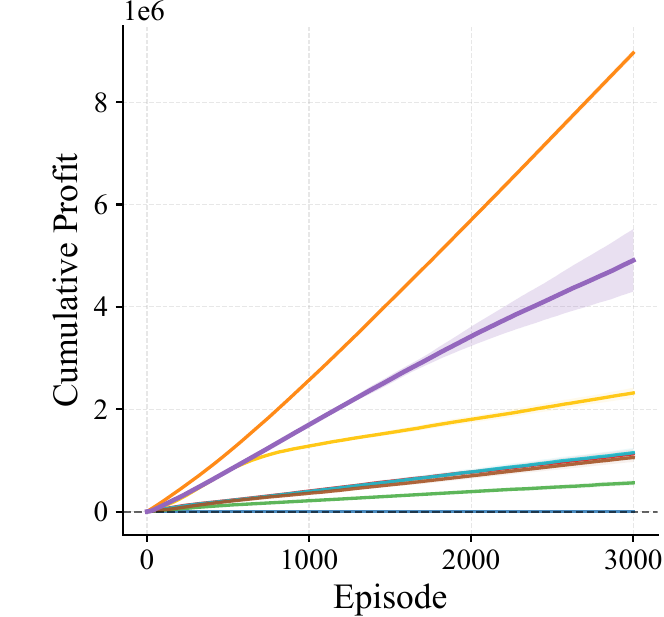}\vspace{-.5em}
        \subcaption{}\label{fig:heuristics3}
    \end{minipage}\hfill
    \begin{minipage}{0.47\textwidth}
        \centering
        \includegraphics[width=0.9\textwidth]{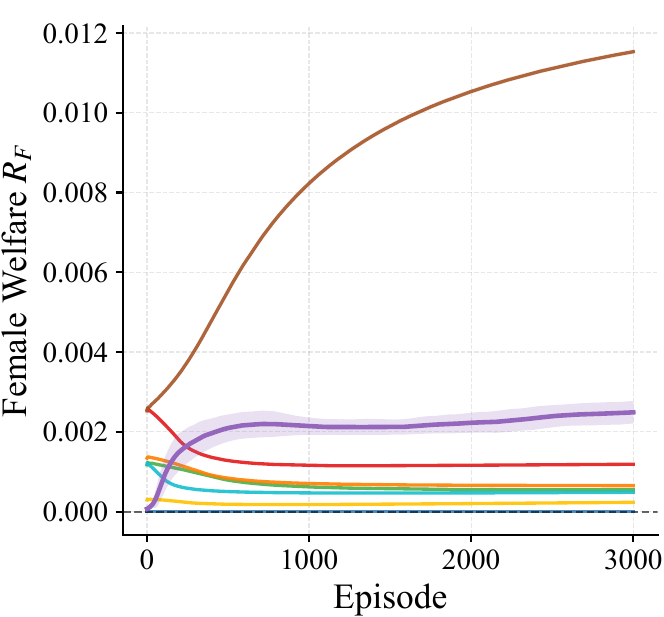}\vspace{-.5em}
        \subcaption{}\label{fig:heuristics4}
    \end{minipage}\\[0.5em]
    \includegraphics[width=\textwidth]{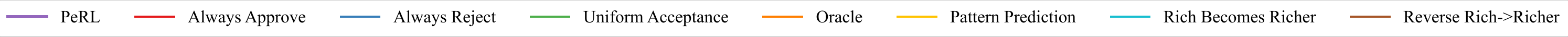}
    \caption{\textbf{Diagnosing \framework{} with rule-based policies.} (a) wealth gap $\Delta_\mu^t$, (b) inequality ratio $\rho^t$, (c) cumulative profit $\mathrm{CP}^t$, (d) blue-group welfare $R_F^t$, over $3{,}000$ deployment episodes, mean over $10$ seeds. PeRL under FL and Equality of Outcome is overlaid for reference; its band is $\pm 1$ std across seeds.}\label{fig:heuristics}
\end{figure*}

We deploy the seven rule-based policies of Appendix~\ref{app:heuristic} for $3{,}000$ episodes and $10$ seeds each, under the learned agents' deployment setting. None of them adapts, so what happens to the population is the simulator's response to a fixed rule. Figure~\ref{fig:heuristics} reports the wealth gap, inequality ratio and cumulative profit of Section~\ref{sec:q1}, and the blue group's welfare $R_F^t = (\mu_F^t - \mu_F^0)/t$, with PeRL under FL and Equality of Outcome (Table~\ref{tab:app_rl_vs_perl_main}) overlaid for reference. Seed variation of the rules is negligible at this population size, so we quote means.

\paragraph{Wealth gap.}
Under Always Reject no loan is issued and the gap stays at its initial $13.2$. Every other rule widens it without saturating. The biased rules reach the largest gaps, $3{,}820$ for Rich Becomes Richer and $3{,}221$ in the opposite direction for Reverse Rich Becomes Richer. The group-blind rules widen it too: Always Approve to $3{,}263$, Oracle to $1{,}517$, Uniform Acceptance to $1{,}379$. This is the Matthew effect of the arrival process (Section~\ref{sec:dynamics}): the red group starts wealthier, applies more, receives more loans and pulls further ahead, and a fixed rule has nothing that counteracts it. Pattern Prediction ends at $3{,}627$. PeRL's gap never exceeds $46.5$, below every rule except Always Reject.

\paragraph{Inequality ratio.}
The group-blind rules drift to $\rho \approx 9$--$10$ (Always Approve $10.2$, Uniform Acceptance $9.4$, Oracle $8.8$): the red group's wealth grows about nine times faster although the ADM ignores group membership. Rich Becomes Richer reaches $27$ and Reverse Rich Becomes Richer $0.06$. The two are mirror images in what the ADM does (approval-rate disparity $\pm 0.50$) but not in outcome, because the blue group also starts poorer and faces the arrival process working against it. Pattern Prediction peaks at $57$ around episode $1{,}900$ and ends at $53$. PeRL settles at $1.05$ after dipping to $0.87$ around episode $770$. $\rho$ is undefined under Always Reject.

\paragraph{Profit.}
Oracle earns the most, $8.96$M, since it alone sees each applicant's default risk. Its break-even point $d_i < 0.153$ is close to the mean default probability of $0.15$, so approving everyone is only marginally profitable: Always Approve earns $1.13$M, the two biased rules $1.15$M and $1.06$M, Uniform Acceptance $0.56$M. Pattern Prediction reaches $2.32$M, starting well but flattening after about $700$ episodes as the population drifts away from the one its scorer was fit on. PeRL earns $4.91 \pm 0.58$M, second only to Oracle.

\paragraph{Female welfare.}
Under every group-blind rule $R_F^t$ falls over deployment, from $0.0022$ at episode $100$ to $0.0012$ under Always Approve and to about $0.0006$ under Oracle and Uniform Acceptance: as the red group pulls ahead, the blue group applies less and receives a shrinking share of the loans. It keeps rising only under Reverse Rich Becomes Richer, to $0.0115$, at the cost of the inverted inequality above. Under PeRL it rises to $0.0022$ by episode $600$ and $0.0025$ by the end, twice Always Approve, with $\rho$ near $1$.

\paragraph{Pattern Prediction.}
The scorer-based rule is the closest analogue to a deployed credit model, and it is the least fair rule without an explicit bias: approval-rate disparity $0.72$, the second-largest wealth gap and $\rho$ above $50$. Its group coefficient $\phi_g$ encodes the correlation of gender and income in the Adult data, so it approves red applicants at a far higher rate; the resulting loans raise red wealth and hence red scores, while blue scores rise more slowly. A static learned rule amplifies the drift it induces.

\paragraph{What the diagnostics establish.}
The rules behave as their definitions predict: nothing moves under Always Reject, the biased rules bias, and Oracle is the most profitable. Beyond that, even group-blind lending turns the initial gap into a nine-fold difference in growth rates, so equal treatment does not yield equal outcomes, and the most profitable rule ends at $\rho \approx 9$. PeRL is the only policy in the figure that keeps $\rho$ near $1$ and the gap small, and it earns more than every rule except Oracle.

\subsection{Passive vs. Performative Models of RL}
\label{sec:passive_vs_performative}
No fixed rule keeps profit and fairness together once the population responds to its decisions (Appendix~\ref{app:heuristic_results}). We now ask whether a learning agent can, and whether it matters that the agent models that response. We compare RL, which follows the ordinary policy gradient and treats the population as fixed, with PeRL, which differentiates through the population's response to its own policy (Appendix~\ref{appendix:hyperparameters}). Each is pre-trained for $1{,}000$ episodes in its own environment, static for RL and performative for PeRL, and both are deployed identically for $3{,}000$ episodes in all six scenarios of Table~\ref{tab:fairness_rewards}, with $10$ seeds. The tables abbreviate Equality of Outcome and Equality of Opportunity as OUT and OPP. We report the macro metrics of Section~\ref{sec:experimental_analysis} and the approval-rate disparity at the end of deployment, then follow each metric over deployment. Throughout this section the multiplier is learned by dual ascent; the fixed-multiplier variant of PeRL under FL and Equality of Outcome reported in Table~\ref{tab:app_rl_vs_perl_main} is examined in Appendix~\ref{app:balanced}.

\subsubsection{PeRL versus RL}
Table~\ref{tab:app_rl_vs_perl} places the two agents side by side in every cell. Tables~\ref{tab:app_rl_by_perspective} and~\ref{tab:app_perl_by_perspective} give, for each agent, the best constraint under each perspective, and Tables~\ref{tab:app_rl_by_constraint} and~\ref{tab:app_perl_by_constraint} the best perspective under each constraint.

\begin{table}[t]
\centering
\caption{\textbf{RL vs.\ PeRL across all six scenarios} at the end of deployment (mean $\pm$ std over $10$ seeds). For each metric the better of the two agents is in bold: higher profit, smaller wealth gap, smaller approval-rate disparity, $\rho$ closer to $1$. $^\dagger$: RL approves nearly every applicant of both groups in this cell, so the disparity is small because approval is near-universal, not because the groups are balanced.}
\label{tab:app_rl_vs_perl}
\resizebox{\textwidth}{!}{%
\begin{tabular}{llcccccccc}
\toprule
\multirow{2}{*}{\textbf{Perspective}} & \multirow{2}{*}{\textbf{Constraint}}
    & \multicolumn{2}{c}{\textbf{Cumulative Profit ($\times10^6$)}}
    & \multicolumn{2}{c}{\textbf{$|$Wealth Gap$|$}}
    & \multicolumn{2}{c}{\textbf{$|$Approval Rate Disparity$|$}}
    & \multicolumn{2}{c}{\textbf{Inequality Ratio $\rho$}} \\
\cmidrule(lr){3-4}\cmidrule(lr){5-6}\cmidrule(lr){7-8}\cmidrule(lr){9-10}
    & & RL & PeRL & RL & PeRL & RL & PeRL & RL & PeRL \\
\midrule
OUT & FL & 0.374 $\pm$ 0.385 & $\mathbf{1.304 \pm 0.974}$ & 745.0 $\pm$ 1288.3 & $\mathbf{71.9 \pm 298.0}$ & 0.0135 $\pm$ 0.5070 & $\mathbf{0.0119 \pm 0.2550}$ & 0.190 $\pm$ 0.132 & $\mathbf{1.059 \pm 0.094}$ \\
OUT & RMM & 1.096 $\pm$ 0.096 & $\mathbf{1.267 \pm 0.058}$ & 3190.0 $\pm$ 68.9 & $\mathbf{3081.9 \pm 26.5}$ & $\mathbf{0.0024 \pm 0.0202}^\dagger$ & 0.1086 $\pm$ 0.0030 & 10.198 $\pm$ 0.032 & $\mathbf{0.101 \pm 0.000}$ \\
OUT & SW & 1.103 $\pm$ 0.071 & $\mathbf{4.721 \pm 1.346}$ & 3169.8 $\pm$ 82.5 & $\mathbf{1956.0 \pm 1012.6}$ & $\mathbf{0.0088 \pm 0.0481}$ & 0.2081 $\pm$ 0.1897 & 10.294 $\pm$ 0.082 & $\mathbf{0.138 \pm 0.048}$ \\
\midrule
OPP & FL & 1.131 $\pm$ 0.062 & $\mathbf{7.443 \pm 0.410}$ & 3154.8 $\pm$ 103.9 & $\mathbf{1086.8 \pm 332.2}$ & 0.0109 $\pm$ 0.0459 & $\mathbf{0.0097 \pm 0.0129}$ & 10.328 $\pm$ 0.045 & $\mathbf{3.981 \pm 0.568}$ \\
OPP & RMM & 1.105 $\pm$ 0.100 & $\mathbf{6.711 \pm 1.229}$ & 3177.3 $\pm$ 55.4 & $\mathbf{207.9 \pm 1328.7}$ & $\mathbf{0.0049 \pm 0.0257}$ & 0.0543 $\pm$ 0.0792 & 10.231 $\pm$ 0.043 & $\mathbf{1.229 \pm 0.643}$ \\
OPP & SW & 1.118 $\pm$ 0.099 & $\mathbf{6.757 \pm 1.137}$ & 3115.9 $\pm$ 136.1 & $\mathbf{535.2 \pm 1187.0}$ & $\mathbf{0.0085 \pm 0.0556}$ & 0.0427 $\pm$ 0.0734 & 10.304 $\pm$ 0.079 & $\mathbf{1.776 \pm 0.878}$ \\
\midrule
\multicolumn{2}{l}{\textbf{PeRL better in}} & \multicolumn{2}{c}{6/6} & \multicolumn{2}{c}{6/6} & \multicolumn{2}{c}{2/6} & \multicolumn{2}{c}{6/6} \\
\bottomrule
\end{tabular}}
\end{table}

\paragraph{PeRL is better wherever the objective trades profit against fairness.}
PeRL earns more, leaves a smaller wealth gap and keeps $\rho$ closer to one than RL in all six scenarios: twenty of the twenty-four comparisons in Table~\ref{tab:app_rl_vs_perl}. Under Equality of Opportunity it earns six to seven times RL's profit and cuts the wealth gap by a factor of three to fifteen; under FL with Equality of Outcome it reaches a gap of $72$ against RL's $745$, with $\rho = 1.06$ against $0.19$. The four comparisons RL wins are all on approval-rate disparity, and are explained next.

\paragraph{RL tends to approve most applicants.}
In five of the six scenarios RL ends at nearly the same point: profit $1.10$ to $1.13$M, wealth gap $3{,}116$ to $3{,}190$, $\rho$ between $10.2$ and $10.3$, and approval-rate disparity below $0.011$. These are, within seed noise, the numbers of Always Approve ($1.13$M, $3{,}263$, $10.2$, $0$; Appendix~\ref{app:heuristic_results}). Under RMM and SW in both perspectives, and under FL with Equality of Opportunity, RL approves nearly every applicant of both groups: its disparities are small because approval is near-universal, and the Matthew effect drives the gap and $\rho$ to the values of the approve-all rule. Under FL with Equality of Outcome, the gap penalty is strong enough to move RL, which then under-lends to the red group, so that the blue group grows five times faster ($\rho = 0.19$).

\paragraph{PeRL's results depend on the utility.}
PeRL's profit ranges from $1.27$M to $7.44$M and its wealth gap from $72$ to $3{,}082$ across the six scenarios, as expected of an agent that responds to each constraint. Tables~\ref{tab:app_perl_by_perspective} and~\ref{tab:app_perl_by_constraint} summarise this: FL under Equality of Outcome gives the smallest gap and the $\rho$ closest to one, FL under Equality of Opportunity the highest profit and smallest disparity, and for RMM and SW the Opportunity perspective is better on every metric. Appendix~\ref{app:perl_metric_by_metric} follows each metric over deployment.

\input{iclr_2027/Appendix/Full_Tables}
\subsubsection{How PeRL Trades: Metric by Metric}
\label{app:perl_metric_by_metric}
\begin{figure*}[t!]
    \centering\vspace*{-1em}
    \begin{minipage}{0.45\textwidth}
        \centering
        \includegraphics[width=\textwidth]{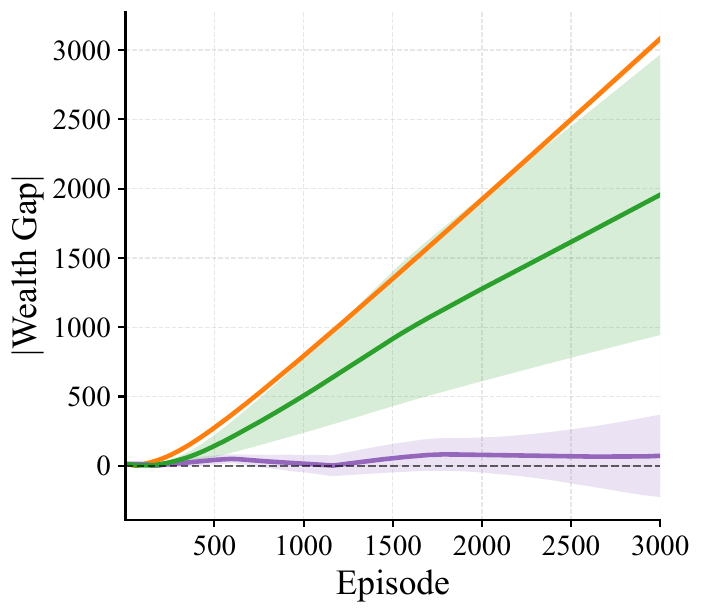}\vspace{-.5em}
        \subcaption{}\label{fig:pepg_sp_wg}
    \end{minipage}\hfill
    \begin{minipage}{0.45\textwidth}
        \centering
        \includegraphics[width=\textwidth]{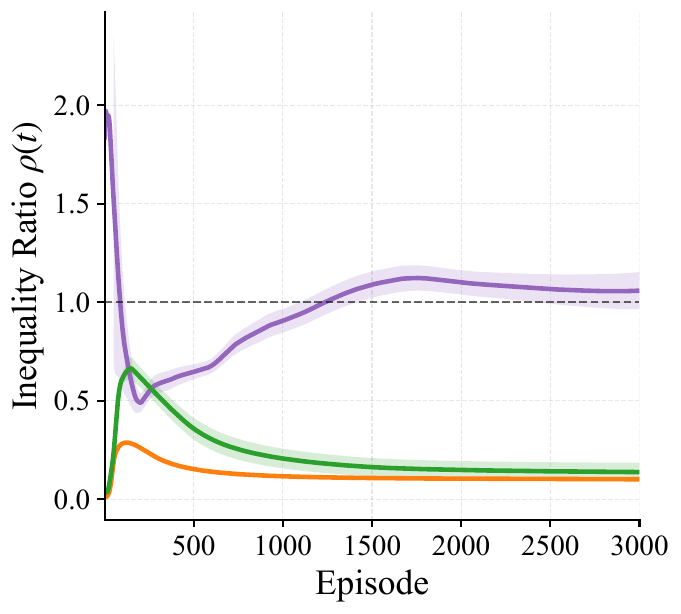}\vspace{-.5em}
        \subcaption{}\label{fig:pepg_sp_ineqr}
    \end{minipage}\\
    \begin{minipage}{0.45\textwidth}
        \centering
        \includegraphics[width=\textwidth]{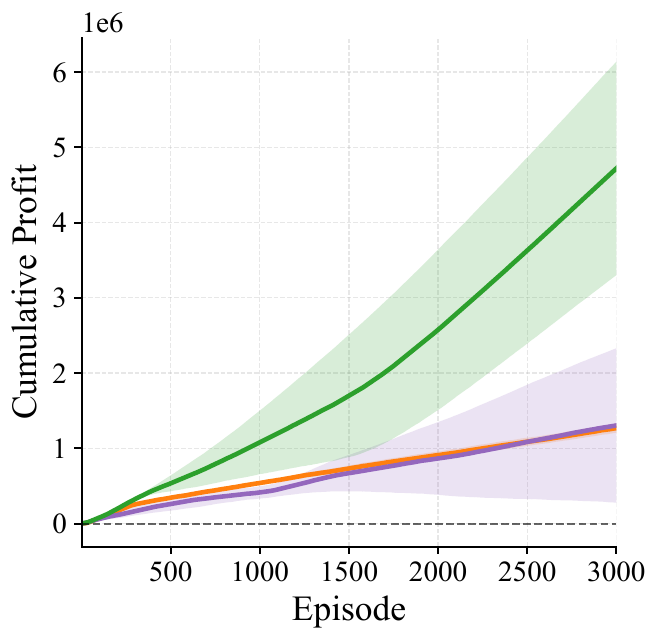}\vspace{-.5em}
        \subcaption{}\label{fig:pepg_sp_profit}
    \end{minipage}\hfill
    \begin{minipage}{0.45\textwidth}
        \centering
        \includegraphics[width=\textwidth]{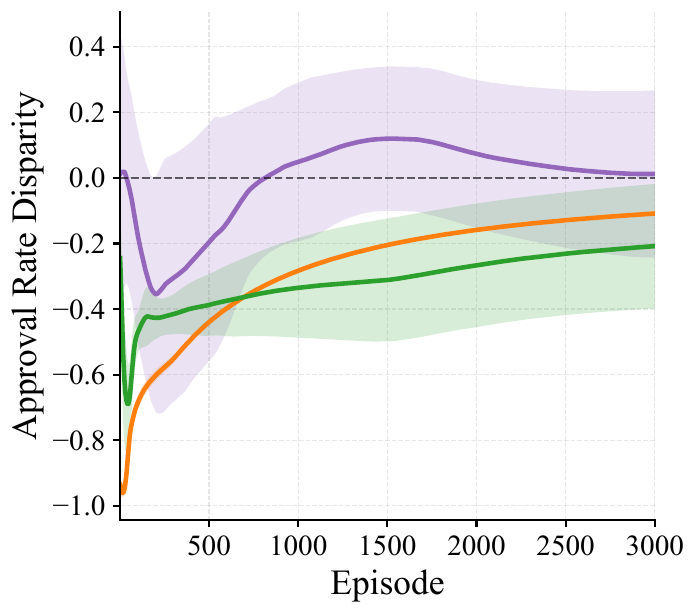}\vspace{-.5em}
        \subcaption{}\label{fig:pepg_sp_disp}
    \end{minipage}\\[0.5em]
    \includegraphics[width=0.45\textwidth]{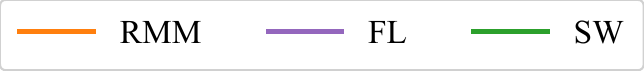}
    \caption{\textbf{PeRL: Equality of Outcome (OUT).} (a) wealth gap, (b) inequality ratio, (c) cumulative profit, (d) approval-rate disparity (red minus blue), for PeRL under FL, RMM and SW with the Equality of Outcome perspective, over $3{,}000$ deployment episodes. Mean $\pm$ std across $10$ seeds.(Section~\ref{sec:experimental_analysis}).}\label{fig:pepg_sp}
\end{figure*}

\begin{figure*}[h!]
    \centering\vspace*{-1em}
    \begin{minipage}{0.45\textwidth}
        \centering
        \includegraphics[width=\textwidth]{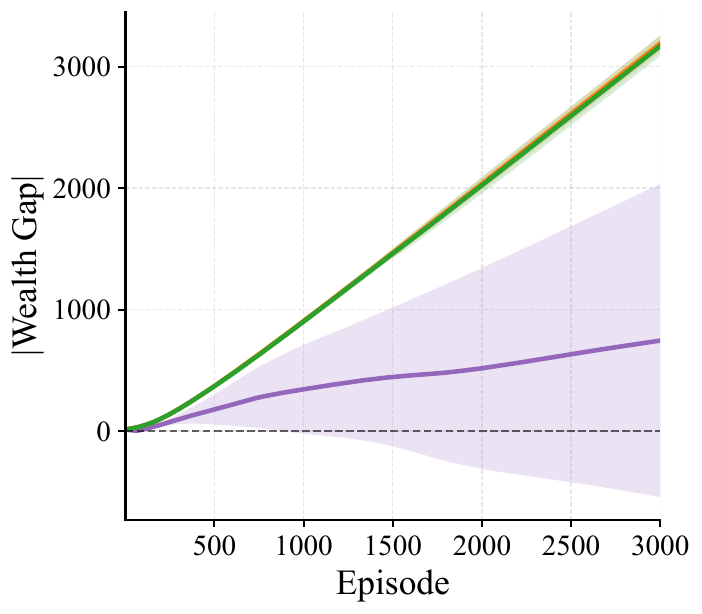}\vspace{-.5em}
        \subcaption{}\label{fig:pg_sp_wg}
    \end{minipage}\hfill
    \begin{minipage}{0.45\textwidth}
        \centering
        \includegraphics[width=\textwidth]{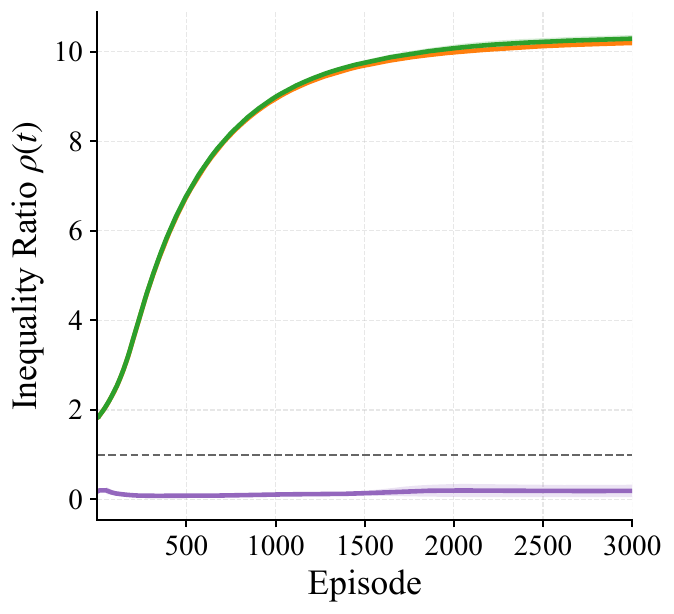}\vspace{-.5em}
        \subcaption{}\label{fig:pg_sp_ineqr}
    \end{minipage}\\
    \begin{minipage}{0.45\textwidth}
        \centering
        \includegraphics[width=\textwidth]{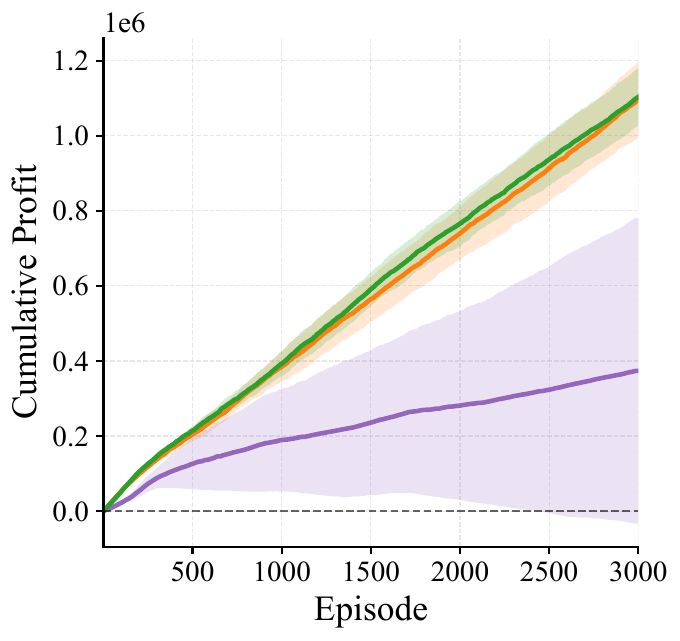}\vspace{-.5em}
        \subcaption{}\label{fig:pg_sp_profit}
    \end{minipage}\hfill
    \begin{minipage}{0.45\textwidth}
        \centering
        \includegraphics[width=\textwidth]{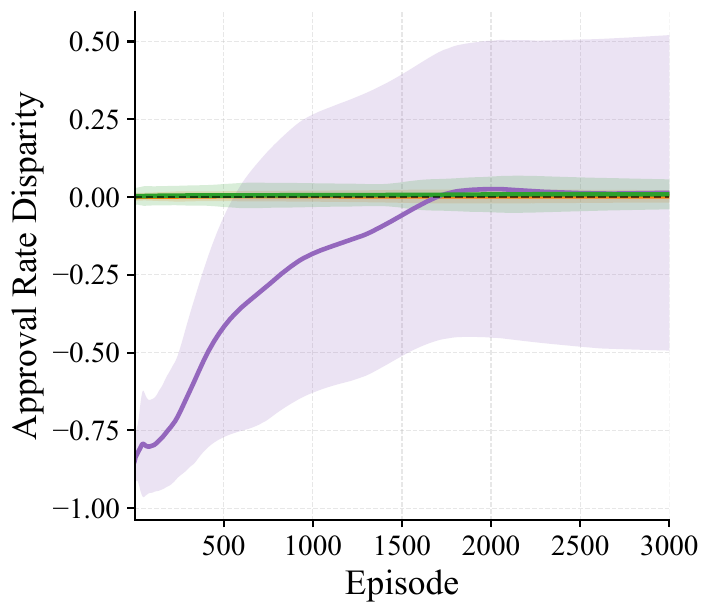}\vspace{-.5em}
        \subcaption{}\label{fig:pg_sp_disp}
    \end{minipage}\\[0.5em]
    \includegraphics[width=0.45\textwidth]{iclr_2027/iclr_2027_plots/trajectories_run11/trajectories_reward_legend.pdf}
    \caption{\textbf{RL: Equality of Outcome (OUT).} (a) wealth gap, (b) inequality ratio, (c) cumulative profit, (d) approval-rate disparity (red minus blue), for RL under FL, RMM and SW with the Equality of Outcome perspective, over $3{,}000$ deployment episodes. Mean $\pm$ std across $10$ seeds. (Section~\ref{sec:experimental_analysis}).}\label{fig:pg_sp}
\end{figure*}

\begin{figure*}[t!]
    \centering\vspace*{-1em}
    \begin{minipage}{0.45\textwidth}
        \centering
        \includegraphics[width=\textwidth]{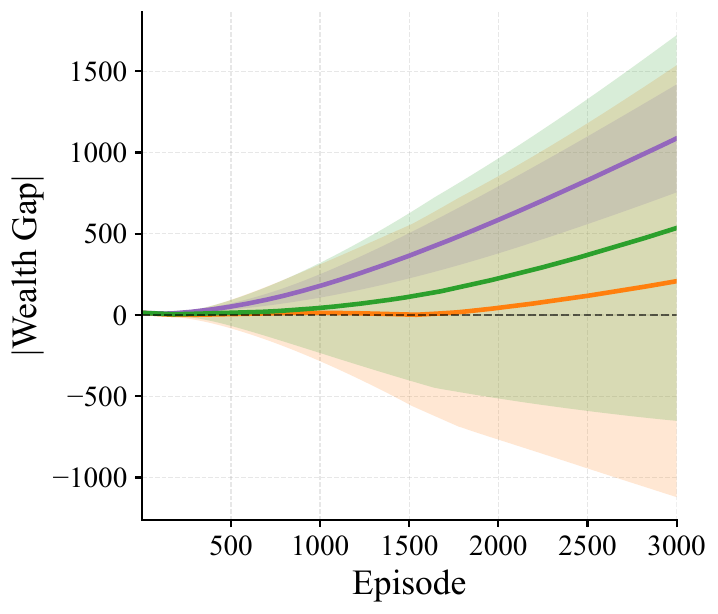}\vspace{-.5em}
        \subcaption{}\label{fig:pepg_eo_wg}
    \end{minipage}\hfill
    \begin{minipage}{0.45\textwidth}
        \centering
        \includegraphics[width=\textwidth]{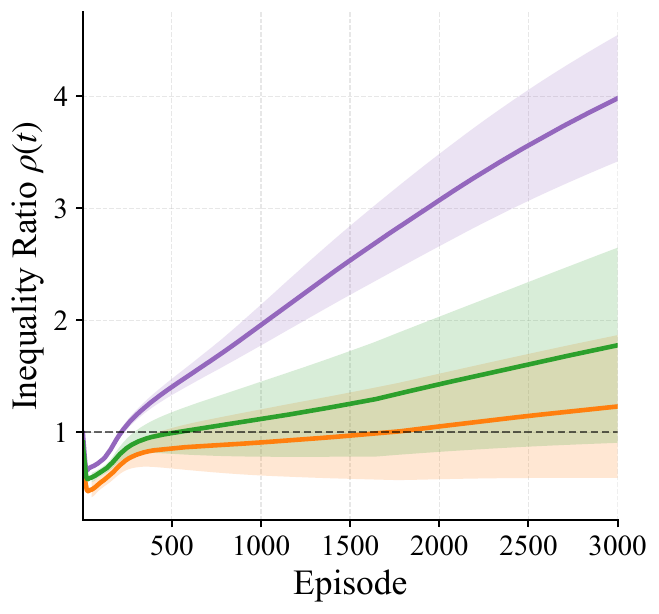}\vspace{-.5em}
        \subcaption{}\label{fig:pepg_eo_ineqr}
    \end{minipage}\\
    \begin{minipage}{0.45\textwidth}
        \centering
        \includegraphics[width=\textwidth]{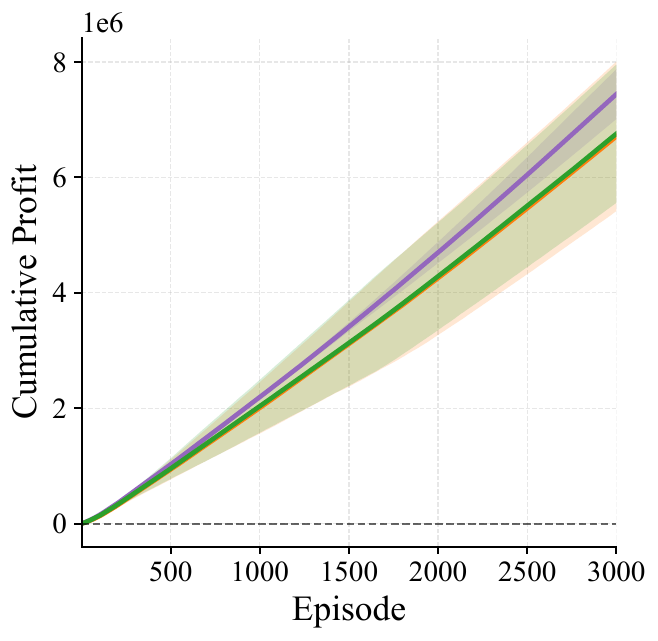}\vspace{-.5em}
        \subcaption{}\label{fig:pepg_eo_profit}
    \end{minipage}\hfill
    \begin{minipage}{0.45\textwidth}
        \centering
        \includegraphics[width=\textwidth]{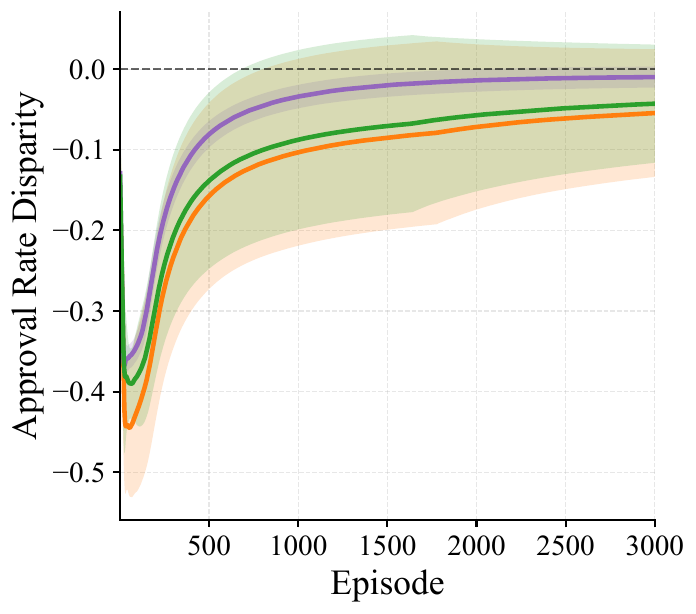}\vspace{-.5em}
        \subcaption{}\label{fig:pepg_eo_disp}
    \end{minipage}\\[0.5em]
    \includegraphics[width=0.45\textwidth]{iclr_2027/iclr_2027_plots/trajectories_run11/trajectories_reward_legend.pdf}
    \caption{\textbf{PeRL: Equality of Opportunity (OPP).} (a) wealth gap, (b) inequality ratio, (c) cumulative profit, (d) approval-rate disparity (red minus blue), for PeRL under FL, RMM and SW with the Equality of Opportunity perspective, over $3{,}000$ deployment episodes. Mean $\pm$ std across $10$ seeds.(Section~\ref{sec:experimental_analysis}).}\label{fig:pepg_eo}
\end{figure*}

\begin{figure*}[h!]
    \centering\vspace*{-1em}
    \begin{minipage}{0.45\textwidth}
        \centering
        \includegraphics[width=\textwidth]{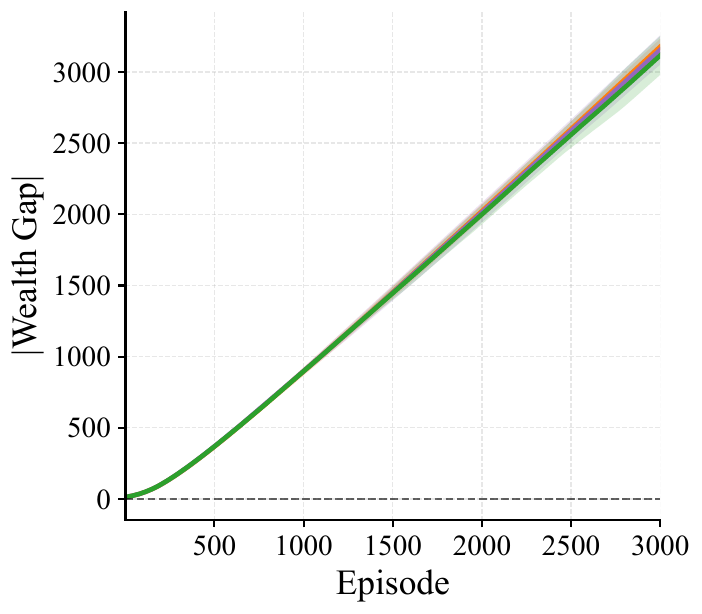}\vspace{-.5em}
        \subcaption{}\label{fig:pg_eo_wg}
    \end{minipage}\hfill
    \begin{minipage}{0.45\textwidth}
        \centering
        \includegraphics[width=\textwidth]{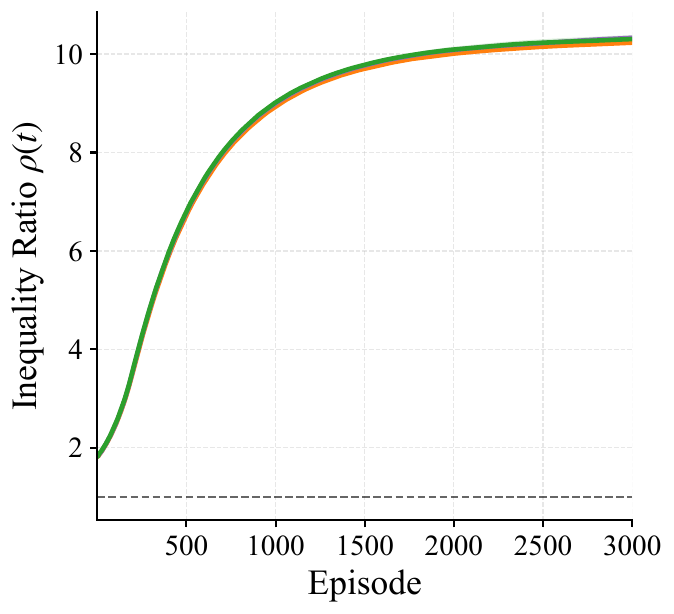}\vspace{-.5em}
        \subcaption{}\label{fig:pg_eo_ineqr}
    \end{minipage}\\
    \begin{minipage}{0.45\textwidth}
        \centering
        \includegraphics[width=\textwidth]{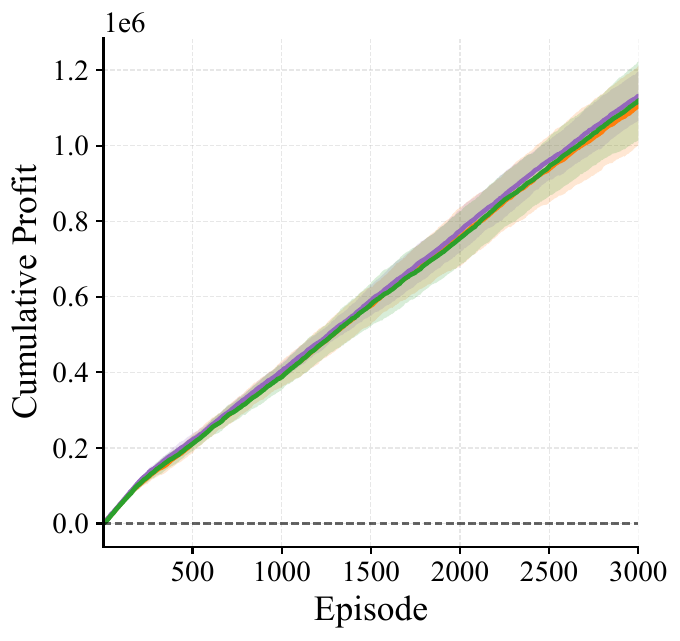}\vspace{-.5em}
        \subcaption{}\label{fig:pg_eo_profit}
    \end{minipage}\hfill
    \begin{minipage}{0.45\textwidth}
        \centering
        \includegraphics[width=\textwidth]{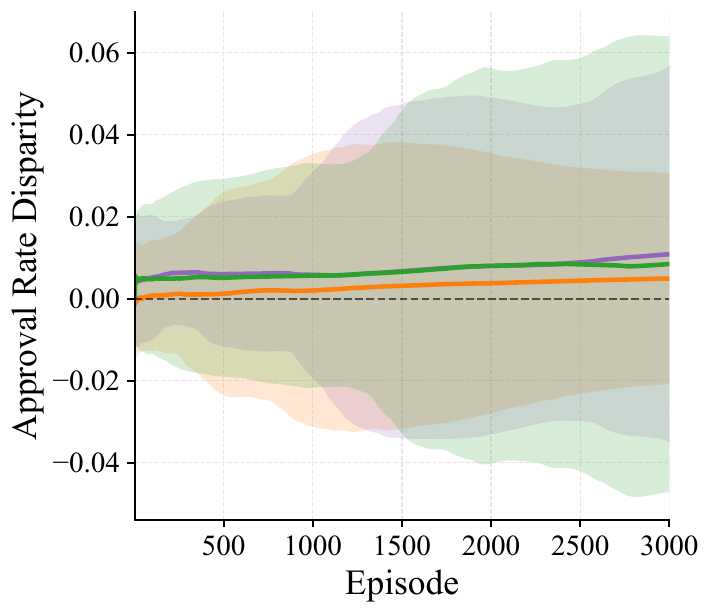}\vspace{-.5em}
        \subcaption{}\label{fig:pg_eo_disp}
    \end{minipage}\\[0.5em]
    \includegraphics[width=0.45\textwidth]{iclr_2027/iclr_2027_plots/trajectories_run11/trajectories_reward_legend.pdf}
    \caption{\textbf{RL: Equality of Opportunity (OPP).} (a) wealth gap, (b) inequality ratio, (c) cumulative profit, (d) approval-rate disparity (red minus blue), for RL under FL, RMM and SW with the Equality of Opportunity perspective, over $3{,}000$ deployment episodes. Mean $\pm$ std across $10$ seeds. (Section~\ref{sec:experimental_analysis}).}\label{fig:pg_eo}
\end{figure*}

Table~\ref{tab:app_rl_vs_perl} reports end points; this section follows each metric over deployment for PeRL (Figures~\ref{fig:pepg_sp} and~\ref{fig:pepg_eo}; RL in Figures~\ref{fig:pg_sp} and~\ref{fig:pg_eo}), with the three constraints overlaid in each panel. Two patterns recur. Every PeRL cell begins deployment by favouring the blue group: $\gamma^t$ is reset to $\gamma^0$ at deployment, and early on the fairness term outweighs profit. What follows depends on whether the constraint pushes back once the blue group catches up: FL penalises the gap in either direction, while RMM and SW do not, and under Equality of Outcome the arrival dynamics (Section~\ref{sec:dynamics}) then amplify whichever group is ahead.

\paragraph{Wealth gap.}
Under Equality of Outcome the constraints separate completely (Figure~\ref{fig:pepg_sp_wg}). FL keeps $|\Delta_\mu^t|$ below $100$ throughout: the gap closes by episode $170$, opens in the blue group's favour to $50$ around episode $600$, closes again at episode $1{,}150$, and ends at $72$ with the red group ahead, after peaking at $82$. RMM and SW open the gap in the blue group's favour and never close it: RMM to $3{,}082$ with negligible seed variation, close to the magnitude RL's approve-all behaviour produces in the other direction, and SW to $1{,}956$ ($\pm 1{,}013$). Under Equality of Opportunity (Figure~\ref{fig:pepg_eo_wg}) the gap stays below $150$ for the first $900$ (FL), $1{,}700$ (SW) and $2{,}700$ (RMM) episodes, and then grows in the red group's favour, to $1{,}087$ for FL, $535$ for SW and $208$ for RMM. The RMM and SW bands are the widest in the figure ($\pm 1{,}329$ and $\pm 1{,}187$), because seeds differ in which group ends up ahead. RL follows the approve-all diagonal in five scenarios; under FL with Equality of Outcome it opens a gap of $745$ in the blue group's favour by under-lending to the red group.

\paragraph{Inequality ratio.}
$\rho^t$ makes the direction explicit (Figures~\ref{fig:pepg_sp_ineqr} and~\ref{fig:pepg_eo_ineqr}; values in the first few episodes are ratios of very small changes). Under Equality of Outcome, FL overshoots and recovers: it falls to $0.49$ by episode $200$, crosses $1$ at episode $1{,}235$, peaks at $1.12$ near episode $1{,}750$ and ends at $1.06$, pulled back by the two-sided gap penalty. RMM and SW have no such pull: RMM's $\rho$ peaks at $0.29$ and decays to $0.10$, SW's peaks at $0.66$ around episode $150$ and decays to $0.14$, so the blue group grows seven to ten times faster, a mirror image of the fixed rules (Appendix~\ref{app:heuristic_results}). Under Equality of Opportunity all three cross $1$ and keep rising: FL at episode $220$, to $3.98$; SW at $530$, to $1.78$; RMM at $1{,}730$, to $1.23$. Equalising true positive rates constrains who is approved among the creditworthy, not how much wealth each group accumulates, so the Matthew effect reopens the growth-rate gap once the constraint is met; the blue group's welfare $R_F^t$ (not shown) peaks between episodes $450$ and $1{,}800$ and then declines as its application rate falls.

\paragraph{Profit.}
Profit separates by perspective (Figures~\ref{fig:pepg_sp_profit} and~\ref{fig:pepg_eo_profit}). Under Equality of Opportunity all three constraints earn $6.7$ to $7.4$M, six to seven times RL, at a nearly constant rate (about $1.0$M by episode $500$ and $2.0$ to $2.2$M by $1{,}000$): approving the creditworthy of both groups at similar rates is close to what a profit-maximising policy does. Under Equality of Outcome FL and RMM earn $1.30$M ($\pm 0.97$M) and $1.27$M, a fifth of their Opportunity counterparts, at a near-linear rate. SW under Outcome earns $4.72$M on a convex curve ($0.53$M at episode $500$, $1.08$M at $1{,}000$, $2.57$M at $2{,}000$), accelerating as the blue group's wealth, and with it its application rate, grows.

\paragraph{Approval-rate disparity.}
The cumulative disparity $\mathrm{app}_{\mathrm{red}} - \mathrm{app}_{\mathrm{blue}}$ (Figures~\ref{fig:pepg_sp_disp} and~\ref{fig:pepg_eo_disp}; negative when the blue group is approved at the higher rate) shows the early tilt: at episode $100$ every PeRL cell favours the blue group, by $0.17$ (FL) to $0.69$ (RMM) under Outcome and by $0.34$ to $0.42$ under Opportunity. Under Opportunity it then decays to between $0.01$ and $0.05$ as the true positive rates equalise and $\gamma^t$ falls. Under Outcome, FL reaches $-0.35$ around episode $200$ and $+0.12$ around episode $1{,}500$ before ending at $0.01$, the same restoring pattern as its gap, while RMM and SW end at $-0.11$ and $-0.21$. RL's disparities stay within $0.01$ of zero in its approve-all scenarios; under FL with Outcome it starts at $-0.80$ and comes within $0.01$ of zero only after about $1{,}650$ episodes.

\paragraph{Summary.}
Under Equality of Outcome the constraint sets the profile: FL closes the gap and keeps $\rho$ near $1$, RMM and SW let the blue group overtake, and only SW turns that into profit. Under Equality of Opportunity the perspective sets it: all three constraints earn the same high profit and differ mainly on the wealth axes, where RMM and SW end closest to balance and FL furthest from it. A ADM prioritising the wealth gap is served by FL under Outcome, at a cost in profit; one prioritising profit, by any constraint under Opportunity, at a cost in wealth equality. Appendix~\ref{app:pareto} compares the six profiles, and Appendix~\ref{app:balanced} examines a single policy that delivers both.

\subsubsection{Macro-analysis: Pareto Frontiers of Utilities}
\label{app:pareto}
\begin{figure*}[t!]
    \centering\vspace*{-1em}
    \begin{minipage}{0.45\textwidth}
        \centering
        \includegraphics[width=\textwidth]{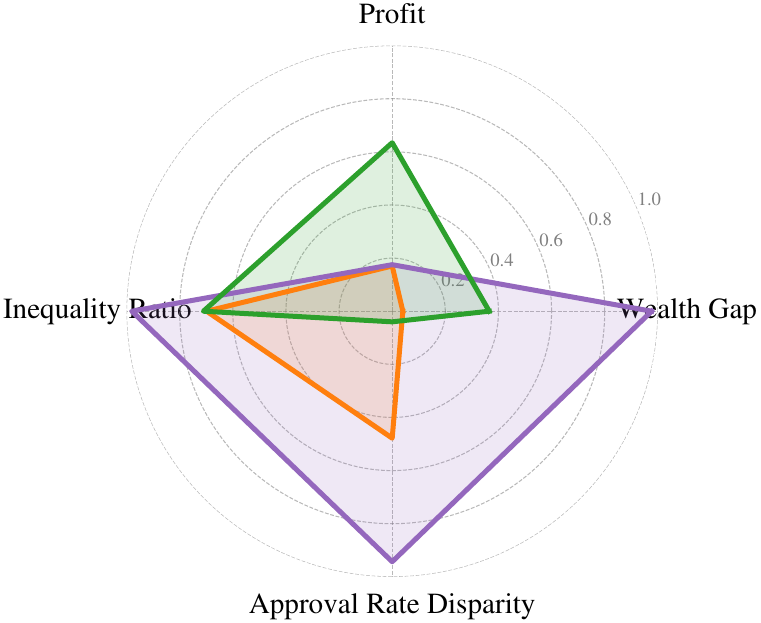}\vspace{-.5em}
        \subcaption{Equality of Outcome (OUT)}\label{fig:radar_sp}
    \end{minipage}\hfill
    \begin{minipage}{0.45\textwidth}
        \centering
        \includegraphics[width=\textwidth]{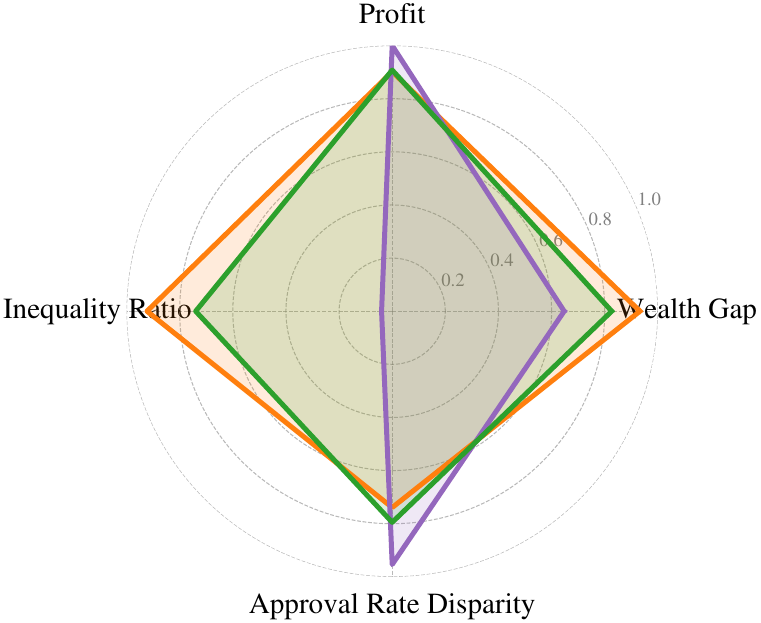}\vspace{-.5em}
        \subcaption{Equality of Opportunity (OPP)}\label{fig:radar_eo}
    \end{minipage}\\[0.5em]
    \includegraphics[width=0.45\textwidth]{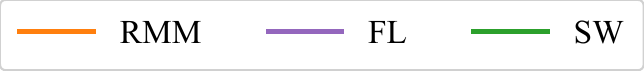}
    \caption{\textbf{PeRL: performance profiles.} End-of-deployment values of the four metrics of Table~\ref{tab:app_rl_vs_perl} for PeRL under FL, RMM and SW, one chart per fairness perspective. Each axis is normalised over all six scenarios, with $1$ the best value observed and $0$ the worst, and oriented so that outward is better (profit as is; wealth gap, $|\rho - 1|$ and approval-rate disparity inverted). The two charts share one scale, so polygons are comparable across them. Larger and more even is better.}\label{fig:radar}
\end{figure*}


Figure~\ref{fig:radar} folds the four metrics of Table~\ref{tab:app_rl_vs_perl} into one profile per PeRL cell; RL is omitted, since its five approve-all scenarios give nearly the same polygon. Each axis is normalised over the six scenarios, so the rim is the best value any cell reached and the centre the worst, and the lower-is-better metrics are inverted so that outward is always better. Two cautions apply. The scale is relative to these six scenarios, not absolute. And the inequality axis uses the linear distance $|\rho - 1|$, so $\rho = 0.10$ (the blue group growing ten times faster) sits at $0.70$ while $\rho = 3.98$ sits at the centre: the axis compresses inequality in the blue group's favour.

\paragraph{Equality of Outcome.}
The three polygons do not overlap (Figure~\ref{fig:radar_sp}). FL reaches the rim on all three fairness axes (wealth gap $0.98$, inequality $0.98$, disparity $0.94$) and the centre on profit ($0.18$, $1.30$M): it is the fairest cell in the grid on every fairness measure. RMM sits at the centre on wealth gap ($0.04$, the largest gap of the six scenarios, $3{,}082$) and at $0.48$ on disparity, with a profit no higher than FL's; its $0.70$ on inequality is the compression just described. SW is the only Outcome cell with a substantial profit vertex ($0.63$, $4.72$M), but it is the worst cell in the grid on disparity ($0.04$, a cumulative disparity of $0.21$ in the blue group's favour). Under Outcome, then, the choice is between FL's fairness and SW's profit.

\paragraph{Equality of Opportunity.}
All three polygons reach the rim on profit ($0.90$ to $1.00$, i.e.\ $6.7$M to $7.4$M; Figure~\ref{fig:radar_eo}), because equalising true positive rates costs the ADM little (Appendix~\ref{app:perl_metric_by_metric}); the wealth axes separate them. RMM has the largest and most even polygon in either chart: wealth gap $0.93$, profit $0.90$, inequality $0.92$, disparity $0.74$, i.e.\ a gap of $208$, $6.71$M, $\rho = 1.23$ and a disparity of $0.054$. SW has a slightly smaller version of the same shape ($0.83$, $0.91$, $0.74$, $0.79$). FL is the most profitable cell ($1.00$) and nearly the best on disparity ($0.95$), but sits at the centre on inequality: $\rho = 3.98$, the largest inequality in the red group's favour of any cell. Under Opportunity the perspective secures profit and disparity, and the constraint decides the wealth axes: RMM and SW cover them, FL does not.

\paragraph{The frontier.}
Treating the six scenarios as candidate policies over the four metrics, two are dominated: RMM and SW under Outcome are each worse than the same constraint under Opportunity on all four. The other four are undominated: FL under Opportunity for profit; SW under Opportunity for a slightly higher profit and lower disparity than RMM, at a wider gap; RMM under Opportunity for the wealth axes among the Opportunity scenarios; and FL under Outcome for the best wealth gap, inequality ratio and disparity of any cell. Since our objective is the fairness outcome, with profit a constraint rather than a target, FL under Outcome is the reference point. It keeps the two groups growing at the same rate, holds the gap near zero and treats the groups alike at the point of decision, and its $\rho$ has settled by the end of deployment, whereas RMM under Opportunity, its nearest competitor on the wealth axes, crosses $1$ only around episode $1{,}730$ and is still rising. What FL under Outcome lacks is profit, a fifth of the Opportunity scenarios'. The frontier thus has FL at both ends: the same constraint gives the most profitable cell under Opportunity ($7.44$M) and the fairest under Outcome ($1.30$M), at opposite corners of the radar. The two scenarios differ only in what the constraint is applied to and in how the multiplier evolves during deployment (Appendix~\ref{app:lambda_trajectories}). Appendix~\ref{app:balanced} therefore keeps the FL constraint and the Outcome perspective but fixes the multiplier, to see whether one policy can take the fairness of the first cell and the profit of the second.

\subsubsection{Micro-analysis: Who Receives the Loans}
\label{app:micro}
\label{sec:inclusivity}

\begin{figure*}[t!]
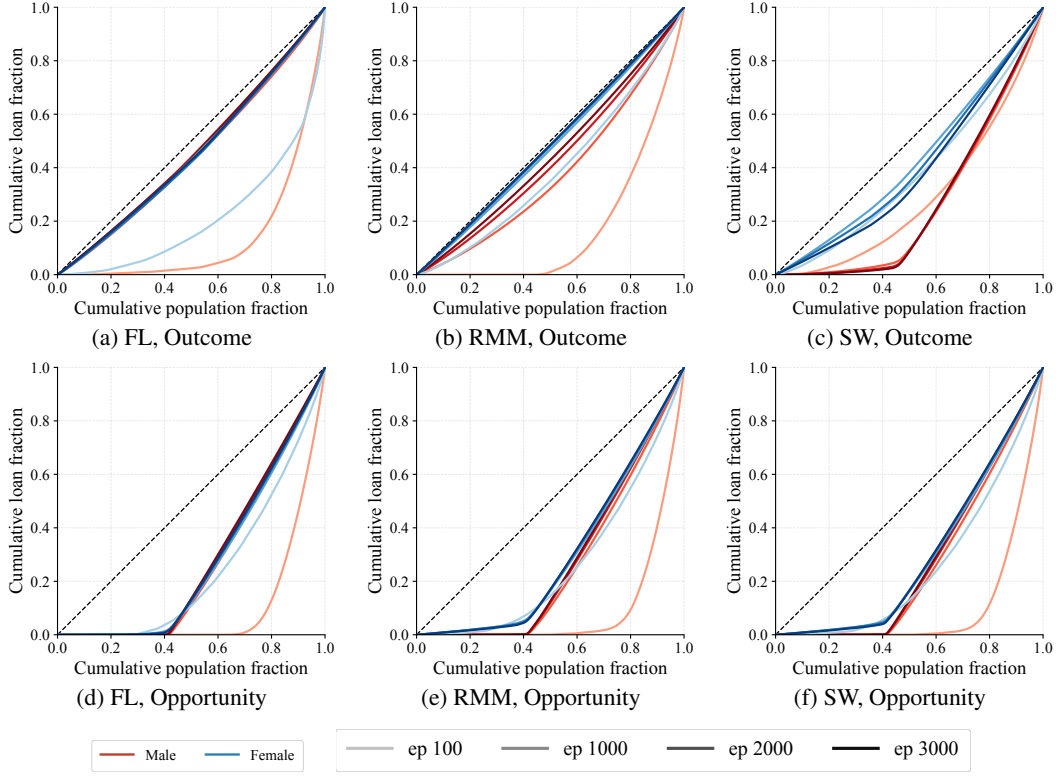

    \centering\vspace*{-1em}
    \begin{minipage}{0.32\textwidth}
        \centering
        \includegraphics[width=\textwidth]{iclr_2027/iclr_2027_plots/lorenz_run11/lorenz_pepg_social_fairness_lagrangian.pdf}\vspace{-.5em}
        \subcaption{FL, Outcome}\label{fig:lorenz_out_fl}
    \end{minipage}\hfill
    \begin{minipage}{0.32\textwidth}
        \centering
        \includegraphics[width=\textwidth]{iclr_2027/iclr_2027_plots/lorenz_run11/lorenz_pepg_social_rawlsian_maximin.pdf}\vspace{-.5em}
        \subcaption{RMM, Outcome}\label{fig:lorenz_out_rmm}
    \end{minipage}\hfill
    \begin{minipage}{0.32\textwidth}
        \centering
        \includegraphics[width=\textwidth]{iclr_2027/iclr_2027_plots/lorenz_run11/lorenz_pepg_social_social_welfare.pdf}\vspace{-.5em}
        \subcaption{SW, Outcome}\label{fig:lorenz_out_sw}
    \end{minipage}\\
    \begin{minipage}{0.32\textwidth}
        \centering
        \includegraphics[width=\textwidth]{iclr_2027/iclr_2027_plots/lorenz_run11/lorenz_pepg_eo_fairness_lagrangian.pdf}\vspace{-.5em}
        \subcaption{FL, Opportunity}\label{fig:lorenz_opp_fl}
    \end{minipage}\hfill
    \begin{minipage}{0.32\textwidth}
        \centering
        \includegraphics[width=\textwidth]{iclr_2027/iclr_2027_plots/lorenz_run11/lorenz_pepg_eo_rawlsian_maximin.pdf}\vspace{-.5em}
        \subcaption{RMM, Opportunity}\label{fig:lorenz_opp_rmm}
    \end{minipage}\hfill
    \begin{minipage}{0.32\textwidth}
        \centering
        \includegraphics[width=\textwidth]{iclr_2027/iclr_2027_plots/lorenz_run11/lorenz_pepg_eo_social_welfare.pdf}\vspace{-.5em}
        \subcaption{SW, Opportunity}\label{fig:lorenz_opp_sw}
    \end{minipage}\\[0.5em]
    \includegraphics[width=0.2\textwidth]{iclr_2027/iclr_2027_plots/lorenz_run11/lorenz_group_legend.pdf}\hspace{1em}
    \includegraphics[width=0.6\textwidth]{iclr_2027/iclr_2027_plots/lorenz_run11/lorenz_episode_shade_legend.pdf}
    \caption{\textbf{PeRL: Lorenz curves of cumulative loans.} Cumulative loan fraction against cumulative population fraction within each group (red: male, blue: female), at episodes $100$ to $3{,}000$ (light to dark), for PeRL under FL, RMM and SW with the Equality of Outcome (top) and Equality of Opportunity (bottom) perspectives. Curves are averaged over $10$ seeds. The dashed diagonal is perfect within-group equality.}\label{fig:lorenz}
\end{figure*}

\begin{figure*}[t!]
    \centering\vspace*{-1em}
    \begin{minipage}{0.32\textwidth}
        \centering
        \includegraphics[width=\textwidth]{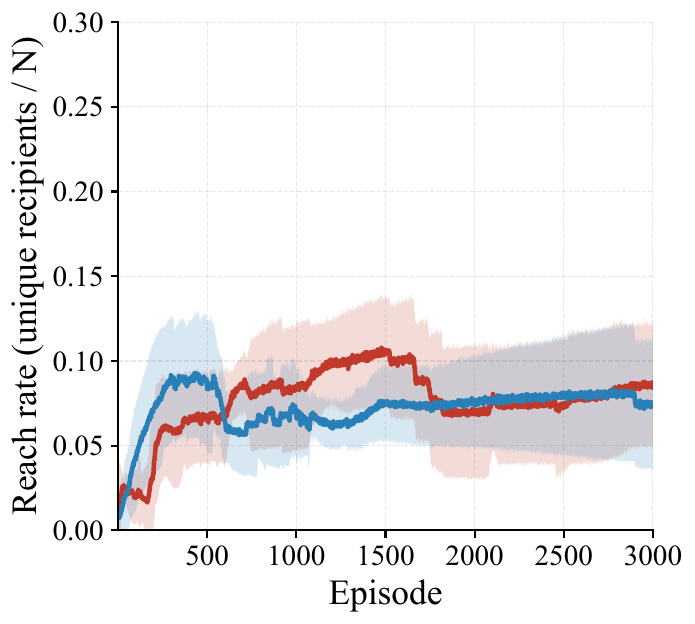}\vspace{-.5em}
        \subcaption{FL, Outcome}\label{fig:reach_out_fl}
    \end{minipage}\hfill
    \begin{minipage}{0.32\textwidth}
        \centering
        \includegraphics[width=\textwidth]{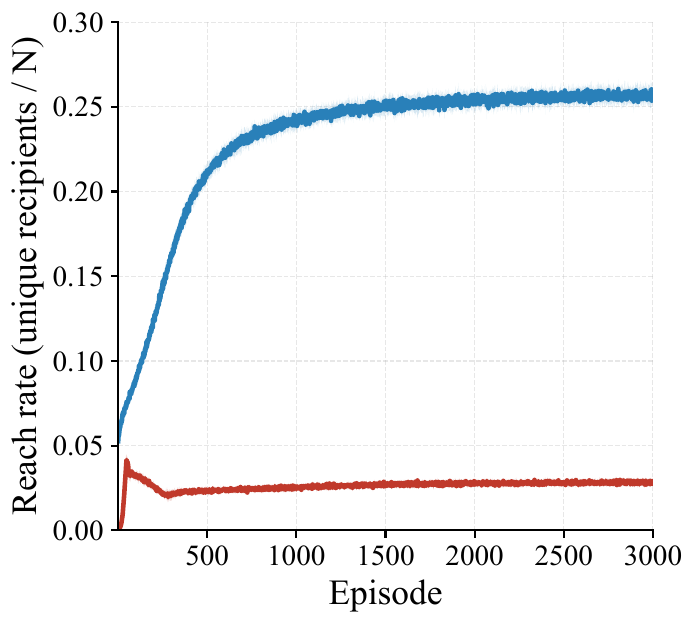}\vspace{-.5em}
        \subcaption{RMM, Outcome}\label{fig:reach_out_rmm}
    \end{minipage}\hfill
    \begin{minipage}{0.32\textwidth}
        \centering
        \includegraphics[width=\textwidth]{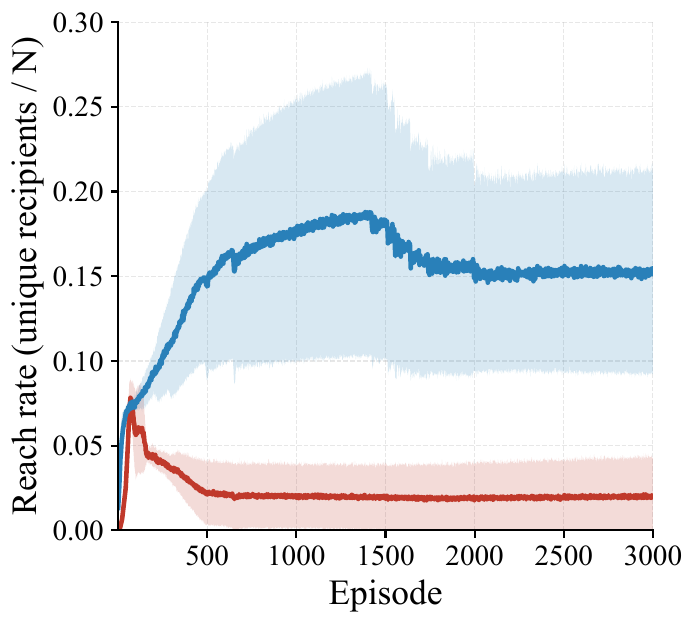}\vspace{-.5em}
        \subcaption{SW, Outcome}\label{fig:reach_out_sw}
    \end{minipage}\\
    \begin{minipage}{0.32\textwidth}
        \centering
        \includegraphics[width=\textwidth]{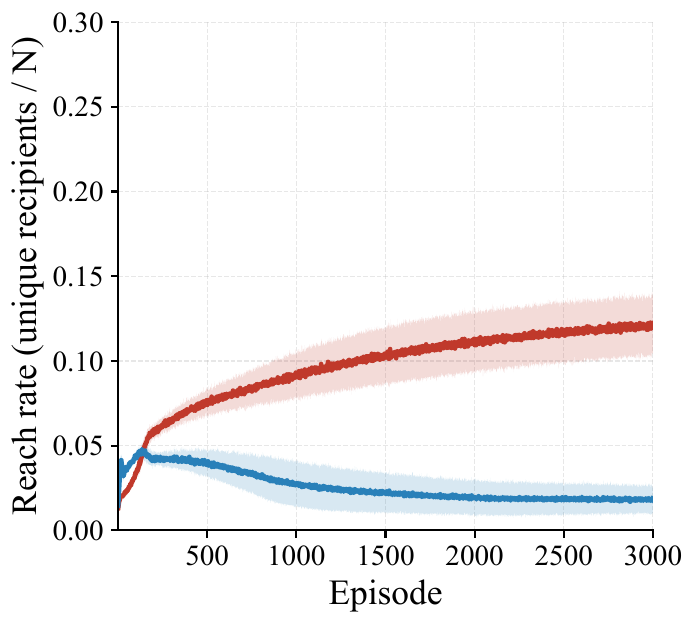}\vspace{-.5em}
        \subcaption{FL, Opportunity}\label{fig:reach_opp_fl}
    \end{minipage}\hfill
    \begin{minipage}{0.32\textwidth}
        \centering
        \includegraphics[width=\textwidth]{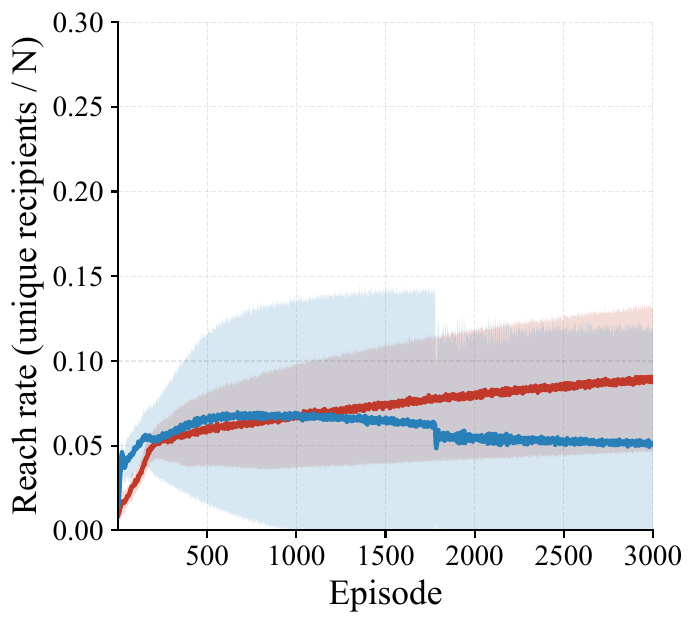}\vspace{-.5em}
        \subcaption{RMM, Opportunity}\label{fig:reach_opp_rmm}
    \end{minipage}\hfill
    \begin{minipage}{0.32\textwidth}
        \centering
        \includegraphics[width=\textwidth]{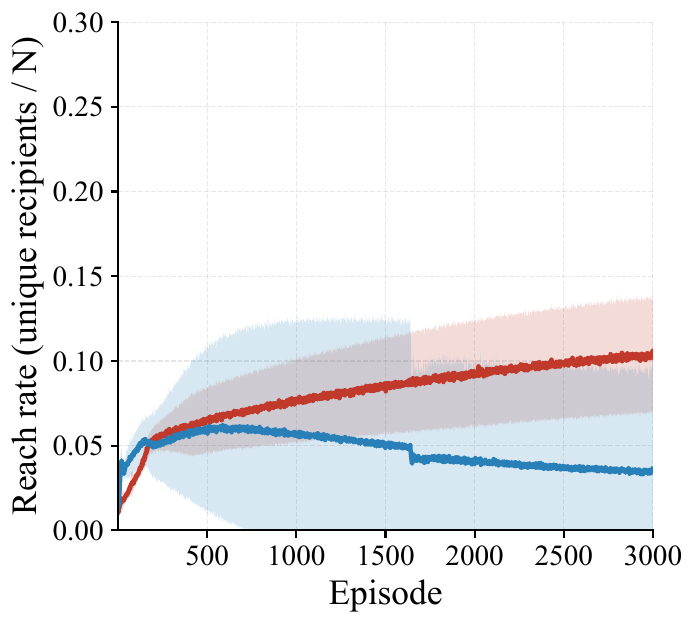}\vspace{-.5em}
        \subcaption{SW, Opportunity}\label{fig:reach_opp_sw}
    \end{minipage}\\[0.5em]
    \includegraphics[width=0.2\textwidth]{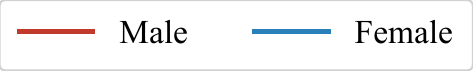}
    \caption{\textbf{How much of each group PeRL reaches over time.} Share of the red (male) and blue (female) groups receiving a loan in each episode; mean $\pm$ one standard deviation over $10$ seeds. Top: Equality of Outcome; bottom: Equality of Opportunity.}\label{fig:reach}
\end{figure*}

\begin{figure*}[t!]
    \centering\vspace*{-1em}
    \begin{minipage}{0.32\textwidth}
        \centering
        \includegraphics[width=\textwidth]{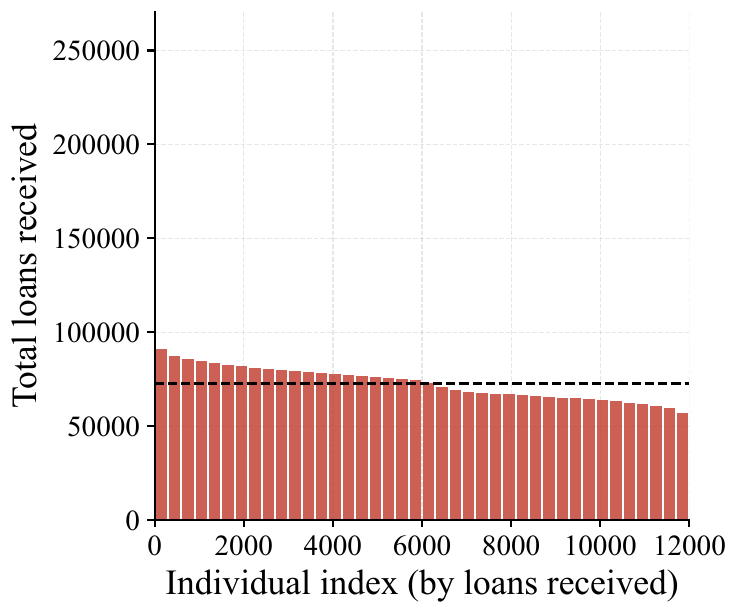}\vspace{-.5em}
        \subcaption{FL, Male}\label{fig:loandist_out_fl_m}
    \end{minipage}\hfill
    \begin{minipage}{0.32\textwidth}
        \centering
        \includegraphics[width=\textwidth]{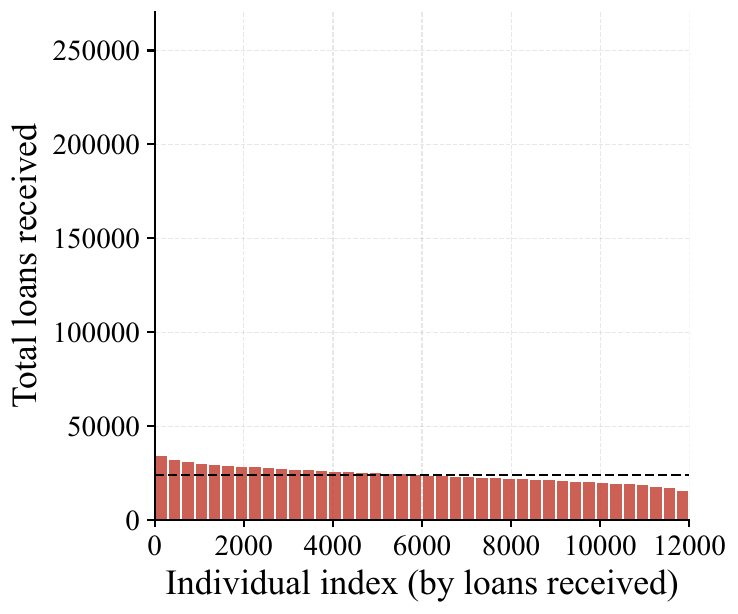}\vspace{-.5em}
        \subcaption{RMM, Male}\label{fig:loandist_out_rmm_m}
    \end{minipage}\hfill
    \begin{minipage}{0.32\textwidth}
        \centering
        \includegraphics[width=\textwidth]{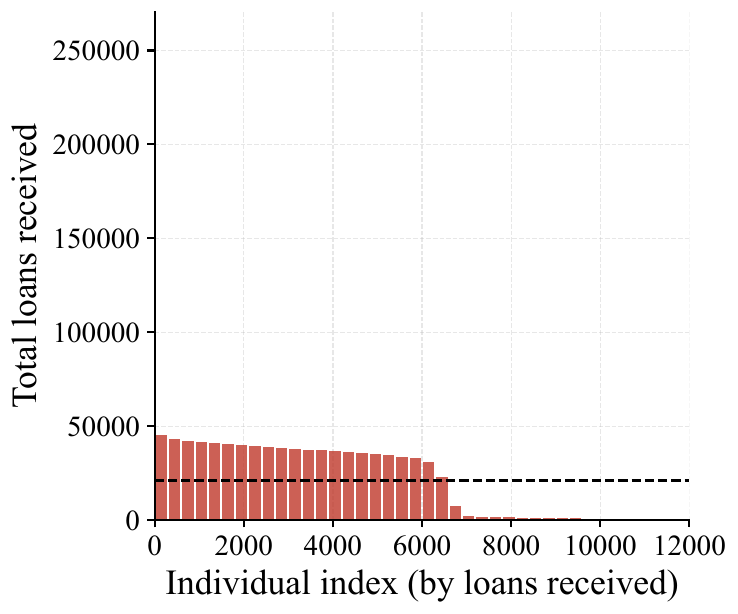}\vspace{-.5em}
        \subcaption{SW, Male}\label{fig:loandist_out_sw_m}
    \end{minipage}\\
    \begin{minipage}{0.32\textwidth}
        \centering
        \includegraphics[width=\textwidth]{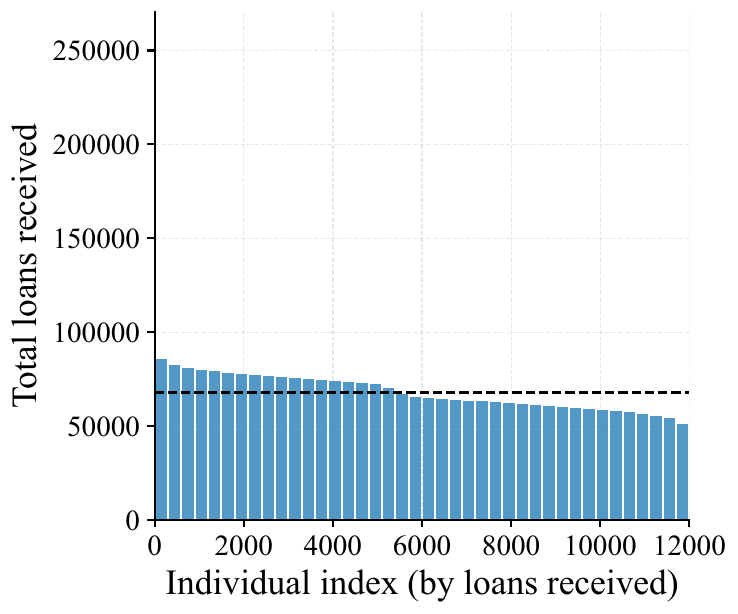}\vspace{-.5em}
        \subcaption{FL, Female}\label{fig:loandist_out_fl_f}
    \end{minipage}\hfill
    \begin{minipage}{0.32\textwidth}
        \centering
        \includegraphics[width=\textwidth]{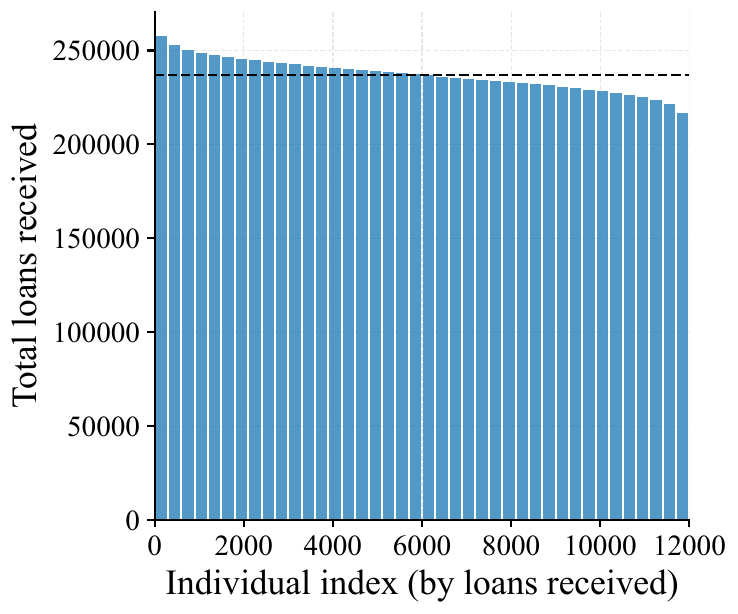}\vspace{-.5em}
        \subcaption{RMM, Female}\label{fig:loandist_out_rmm_f}
    \end{minipage}\hfill
    \begin{minipage}{0.32\textwidth}
        \centering
        \includegraphics[width=\textwidth]{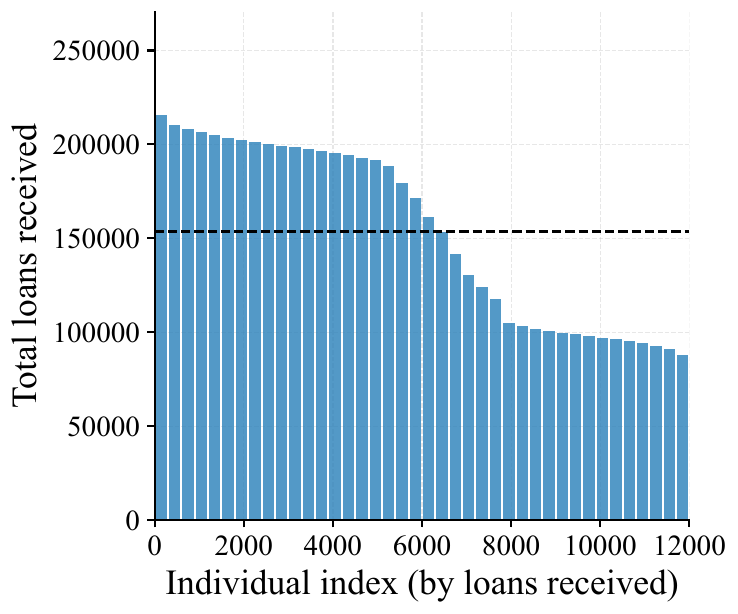}\vspace{-.5em}
        \subcaption{SW, Female}\label{fig:loandist_out_sw_f}
    \end{minipage}\\[0.5em]
    \includegraphics[width=0.3\textwidth]{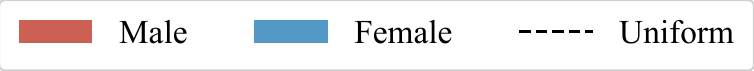}
    \caption{\textbf{Loans received by each individual, PeRL under Equality of Outcome.} Individuals are ordered from most to least served, in bins of $300$; each bar is a bin's total loans over deployment, and the dashed line is what every bin would receive under equal shares. Top: male group; bottom: female group. Mean over $10$ seeds.}\label{fig:loandist_out}
\end{figure*}

\begin{figure*}[t!]
    \centering\vspace*{-1em}
    \begin{minipage}{0.32\textwidth}
        \centering
        \includegraphics[width=\textwidth]{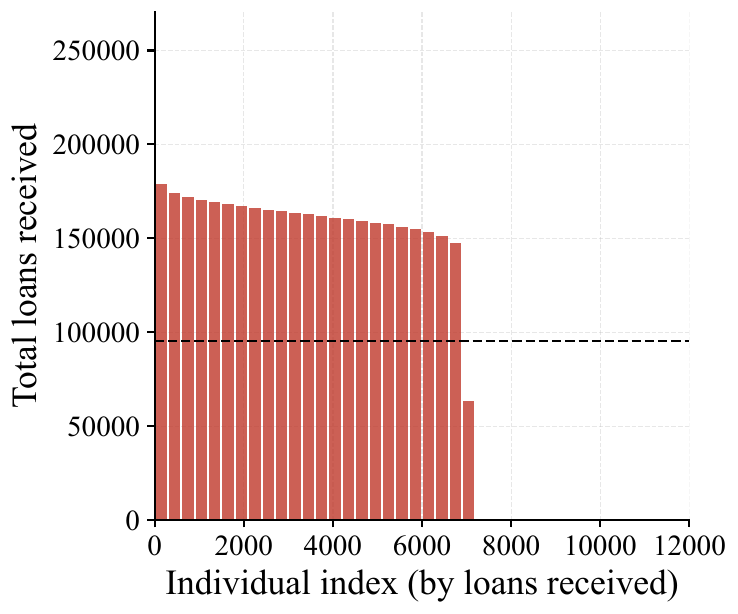}\vspace{-.5em}
        \subcaption{FL, Male}\label{fig:loandist_opp_fl_m}
    \end{minipage}\hfill
    \begin{minipage}{0.32\textwidth}
        \centering
        \includegraphics[width=\textwidth]{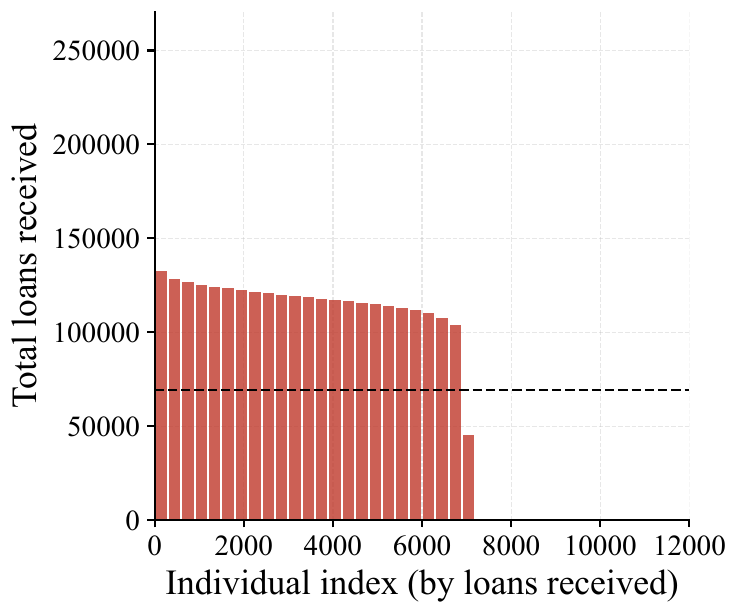}\vspace{-.5em}
        \subcaption{RMM, Male}\label{fig:loandist_opp_rmm_m}
    \end{minipage}\hfill
    \begin{minipage}{0.32\textwidth}
        \centering
        \includegraphics[width=\textwidth]{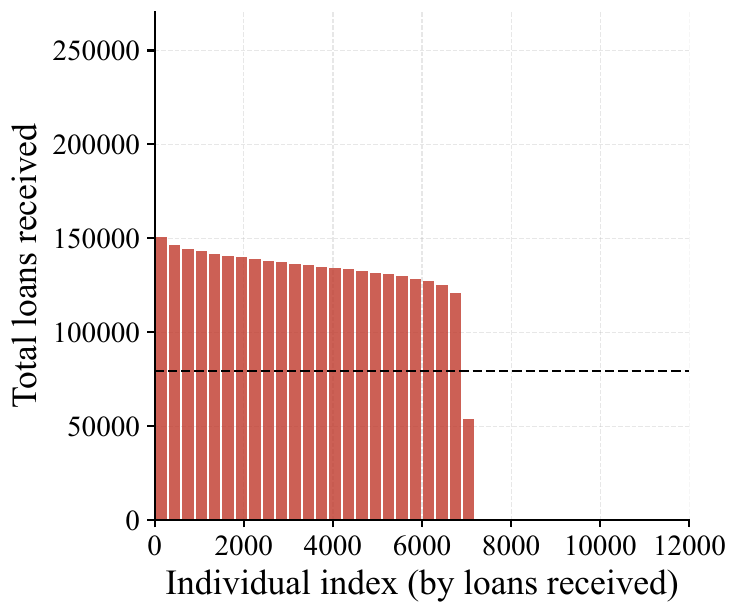}\vspace{-.5em}
        \subcaption{SW, Male}\label{fig:loandist_opp_sw_m}
    \end{minipage}\\
    \begin{minipage}{0.32\textwidth}
        \centering
        \includegraphics[width=\textwidth]{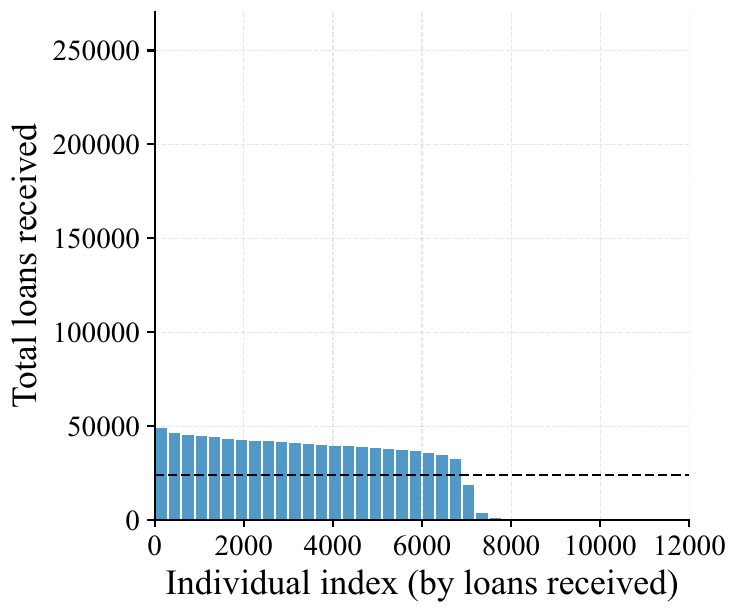}\vspace{-.5em}
        \subcaption{FL, Female}\label{fig:loandist_opp_fl_f}
    \end{minipage}\hfill
    \begin{minipage}{0.32\textwidth}
        \centering
        \includegraphics[width=\textwidth]{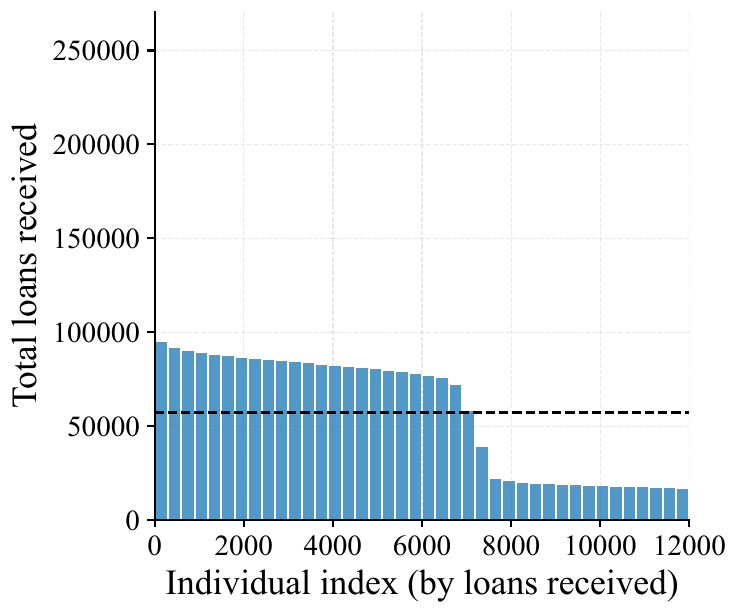}\vspace{-.5em}
        \subcaption{RMM, Female}\label{fig:loandist_opp_rmm_f}
    \end{minipage}\hfill
    \begin{minipage}{0.32\textwidth}
        \centering
        \includegraphics[width=\textwidth]{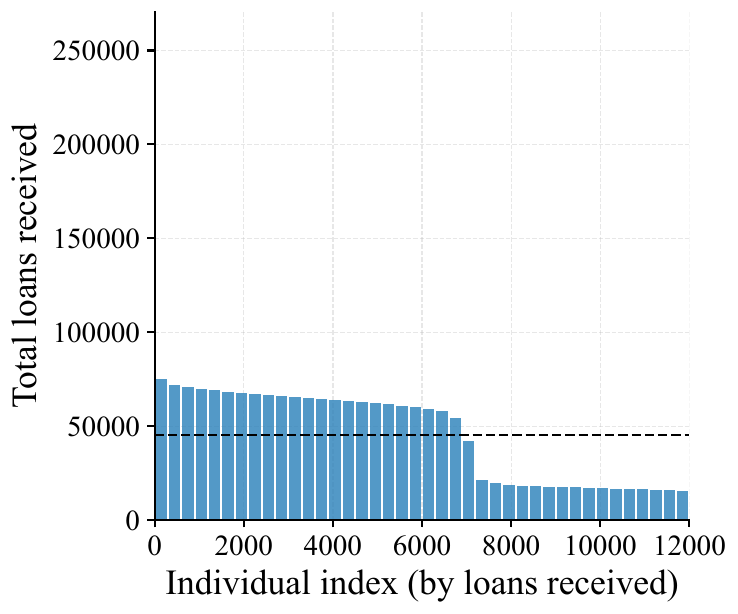}\vspace{-.5em}
        \subcaption{SW, Female}\label{fig:loandist_opp_sw_f}
    \end{minipage}\\[0.5em]
    \includegraphics[width=0.3\textwidth]{iclr_2027/iclr_2027_plots/loandist_run11/loandist_group_legend.pdf}
    \caption{\textbf{Loans received by each individual, PeRL under Equality of Opportunity.} As in Figure~\ref{fig:loandist_out}, on the same axes. The drop to zero on the right of each panel marks the individuals who were never served.}\label{fig:loandist_opp}
\end{figure*}

A wealth gap of zero says that the two groups' mean wealth coincide, not how the loans behind it were spread within each group. For each PeRL cell we record how many loans each of the $12{,}000$ individuals per group received, and read that record in three ways:
\begin{itemize}
    \item \textbf{Lorenz curves} (Figure~\ref{fig:lorenz_main}, also expanded in Figure~\ref{fig:lorenz}), at episodes $100$, $1{,}000$, $2{,}000$ and $3{,}000$: the cumulative share of a group's loans held by its least-served members. The diagonal is equal shares; the further a curve sags below it, the more concentrated the group's credit, and a flat stretch on the left is the part of the group that received almost nothing. The Gini coefficient summarises a curve in one number, $0$ for equal shares.
    \item \textbf{Reach rate} (Figure~\ref{fig:reach}): the share of a group that receives a loan in each episode, i.e.\ the breadth of access over time.
    \item \textbf{Loan distributions} (Figures~\ref{fig:loandist_out} and~\ref{fig:loandist_opp}): total loans per individual at the end of deployment, from the most to the least served, against the equal-share line.
\end{itemize}
Table~\ref{tab:micro} collects the corresponding end-of-deployment numbers.

\begin{table}[t!]
\centering
\caption{\textbf{Who receives the loans under PeRL, at the end of deployment.} Loans per individual; share of the group never served; share of the group's loans held by its most-served $10\%$ ($0.10$ means equal shares); Gini coefficient ($0$ means equal shares); and reach rate over the last $100$ episodes. R: red (male), B: blue (female). Means over $10$ seeds.}
\label{tab:micro}
\small
\begin{tabular}{ll cc cc cc cc cc}
\toprule
& & \multicolumn{2}{c}{Loans / individual} & \multicolumn{2}{c}{Never served} & \multicolumn{2}{c}{Top-10\% share} & \multicolumn{2}{c}{Gini} & \multicolumn{2}{c}{Reach rate} \\
\cmidrule(lr){3-4}\cmidrule(lr){5-6}\cmidrule(lr){7-8}\cmidrule(lr){9-10}\cmidrule(lr){11-12}
Persp. & Constr. & R & B & R & B & R & B & R & B & R & B \\
\midrule
OUT & FL  & 243 & 227 & 0\%  & 0\%  & 0.12 & 0.13 & 0.08 & 0.09 & 0.086 & 0.074 \\
OUT & RMM & 80  & 790 & 0\%  & 0\%  & 0.13 & 0.11 & 0.10 & 0.02 & 0.028 & 0.257 \\
OUT & SW  & 71  & 512 & 6\%  & 0\%  & 0.21 & 0.15 & 0.47 & 0.22 & 0.020 & 0.152 \\
\midrule
OPP & FL  & 318 & 80  & 24\% & 21\% & 0.18 & 0.20 & 0.43 & 0.44 & 0.120 & 0.018 \\
OPP & RMM & 231 & 191 & 26\% & 15\% & 0.19 & 0.18 & 0.44 & 0.40 & 0.089 & 0.051 \\
OPP & SW  & 265 & 152 & 25\% & 17\% & 0.19 & 0.18 & 0.44 & 0.40 & 0.103 & 0.035 \\
\bottomrule
\end{tabular}
\end{table}

\paragraph{Equality of Outcome.}
The three Outcome scenarios reach everyone: no individual goes without a loan under FL or RMM, and only $6\%$ of the red group does under SW (Table~\ref{tab:micro}). They differ in how the loans are shared.

Under FL the two groups end alike on every count. The Lorenz curves (Figure~\ref{fig:lorenz_main}a) start far from the diagonal at episode $100$ (Gini $0.73$ for the red group, $0.58$ for the blue), when few loans have been issued; by episode $1{,}000$ they are at $0.11$ and $0.10$, and they end at $0.08$ and $0.09$. The reach rate (Figure~\ref{fig:reach_out_fl}) rises from $2\%$ of each group per episode to $7$ to $9\%$ by episode $500$ and stays there. The loan distributions are the flattest of the twelve panels (Figures~\ref{fig:loandist_out_fl_m} and~\ref{fig:loandist_out_fl_f}): the most-served tenth of each group holds $12$ to $13\%$ of its loans, and the two groups receive nearly the same volume ($243$ and $227$ loans per individual). FL under Outcome is the one cell that shares loans evenly within each group and in equal volume across the two.

Under RMM each group is served as evenly as under FL, but in very different volumes. The blue group's Lorenz curve is on the diagonal from episode $1{,}000$ on (Gini $0.04$, falling to $0.02$), and the red group's approaches it more slowly ($0.66$ at episode $100$, $0.10$ at the end). The blue group receives $790$ loans per individual and the red $80$, ten to one, and the reach rates show the same split over time: $26\%$ of the blue group per episode against $3\%$ of the red. This is the micro-level counterpart of RMM's inverted inequality ratio ($\rho = 0.10$, Appendix~\ref{app:perl_metric_by_metric}).

SW is the one Outcome cell with within-group concentration. The red group's Lorenz curve (Figure~\ref{fig:lorenz_main}c) has a flat segment over the first $45\%$ of the population and a Gini of $0.47$ from episode $1{,}000$ on: nearly half the red group receives almost nothing (median $35$ loans, mean $71$) while the rest is served evenly. The blue group's curve moves away from the diagonal over deployment (Gini $0.15$ at episode $1{,}000$, $0.22$ at the end), and its distribution slopes down from $640$ to $720$ loans per individual over the most-served $45\%$ to about $300$ over the least-served third (Figure~\ref{fig:loandist_out_sw_f}). The blue reach rate peaks at $19\%$ around episode $1{,}400$ and settles at $15\%$; the red stays at $2\%$.

\paragraph{Equality of Opportunity.}
The three Opportunity scenarios look alike, and unlike any Outcome cell. From episode $1{,}000$ on, every Lorenz curve in the bottom row of Figure~\ref{fig:lorenz_main} is flat over roughly the first $40\%$ of the group and then rises straight to the corner: part of the group receives little or nothing, and the rest is served nearly uniformly. The Gini coefficients are $0.41$ to $0.47$ at episode $1{,}000$ and $0.40$ to $0.44$ at the end. The loan distributions show the same structure as a plateau followed by a drop (Figure~\ref{fig:loandist_opp}). For the red group it is the same in all three scenarios: the most-served $57\%$ sit on a common plateau, the next $15\%$ receive about one loan, and the last $25\%$ none, in every seed ($22$ to $30\%$ per seed). The blue group has the same profile under FL ($21\%$ never served). Under RMM and SW its mean profile shows a second, lower plateau of about $60$ loans over the last $40\%$, an average over seeds: in eight RMM seeds and nine SW seeds, $14$ to $21\%$ of the blue group are never served, while one RMM seed excludes $8\%$ and one seed of each serves the whole group.

What the constraint changes under Opportunity is the volume each group receives, not the shape. The red group's plateau stands at four times the blue group's under FL ($318$ against $80$ loans per individual), $1.7$ times under SW ($265$ against $152$), and close to level under RMM ($231$ against $191$). The reach rates show how this builds up (Figure~\ref{fig:reach}, bottom row): the red group's rises throughout deployment, to $12\%$ under FL, $9\%$ under RMM and $10\%$ under SW, while the blue group's peaks at $5$ to $7\%$ between episodes $130$ and $670$ and then declines, to $2\%$, $5\%$ and $3.5\%$. The blue line starts above the red and falls below it, earliest under FL and latest under RMM.

\paragraph{Summary.}
The three views separate the perspectives more sharply than any aggregate metric. Under Equality of Outcome everyone is served, within-group shares are even, and the constraint sets the balance between groups: equal under FL, ten to one in the blue group's favour under RMM, and, under SW, a blue group served broadly against a red group of which half receives almost nothing. Under Equality of Opportunity roughly a quarter of the red group and a fifth of the blue are never served, the rest are served uniformly, and the constraint only sets the ratio between the groups' volumes. Appendix~\ref{app:balanced} adds the fixed-multiplier FL policy to this comparison.

\subsection{The Best of Both Worlds: An ADM's Policy for Equity and Profit}
\label{app:balanced}
\newcommand{\balanced}{PeRL-$\gamma^\star$}

Appendices~\ref{app:pareto} and~\ref{app:micro} leave the ADM with a choice between two FL scenarios. FL under Equality of Outcome closes the wealth gap, holds $\rho$ near $1$ and serves every individual of both groups, at a profit of $1.30$M. FL under Equality of Opportunity earns $7.44$M, the most of any cell, but leaves a gap of $1{,}087$, $\rho = 3.98$ and a quarter of each group never served. This section describes \balanced, a single policy that keeps the constraint and perspective of the first cell and recovers most of the profit of the second. It is the policy reported for FL under Equality of Outcome in Table~\ref{tab:app_rl_vs_perl_main}.

\paragraph{Method.}
Dual ascent sets the strength of the fairness term from how slack the constraint currently is, not from the trade-off the ADM wants. For FL this drives the multiplier to one of two extremes (Appendix~\ref{app:lambda_trajectories}): under Outcome it rises to the cap and then moves between floor and cap, and under Opportunity it falls to the floor within the first episodes, so the fairness term is either enforced at full strength or switched off. \balanced{} replaces the constraint with a fixed trade-off: PeRL is trained on the reward of Equation~\eqref{eq:app_reward} with $\gamma^t \equiv \gamma^\star$ throughout pre-training and deployment, so that it maximises a single weighted utility of profit and the FL term rather than solving a constrained problem. A fixed weight needs no threshold $\vartheta$, and each value of $\gamma^\star$ selects one point on the trade-off between profit and parity. We take $\gamma^\star = 1.87$, the strength at which the learned FL multipliers of the Outcome and Opportunity scenarios meet during pre-training, a level the dual ascent itself passes through on its way to the fairness-holding and to the profit-earning extreme. Everything else is as in Appendix~\ref{sec:passive_vs_performative}: the PeRL agent, the FL constraint under the Outcome perspective, the hyperparameters of Table~\ref{tab:agent_hyperparams}, and $1{,}000$ pre-training and $3{,}000$ deployment episodes over $10$ seeds.

\begin{table}[t]
\centering
\caption{\textbf{\balanced{} against the two FL scenarios with a learned multiplier.} End-of-deployment metrics, mean $\pm$ standard deviation over $10$ seeds, as in Table~\ref{tab:app_rl_vs_perl}. RMM under Opportunity is included as the Opportunity cell closest to balance. Best value per column in bold.}
\label{tab:balanced_macro}
\footnotesize
\setlength{\tabcolsep}{4pt}
\begin{tabular}{l cccc}
\toprule
\multirow{2}{*}{Policy} & Cumulative Profit & $|$Wealth & $|$Approval Rate & Inequality \\
 & ($\times 10^6$) & Gap$|$ & Disparity$|$ & Ratio $\rho$ \\
\midrule
FL / OUT   & $1.304 \pm 0.974$ & $71.9 \pm 298.0$ & $0.0119 \pm 0.2550$ & $1.059 \pm 0.094$ \\
FL / OPP   & $\mathbf{7.443 \pm 0.410}$ & $1086.8 \pm 332.2$ & $0.0097 \pm 0.0129$ & $3.981 \pm 0.568$ \\
RMM / OPP  & $6.711 \pm 1.229$ & $207.9 \pm 1328.7$ & $0.0543 \pm 0.0792$ & $1.229 \pm 0.643$ \\
\balanced{} & $4.909 \pm 0.579$ & $\mathbf{46.5 \pm 82.1}$ & $\mathbf{0.0039 \pm 0.1114}$ & $\mathbf{1.045 \pm 0.032}$ \\
\bottomrule
\end{tabular}
\end{table}

\begin{figure}[t!]
    \centering
    \includegraphics[width=0.5\textwidth]{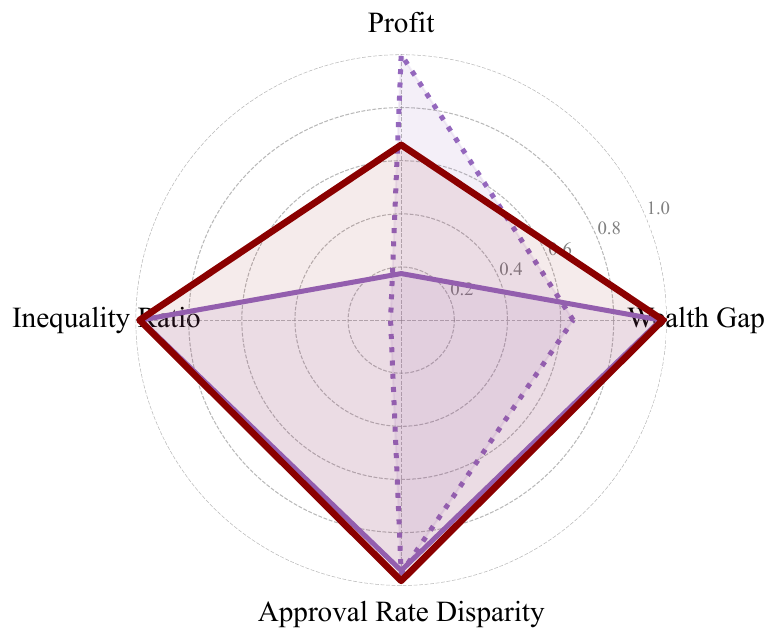}\\[0.3em]
    \includegraphics[width=0.3\textwidth]{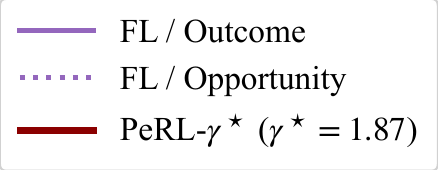}
    \caption{\textbf{\balanced{} against the two learned-multiplier FL scenarios.} Axes as in Figure~\ref{fig:radar}, rescaled to include \balanced; outward is better on every axis.}\label{fig:radar_balanced}
\end{figure}

\paragraph{Aggregate outcomes.}
Table~\ref{tab:balanced_macro} and Figure~\ref{fig:radar_balanced} compare \balanced{} with the two FL scenarios. It is the best of the seven policies on all three fairness metrics: a wealth gap of $46.5$, below FL under Outcome's $71.9$ and four to twenty-three times below the Opportunity scenarios ($208$ to $1{,}087$); $\rho = 1.045$, the closest to $1$ in the grid; and an approval-rate disparity of $0.004$, the smallest of any cell. Its profit is $4.91$M, $3.8$ times FL under Outcome and two thirds of both FL under Opportunity ($7.44$M) and the unconstrained \textsc{MaxUtil} agent ($7.50$M).

The three policies differ in what they do with an application. FL under Outcome approves three quarters of each group's applications ($0.75$ and $0.74$), and its loans default at close to the population rate ($14.5\%$): it holds the gap by lending to nearly everyone who applies, earning $0.27$ per loan. FL under Opportunity approves $65\%$ of each group's applications, with $11\%$ of its loans defaulting, and earns $1.56$ per loan; but since its constraint does not depend on how many apply, the red group ends up filing four times as many applications as the blue ($5.8$M against $1.5$M over deployment) and receiving four times the loans. \balanced{} is selective like the Opportunity cell and even-handed like the Outcome cell. It approves $59\%$ of the red group's applications and $58\%$ of the blue group's, $11$ to $12\%$ of its loans default, and it earns $1.26$ per loan; with the same approval rate in both groups, both keep applying at the same volume ($3.5$M and $3.3$M applications), receive the same volume of credit ($2.1$M and $1.9$M loans), and grow their wealth at the same rate ($R_M = 0.0026$, $R_F = 0.0025$). The profit comes from lending to fewer, safer applicants; the fairness from doing so identically in both groups.

\begin{table}[t!]
\centering
\caption{\textbf{\balanced{} at the micro level.} As in Table~\ref{tab:micro}, for \balanced{} and the two learned-multiplier FL scenarios. R: red (male), B: blue (female).}
\label{tab:balanced_micro}
\small
\begin{tabular}{l cc cc cc cc cc}
\toprule
& \multicolumn{2}{c}{Loans / individual} & \multicolumn{2}{c}{Never served} & \multicolumn{2}{c}{Top-10\% share} & \multicolumn{2}{c}{Gini} & \multicolumn{2}{c}{Reach rate} \\
\cmidrule(lr){2-3}\cmidrule(lr){4-5}\cmidrule(lr){6-7}\cmidrule(lr){8-9}\cmidrule(lr){10-11}
Policy & R & B & R & B & R & B & R & B & R & B \\
\midrule
FL / Outcome     & 243 & 227 & 0\%  & 0\%  & 0.12 & 0.13 & 0.08 & 0.09 & 0.086 & 0.074 \\
FL / Opportunity & 318 & 80  & 24\% & 21\% & 0.18 & 0.20 & 0.43 & 0.44 & 0.120 & 0.018 \\
\balanced{}      & 174 & 161 & 1\%  & 1\%  & 0.21 & 0.20 & 0.38 & 0.36 & 0.072 & 0.069 \\
\bottomrule
\end{tabular}
\end{table}

\begin{figure*}[t!]
    \centering\vspace*{-1em}
    \begin{minipage}{0.24\textwidth}
        \centering
        \includegraphics[width=\textwidth]{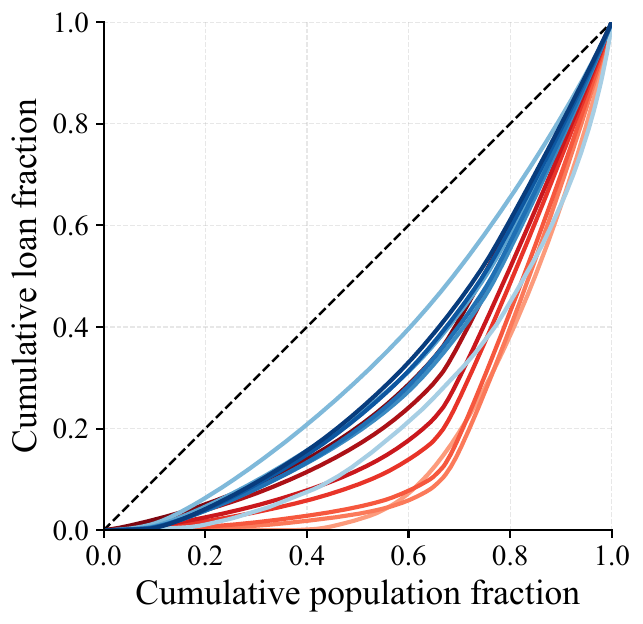}\vspace{-.5em}
        \subcaption{Lorenz curves}\label{fig:balanced_lorenz}
    \end{minipage}\hfill
    \begin{minipage}{0.24\textwidth}
        \centering
        \includegraphics[width=\textwidth]{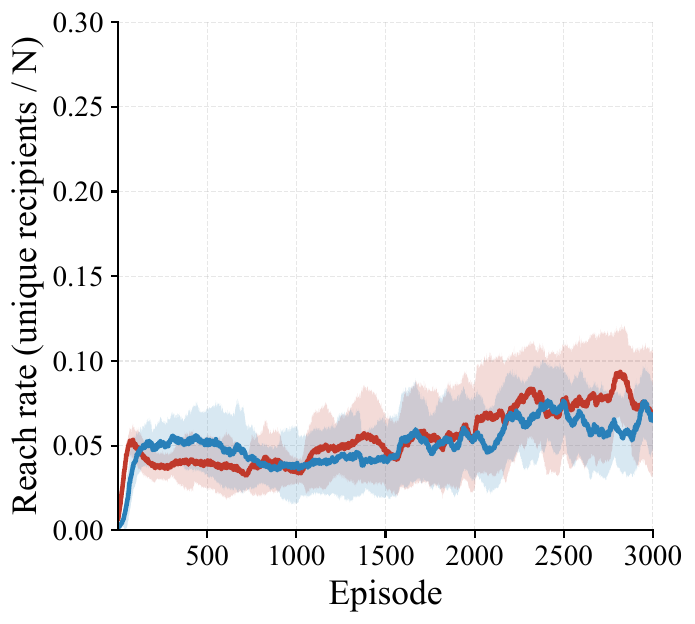}\vspace{-.5em}
        \subcaption{Reach rate}\label{fig:balanced_reach}
    \end{minipage}\hfill
    \begin{minipage}{0.24\textwidth}
        \centering
        \includegraphics[width=\textwidth]{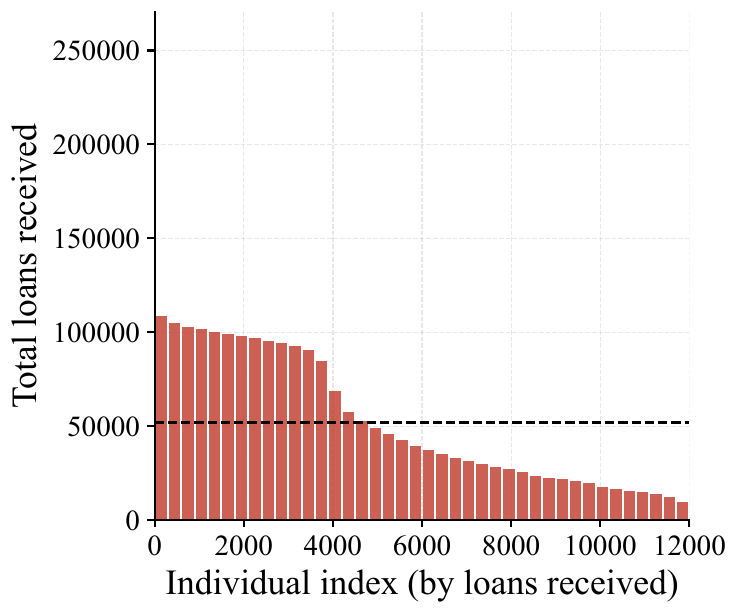}\vspace{-.5em}
        \subcaption{Loans, male}\label{fig:balanced_loandist_m}
    \end{minipage}\hfill
    \begin{minipage}{0.24\textwidth}
        \centering
        \includegraphics[width=\textwidth]{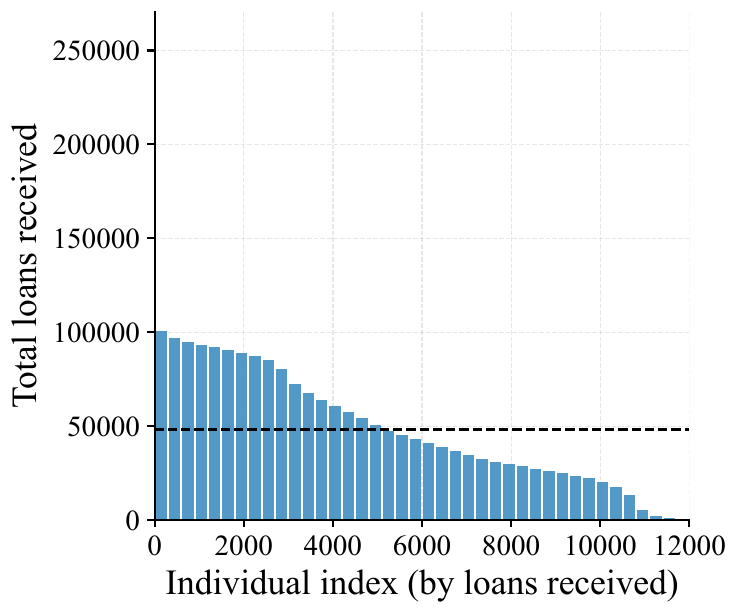}\vspace{-.5em}
        \subcaption{Loans, female}\label{fig:balanced_loandist_f}
    \end{minipage}\\[0.5em]
    \includegraphics[width=0.2\textwidth]{iclr_2027/iclr_2027_plots/lorenz_run11/lorenz_group_legend.pdf}\hspace{1em}
    \includegraphics[width=0.6\textwidth]{iclr_2027/iclr_2027_plots/lorenz_run11/lorenz_episode_shade_legend.pdf}
    \caption{\textbf{Who receives the loans under \balanced.} (a) Lorenz curves, (b) reach rate and (c, d) loans received by each individual, read as in Figure~\ref{fig:lorenz_main} and Appendix~\ref{app:micro}, on the same axes. Red: male group; blue: female group.}\label{fig:balanced_micro}
\end{figure*}

\paragraph{Who receives the loans.}
Figure~\ref{fig:balanced_micro} and Table~\ref{tab:balanced_micro} place \balanced{} between the two FL scenarios. It is as inclusive as the Outcome cell: $1\%$ of each group is never served, against $0\%$ under FL/Outcome and $21$ to $24\%$ under FL/Opportunity, and the two groups receive nearly the same volume ($174$ and $161$ loans per individual). After the first $100$ episodes its reach rate stays between $3.5$ and $9\%$ of each group, the two groups' curves overlapping throughout and both rising to about $7\%$ by the end. It differs from FL under Outcome in how evenly loans are spread within each group. Its Lorenz curves end at a Gini of $0.38$ (red) and $0.36$ (blue), between FL under Outcome ($0.08$, $0.09$) and FL under Opportunity ($0.43$, $0.44$); its loan distributions show a most-served third of each group at roughly twice the equal-share line, then a step and a long decline toward zero, and the top tenth of each group holds $20$ to $21\%$ of its loans. These curves are still moving at episode $3{,}000$, more than any in Figure~\ref{fig:lorenz_main}: the red group's Gini falls from $0.65$ at episode $100$ to $0.50$ at $2{,}000$ and $0.38$ at $3{,}000$, and the blue group's, at $0.38$ at episode $1{,}000$ and $0.41$ at $2{,}000$, ends at $0.36$.

\paragraph{What the ADM gets.}
\balanced{} delivers the Outcome cell's fairness at the aggregate level while earning nearly four times its profit, and its inclusivity too: nobody is left out, and both groups are served in equal measure. What it does not reproduce is the Outcome cell's within-group evenness: its loans reach everyone but not in equal amounts, at a concentration between the two FL scenarios. A ADM whose objective is the between-group outcome, with a profit that must not collapse, is better served by \balanced{} than by either FL cell; a ADM who also requires every individual to be served alike still needs the learned multiplier and its cost.

\subsection{Why a Policy-Gradient Learner: PG versus SAC}
\label{app:sac}

\begin{figure*}[t!]
    \centering\vspace*{-1em}
    \begin{minipage}{0.24\textwidth}
        \centering
        \includegraphics[width=\textwidth]{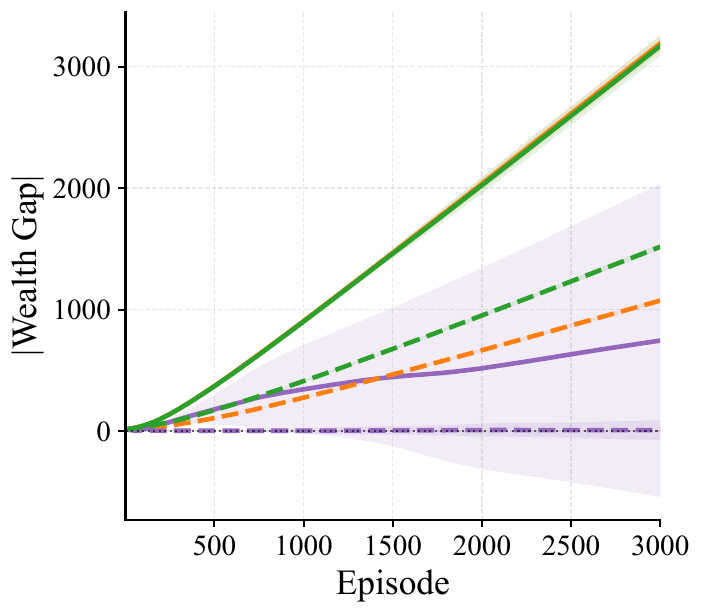}\vspace{-.5em}
        \subcaption{}\label{fig:sac_out_wg}
    \end{minipage}\hfill
    \begin{minipage}{0.24\textwidth}
        \centering
        \includegraphics[width=\textwidth]{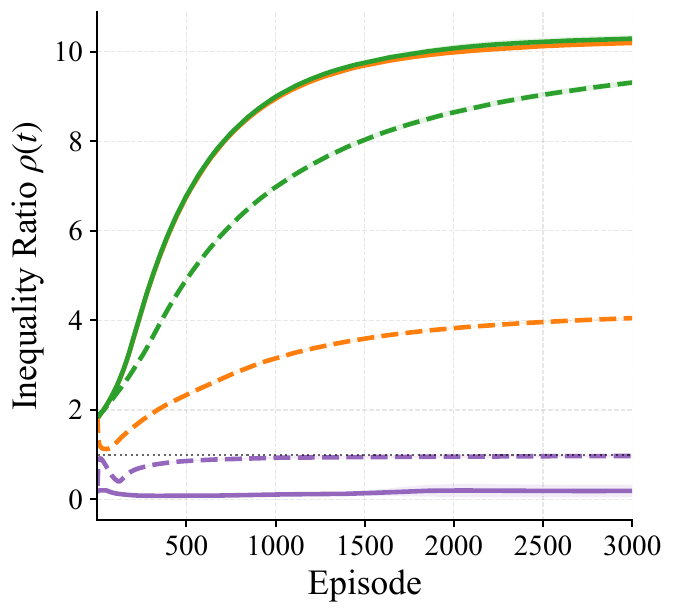}\vspace{-.5em}
        \subcaption{}\label{fig:sac_out_rho}
    \end{minipage}\hfill
    \begin{minipage}{0.24\textwidth}
        \centering
        \includegraphics[width=\textwidth]{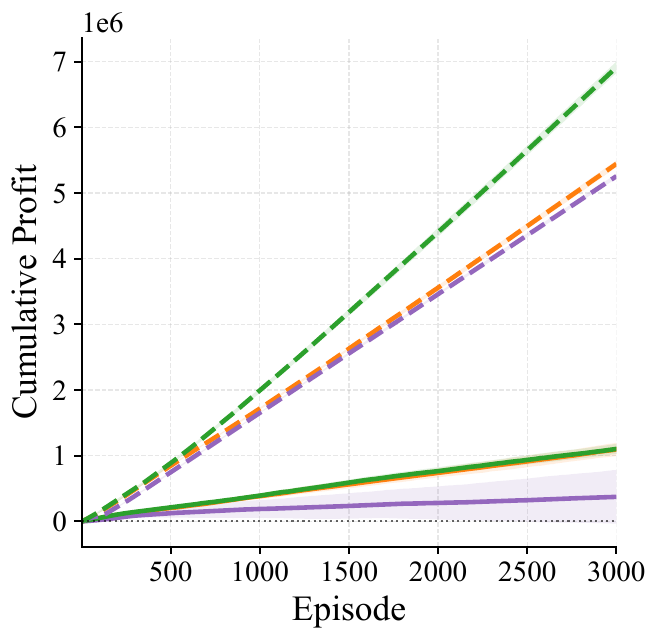}\vspace{-.5em}
        \subcaption{}\label{fig:sac_out_profit}
    \end{minipage}\hfill
    \begin{minipage}{0.24\textwidth}
        \centering
        \includegraphics[width=\textwidth]{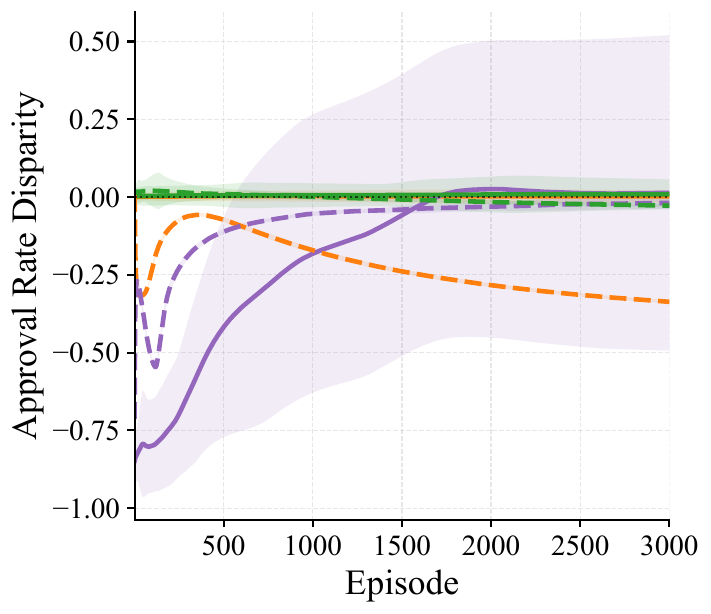}\vspace{-.5em}
        \subcaption{}\label{fig:sac_out_disp}
    \end{minipage}\\[0.5em]
    \includegraphics[width=0.25\textwidth]{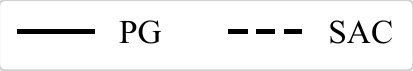}\hspace{1em}
    \includegraphics[width=0.4\textwidth]{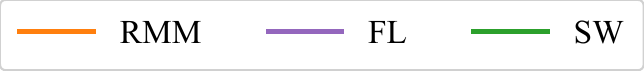}
    \caption{\textbf{PG (solid) and SAC (dashed) under Equality of Outcome.} (a) Wealth gap, (b) inequality ratio, (c) cumulative profit and (d) approval-rate disparity (red minus blue) over deployment, for each fairness constraint. Mean over $10$ seeds; bands are one standard deviation across seeds, bootstrapped for $\rho$.}\label{fig:sac_out}
\end{figure*}

\begin{figure*}[t!]
    \centering\vspace*{-1em}
    \begin{minipage}{0.24\textwidth}
        \centering
        \includegraphics[width=\textwidth]{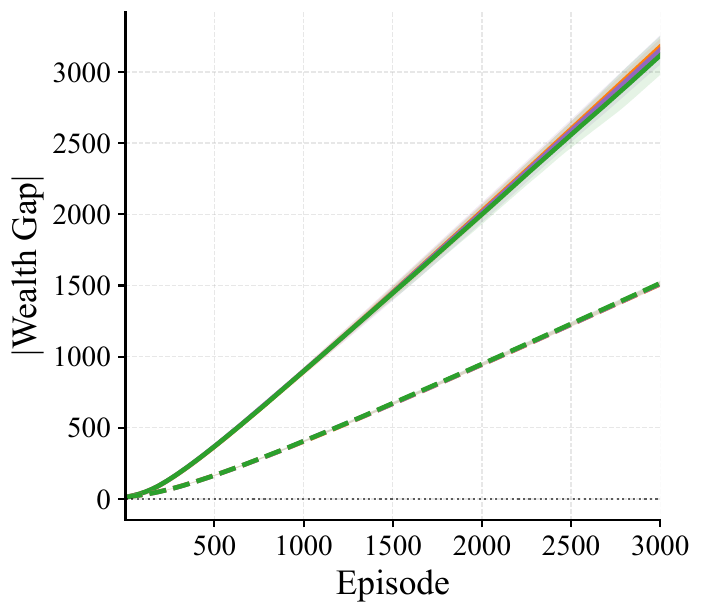}\vspace{-.5em}
        \subcaption{}\label{fig:sac_opp_wg}
    \end{minipage}\hfill
    \begin{minipage}{0.24\textwidth}
        \centering
        \includegraphics[width=\textwidth]{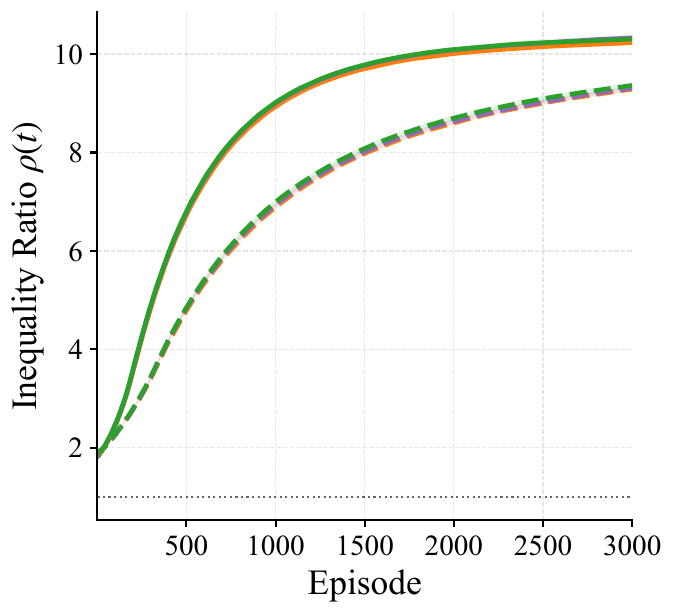}\vspace{-.5em}
        \subcaption{}\label{fig:sac_opp_rho}
    \end{minipage}\hfill
    \begin{minipage}{0.24\textwidth}
        \centering
        \includegraphics[width=\textwidth]{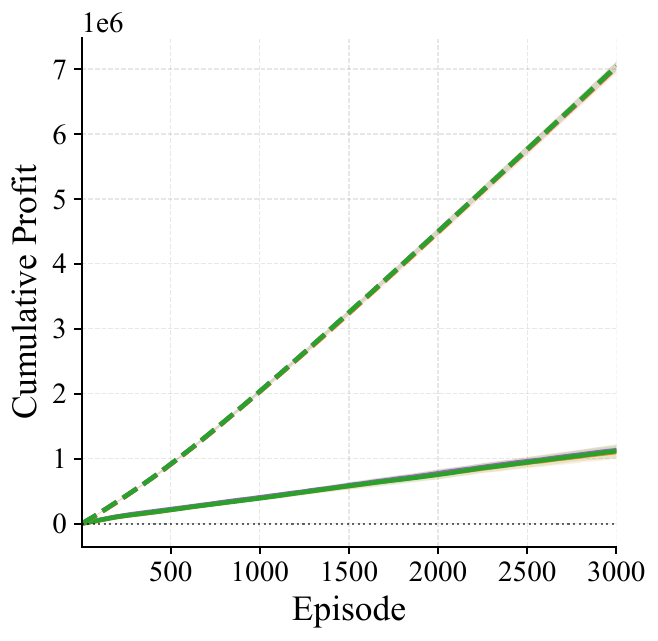}\vspace{-.5em}
        \subcaption{}\label{fig:sac_opp_profit}
    \end{minipage}\hfill
    \begin{minipage}{0.24\textwidth}
        \centering
        \includegraphics[width=\textwidth]{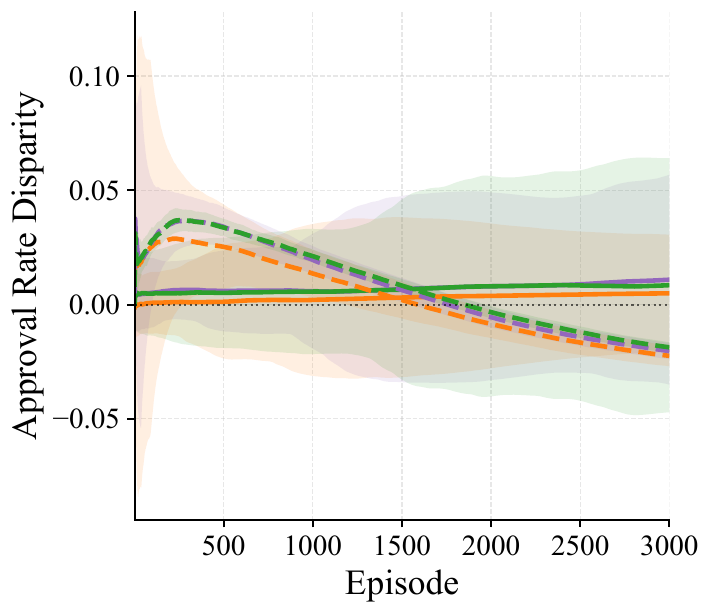}\vspace{-.5em}
        \subcaption{}\label{fig:sac_opp_disp}
    \end{minipage}\\[0.5em]
    \includegraphics[width=0.25\textwidth]{iclr_2027/iclr_2027_plots/pg_vs_sac_run11/pgvsac_agent_legend.pdf}\hspace{1em}
    \includegraphics[width=0.4\textwidth]{iclr_2027/iclr_2027_plots/pg_vs_sac_run11/pgvsac_reward_legend.pdf}
    \caption{\textbf{PG (solid) and SAC (dashed) under Equality of Opportunity.} Panels as in Figure~\ref{fig:sac_out}.}\label{fig:sac_opp}
\end{figure*}

The passive learner of the main text is vanilla policy gradient (PG) on a Beta policy, and PeRL is its performative counterpart~\citep{basu2025performativepolicygradientoptimality}, which differentiates through the population's response to the policy. A natural question is whether a stronger passive learner would make a better baseline. We therefore trained Soft Actor-Critic (SAC) through the same pipeline: the same policy network, observation, per-applicant reward and dual ascent as PG, with the update replaced by off-policy soft actor-critic (twin $Q$ critics with Polyak targets, automatic temperature tuning, a replay buffer of $10{,}000$ cohorts, discount factor $\zeta_{\mathrm{discount}} = 0.99$, Polyak rate $0.005$, learning rate $10^{-3}$). Since there is no per-applicant next state, the critic bootstraps on the mean soft value of the next cohort. SAC is run in all six scenarios with $1{,}000$ pre-training and $3{,}000$ deployment episodes and $10$ seeds. Table~\ref{tab:sac} compares the two learners at the end of deployment, and Figures~\ref{fig:sac_out} and~\ref{fig:sac_opp} over deployment.

\begin{table}[t!]
\centering
\caption{\textbf{PG versus SAC at the end of deployment.} Approval rate of the red and blue group, cumulative profit, $|$wealth gap$|$ and inequality ratio $\rho$, as in Table~\ref{tab:app_rl_vs_perl}; means over $10$ seeds.}
\label{tab:sac}
\footnotesize
\setlength{\tabcolsep}{4pt}
\begin{tabular}{ll cc cc cc cc}
\toprule
& & \multicolumn{2}{c}{Approval rate R / B} & \multicolumn{2}{c}{Profit ($\times 10^6$)} & \multicolumn{2}{c}{$|$Wealth Gap$|$} & \multicolumn{2}{c}{Inequality Ratio $\rho$} \\
\cmidrule(lr){3-4}\cmidrule(lr){5-6}\cmidrule(lr){7-8}\cmidrule(lr){9-10}
Persp. & Constr. & PG & SAC & PG & SAC & PG & SAC & PG & SAC \\
\midrule
OUT & FL  & 0.32 / 0.31 & 0.53 / 0.55 & 0.37 & 5.26 & 745  & 8    & 0.19  & 0.97 \\
OUT & RMM & 0.98 / 0.98 & 0.53 / 0.87 & 1.10 & 5.44 & 3190 & 1074 & 10.20 & 4.05 \\
OUT & SW  & 0.97 / 0.97 & 0.53 / 0.56 & 1.10 & 6.91 & 3170 & 1517 & 10.29 & 9.31 \\
\midrule
OPP & FL  & 0.97 / 0.96 & 0.53 / 0.55 & 1.13 & 7.04 & 3155 & 1511 & 10.33 & 9.30 \\
OPP & RMM & 0.98 / 0.97 & 0.53 / 0.55 & 1.11 & 7.03 & 3177 & 1509 & 10.23 & 9.29 \\
OPP & SW  & 0.96 / 0.95 & 0.53 / 0.55 & 1.12 & 7.04 & 3116 & 1516 & 10.30 & 9.36 \\
\bottomrule
\end{tabular}
\end{table}

\paragraph{What SAC learns.}
SAC learns a score cutoff. In all six scenarios it approves $53\%$ of the red group's applications, and in five of them $55\%$ of the blue group's, whatever the constraint or perspective. The approved half is the lower-risk half: $11$ to $12\%$ of SAC's loans default, against $15\%$ in the population, and the loans an individual receives track their wealth, with a correlation of $0.81$ (Outcome) and $0.85$ (Opportunity) between wealth at episode $100$ and loans received by episode $3{,}000$, against $0.31$ and $0.32$ for PG. The fairness term plays little part: in five of the six scenarios the multiplier is at its floor for $75$ to $99\%$ of deployment, and four scenarios (SW under Outcome and all three Opportunity scenarios) end at the same point, $6.9$ to $7.0$M in profit, a gap of $1{,}509$ to $1{,}517$ and $\rho$ of $9.3$ to $9.4$. The two exceptions are Outcome scenarios. Under RMM, the one cell whose multiplier stays off the floor ($4\%$ of episodes at the floor), the blue group's approval rate rises to $0.87$ while the red group's stays at $0.53$. Under FL, the cutoff is applied to both groups alike, which balances the aggregates (gap $8$, $\rho = 0.97$) while loans still follow wealth within each group (Gini $0.41$ in both).

\paragraph{Why PG is the baseline.}
A passive baseline is useful only if it responds to the reward it is given. SAC settles on the same cutoff under four of the six rewards and on small variations of it under the other two: a fixed score threshold of the kind the rules of Appendix~\ref{app:heuristic} define by hand, here recovered by a critic. The threshold denies credit at every step to the lower-scored half of each cohort, whose wealth a loan would move most, and in the four scenarios where SAC settles on it the red group's wealth grows at half the rate it does under PG ($R_M = 0.006$ against $0.012$). PG, in contrast, responds to each reward differently, and it is what PeRL reduces to without the performative term, so the comparison of Appendix~\ref{sec:passive_vs_performative} isolates that term. SAC's higher profit is the profit of a fixed cutoff, and a learner that holds a fixed cutoff does not engage with the question of this paper: what a ADM does when its decisions reshape the population it faces next.

\subsection{Robustness to the Simulator's Settings}
\label{app:ablation}

\begin{figure*}[t!]
    \centering\vspace*{-1em}
    \begin{minipage}{0.24\textwidth}
        \centering
        \includegraphics[width=\textwidth]{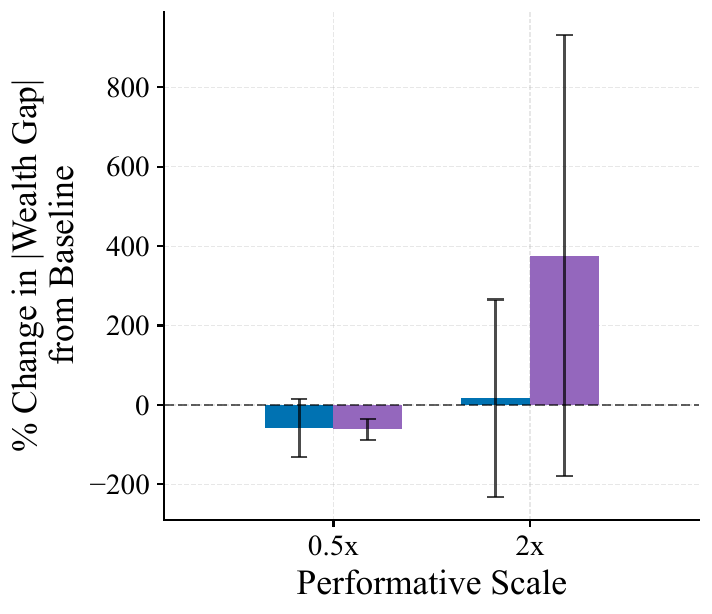}\vspace{-.5em}
        \subcaption{}\label{fig:ablation_perf_wg}
    \end{minipage}\hfill
    \begin{minipage}{0.24\textwidth}
        \centering
        \includegraphics[width=\textwidth]{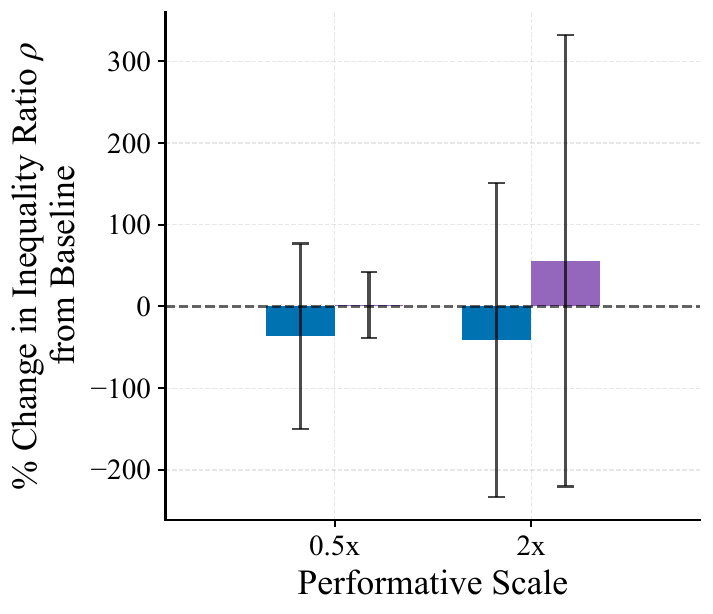}\vspace{-.5em}
        \subcaption{}\label{fig:ablation_perf_rho}
    \end{minipage}\hfill
    \begin{minipage}{0.24\textwidth}
        \centering
        \includegraphics[width=\textwidth]{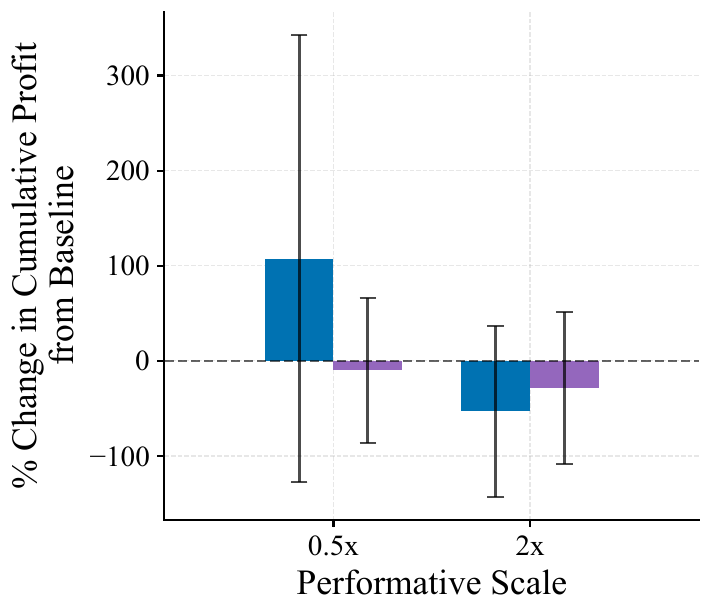}\vspace{-.5em}
        \subcaption{}\label{fig:ablation_perf_profit}
    \end{minipage}\hfill
    \begin{minipage}{0.24\textwidth}
        \centering
        \includegraphics[width=\textwidth]{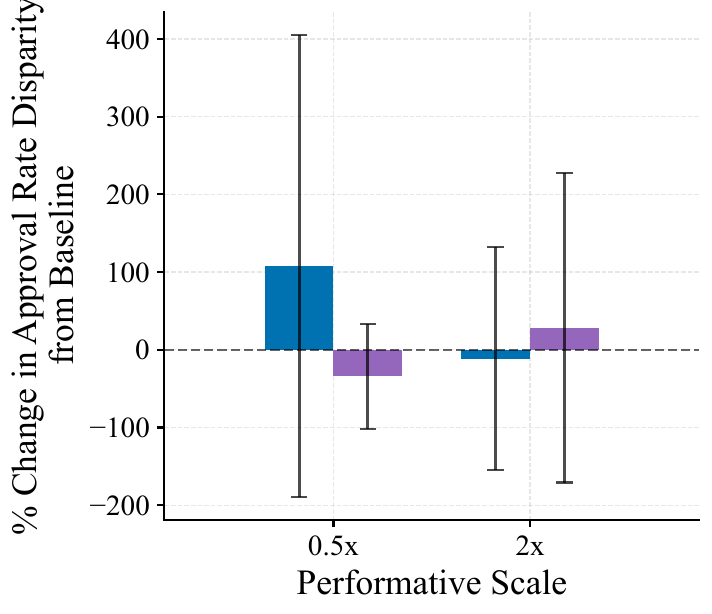}\vspace{-.5em}
        \subcaption{}\label{fig:ablation_perf_disp}
    \end{minipage}\\[0.5em]
    \includegraphics[width=0.25\textwidth]{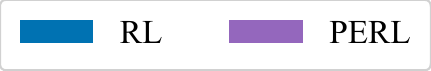}
    \caption{\textbf{Sensitivity to the strength of the population's response.} Change in (a) wealth gap, (b) inequality ratio, (c) cumulative profit and (d) approval-rate disparity at the end of deployment, relative to the default simulator, when the performative channels are halved or doubled; RL and PeRL under FL and Equality of Outcome. Error bars: one standard deviation across $10$ seeds.}\label{fig:ablation_perf}
\end{figure*}

\begin{figure*}[t!]
    \centering\vspace*{-1em}
    \begin{minipage}{0.24\textwidth}
        \centering
        \includegraphics[width=\textwidth]{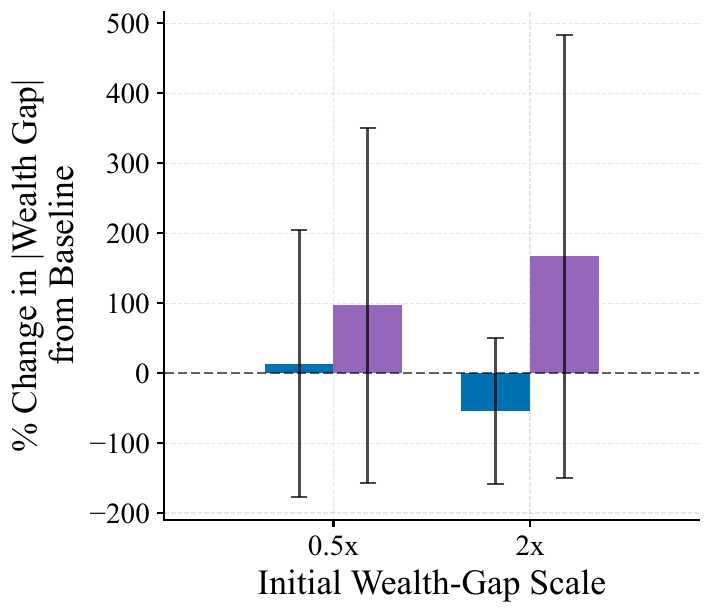}\vspace{-.5em}
        \subcaption{}\label{fig:ablation_gap_wg}
    \end{minipage}\hfill
    \begin{minipage}{0.24\textwidth}
        \centering
        \includegraphics[width=\textwidth]{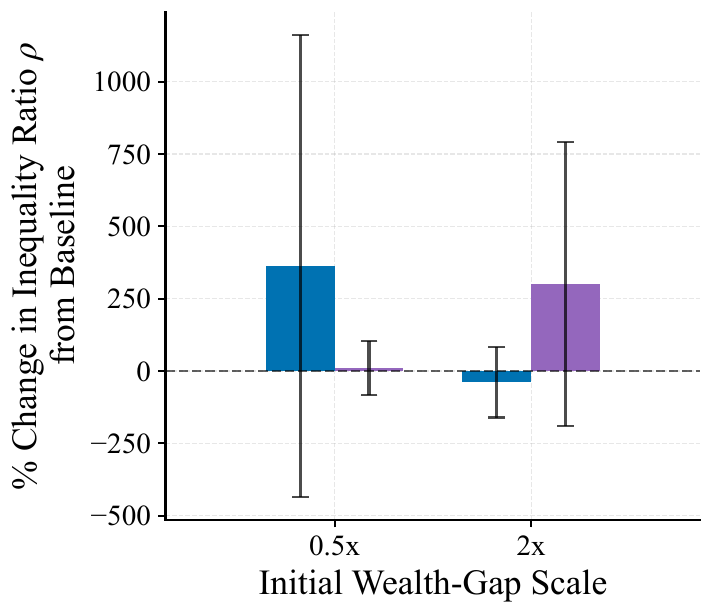}\vspace{-.5em}
        \subcaption{}\label{fig:ablation_gap_rho}
    \end{minipage}\hfill
    \begin{minipage}{0.24\textwidth}
        \centering
        \includegraphics[width=\textwidth]{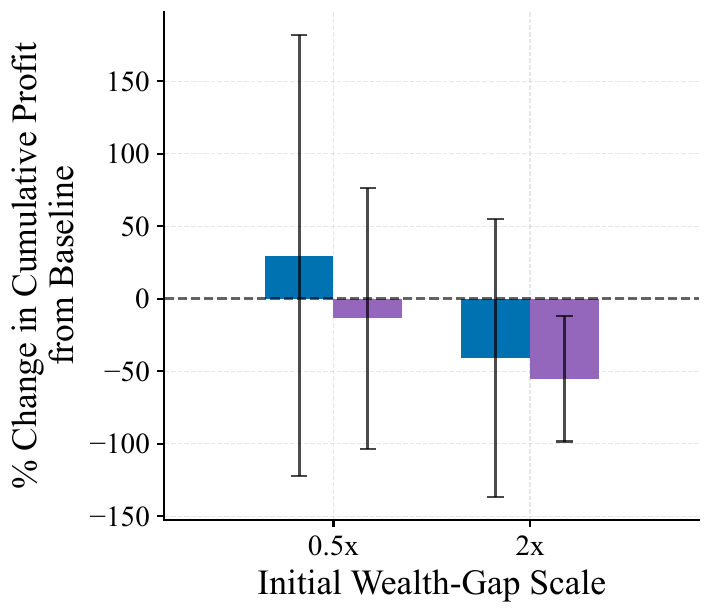}\vspace{-.5em}
        \subcaption{}\label{fig:ablation_gap_profit}
    \end{minipage}\hfill
    \begin{minipage}{0.24\textwidth}
        \centering
        \includegraphics[width=\textwidth]{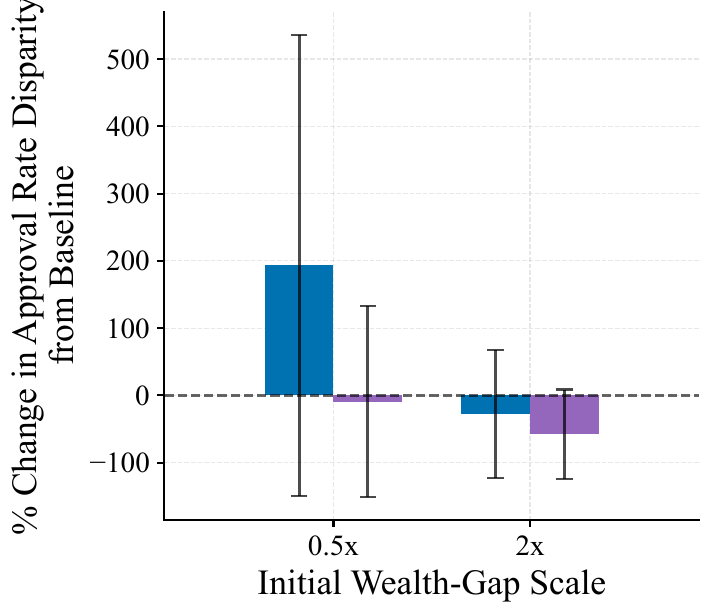}\vspace{-.5em}
        \subcaption{}\label{fig:ablation_gap_disp}
    \end{minipage}\\[0.5em]
    \includegraphics[width=0.25\textwidth]{iclr_2027/iclr_2027_plots/ablation_run11/ablation_agent_legend.pdf}
    \caption{\textbf{Sensitivity to the initial wealth gap.} As in Figure~\ref{fig:ablation_perf}, with the initial gap between the groups halved or doubled.}\label{fig:ablation_gap}
\end{figure*}

Two settings of \framework{} fix the strength of the effects the paper studies: how strongly the population responds to lending, and how unequal the groups are at the start. We vary each around its default for RL and PeRL under FL with Equality of Outcome and the learned multiplier of Appendix~\ref{sec:passive_vs_performative}, with $1{,}000$ pre-training and $3{,}000$ deployment episodes and $10$ seeds per setting.
\begin{itemize}
    \item \textbf{Performative strength} $\times 0.5$ and $\times 2$: one factor scales both channels through which decisions feed back into the population, the Hawkes excitation $\alpha_g$ of an approval and the wealth gain $\kappa_i$ of a repaid loan. At $\times 1$ it is the simulator of Section~\ref{sec:eutopia_main}.
    \item \textbf{Initial wealth gap} $\times 0.5$ and $\times 2$: each group's initial wealth is rescaled by one factor so that $\mu_M^0 - \mu_F^0$ is halved or doubled, keeping the population-mean wealth, and hence the arrival rates, which depend on $\mu_g/\bar\mu$, and each group's within-group distribution unchanged. Creditworthiness, default probabilities and loan amounts are not changed.
\end{itemize}
Figures~\ref{fig:ablation_perf} and~\ref{fig:ablation_gap} report each end-of-deployment metric as the percent change from the same agent's $\times 1$ result. Here the metrics are per-seed final values averaged over seeds, so the $\times 1$ references differ from Table~\ref{tab:app_rl_vs_perl}: PeRL's wealth gap is $252$ (the mean of the per-seed gaps, against the gap of the mean, $72$), and $\rho$ is the mean of the per-seed ratios ($1.11$ for PeRL and $0.92$ for RL, against $1.06$ and $0.19$).

\paragraph{Performative strength.}
At half strength PeRL's wealth gap falls by $62\%$ ($96 \pm 30$ against $252 \pm 154$), with every seed ending at $120$ or below, and its disparity by $34\%$, while its profit and $\rho$ stay within about $10\%$ of the default. RL's gap also falls ($-58\%$), while its disparity and profit double, with uncertainties larger than the changes. At twice the strength PeRL's gap rises to $1{,}201 \pm 1{,}193$, nearly five times the default, some seeds holding the gap and others not; its profit falls by $29\%$ and $\rho$ rises to $1.7 \pm 3.0$. RL loses half its profit, and its $\rho$ moves further from $1$.

\paragraph{Initial wealth gap.}
Changing the initial gap in either direction widens PeRL's final gap, to $497 \pm 563$ when it is halved and $673 \pm 683$ when it is doubled, against $252$; its disparity falls ($-9\%$ and $-57\%$) and so does its profit ($-14\%$ and $-55\%$). With the doubled gap $\rho$ reaches $4.5 \pm 5.3$: the red group pulls ahead in some seeds and not in others. RL's changes are within their uncertainty at every setting. Across both settings the seed spreads are wide, and only two shifts clear them: PeRL's smaller gap at half the performative strength and its lower profit at twice the initial gap. The others indicate directions rather than sizes.

\subsection{Trajectories for Lagrangian Multiplier}
\label{app:lambda_trajectories}

\begin{figure}[t!]
    \centering
    \begin{minipage}{0.48\textwidth}
        \centering
        \includegraphics[width=0.95\textwidth]{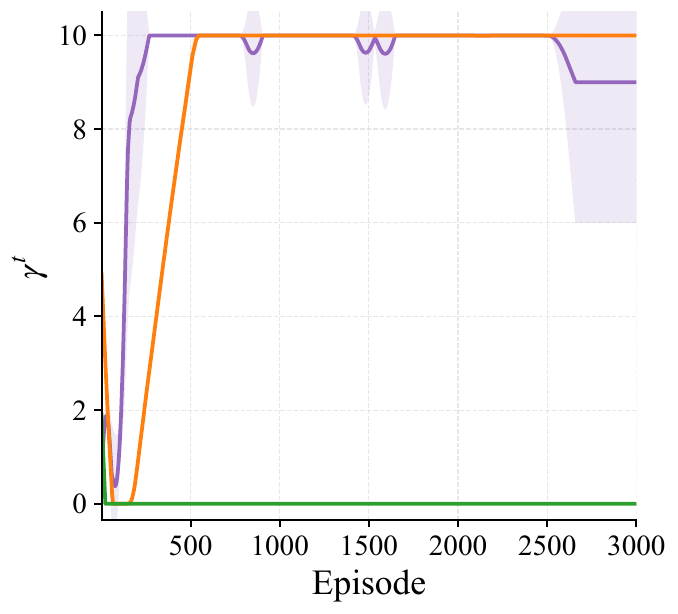}\vspace{-.5em}
        \subcaption{Equality of Outcome}\label{fig:lambda_pg_out}
    \end{minipage}\hfill
    \begin{minipage}{0.48\textwidth}
        \centering
        \includegraphics[width=0.95\textwidth]{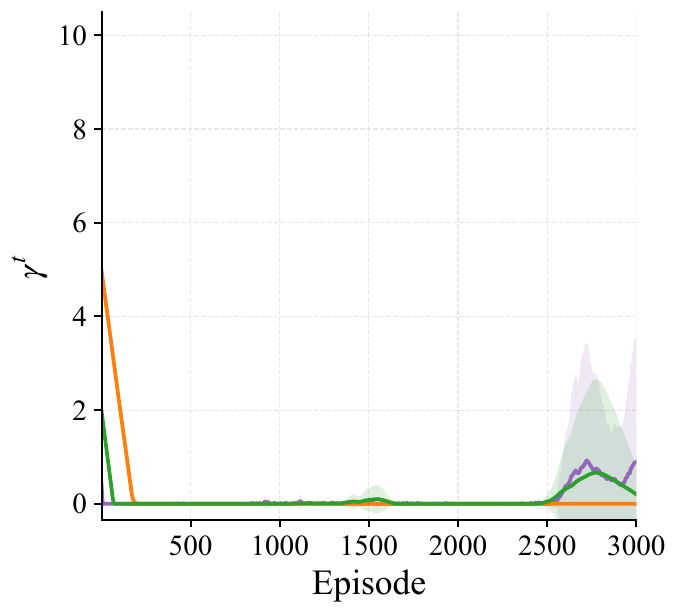}\vspace{-.5em}
        \subcaption{Equality of Opportunity}\label{fig:lambda_pg_opp}
    \end{minipage}\\[0.5em]
    \includegraphics[width=0.4\textwidth]{iclr_2027/iclr_2027_plots/lambda_overlay_run11/lambda_overlay_reward_legend.pdf}
    \caption{\textbf{The multiplier over deployment for RL.} $\gamma^t$ for FL, RMM and SW; mean $\pm$ one standard deviation over $10$ seeds. Each cell starts at its $\gamma^0$ ($0.5$ for FL, $2$ for SW, $5$ for RMM), between the floor $10^{-4}$ and the cap $10$. PeRL's multipliers are in Figure~\ref{fig:lambda_overlay_main}.}\label{fig:lambda_overlay}
\end{figure}

Figures~\ref{fig:lambda_overlay_main} (PeRL) and Figure~\ref{fig:lambda_overlay} (RL) show the multiplier of Algorithm~\ref{alg:dual} in every cell, from its reset value $\gamma^0$ at the start of deployment. Under Equality of Outcome the three constraints behave differently. FL rises from $0.5$ to the cap within the first few hundred episodes (by episode $545$ for PeRL and $264$ for RL); PeRL's multiplier then moves between floor and cap for the rest of deployment, at the cap in about half of all episodes, while RL's stays at the cap in over $90\%$ of episodes. RMM first falls from $5$ to the floor within $40$ to $65$ episodes, then climbs to the cap by episode $700$ ($530$ for RL) and stays there. SW falls from $2$ to the floor within about $100$ episodes and does not leave it.

Under Equality of Opportunity all three multipliers fall to the floor and stay there for both agents: FL within $40$ episodes, SW within $70$ to $125$, and RMM, which starts highest, within $180$ to $380$. The multiplier is at the floor in more than $93\%$ of deployment episodes in every Opportunity cell; RL's small late rises under FL and SW stay below $1$.

The two scenarios whose multiplier ends high, FL and RMM under Outcome, are the two whose constraint is written on the groups' wealth. The four whose multiplier ends at the floor have constraints on true positive rates or on total credit, which the policy satisfies early in deployment.

\subsection{Computational Resources}
\label{app:compute}

All experiments run on CPUs; no GPU is used. The policy networks are small, and the cost of a run is dominated by simulating the environment: $24{,}000$ individuals over $4{,}000$ episodes of $200$ decision steps each. Every run, one agent in one cell with one seed, is an independent job on a compute cluster with $2$ CPU cores and $8$~GB of memory, and takes about one hour; the rule-based policies, which are only deployed, use $1$ core and $4$~GB. The reported experiments comprise $120$ runs of RL and PeRL across the six scenarios, $60$ of SAC, $80$ for the robustness study, $10$ of \balanced, $20$ of the \textsc{MaxUtil} agents and $70$ rule-based deployments, all over $10$ seeds.

%% file: iclr_2027/Appendix/Full_Tables.tex
\begin{table}[t!]
\centering
\caption{\textbf{RL: best fairness constraint under each fairness perspective}, per metric, at the end of deployment (mean $\pm$ std over $10$ seeds; $\rho$ is the ratio of seed-mean growths, std by bootstrap). Bold: best value in the table for that metric. $^\dagger$: RL approves nearly every applicant of both groups in this cell, so the disparity is small because approval is near-universal, not because the groups are balanced.}
\label{tab:app_rl_by_perspective}
\resizebox{\textwidth}{!}{%
\begin{tabular}{lcccccccc}
\toprule
\multirow{2}{*}{\textbf{Perspective}}
    & \multicolumn{2}{c}{\textbf{Cumulative Profit ($\times10^6$)}}
    & \multicolumn{2}{c}{\textbf{$|$Wealth Gap$|$}}
    & \multicolumn{2}{c}{\textbf{$|$Approval Rate Disparity$|$}}
    & \multicolumn{2}{c}{\textbf{Inequality Ratio $\rho$}} \\
\cmidrule(lr){2-3}\cmidrule(lr){4-5}\cmidrule(lr){6-7}\cmidrule(lr){8-9}
    & Best & Mean $\pm$ Std
    & Best & Mean $\pm$ Std
    & Best & Mean $\pm$ Std
    & Best & Mean $\pm$ Std \\
\midrule
OUT & SW & 1.103 $\pm$ 0.071 & FL & $\mathbf{745.0 \pm 1288.3}$ & RMM & $\mathbf{0.0024 \pm 0.0202}^\dagger$ & FL & $\mathbf{0.190 \pm 0.132}$ \\
OPP & FL & $\mathbf{1.131 \pm 0.062}$ & SW & 3115.9 $\pm$ 136.1 & RMM & 0.0049 $\pm$ 0.0257 & RMM & 10.231 $\pm$ 0.043 \\
\bottomrule
\end{tabular}}
\end{table}

\begin{table}[t!]
\centering
\caption{\textbf{PeRL: best fairness constraint under each fairness perspective}, per metric, at the end of deployment (mean $\pm$ std over $10$ seeds; $\rho$ is the ratio of seed-mean growths, std by bootstrap). Bold: best value in the table for that metric.}
\label{tab:app_perl_by_perspective}
\resizebox{\textwidth}{!}{%
\begin{tabular}{lcccccccc}
\toprule
\multirow{2}{*}{\textbf{Perspective}}
    & \multicolumn{2}{c}{\textbf{Cumulative Profit ($\times10^6$)}}
    & \multicolumn{2}{c}{\textbf{$|$Wealth Gap$|$}}
    & \multicolumn{2}{c}{\textbf{$|$Approval Rate Disparity$|$}}
    & \multicolumn{2}{c}{\textbf{Inequality Ratio $\rho$}} \\
\cmidrule(lr){2-3}\cmidrule(lr){4-5}\cmidrule(lr){6-7}\cmidrule(lr){8-9}
    & Best & Mean $\pm$ Std
    & Best & Mean $\pm$ Std
    & Best & Mean $\pm$ Std
    & Best & Mean $\pm$ Std \\
\midrule
OUT & SW & 4.721 $\pm$ 1.346 & FL & $\mathbf{71.9 \pm 298.0}$ & FL & 0.0119 $\pm$ 0.2550 & FL & $\mathbf{1.059 \pm 0.094}$ \\
OPP & FL & $\mathbf{7.443 \pm 0.410}$ & RMM & 207.9 $\pm$ 1328.7 & FL & $\mathbf{0.0097 \pm 0.0129}$ & RMM & 1.229 $\pm$ 0.643 \\
\bottomrule
\end{tabular}}
\end{table}

\begin{table}[t!]
\centering
\caption{\textbf{RL: best fairness perspective under each fairness constraint}, per metric, at the end of deployment (mean $\pm$ std over $10$ seeds). Bold: best value in the table for that metric. $^\dagger$: RL approves nearly every applicant of both groups in this cell, so the disparity is small because approval is near-universal, not because the groups are balanced.}
\label{tab:app_rl_by_constraint}
\resizebox{\textwidth}{!}{%
\begin{tabular}{lcccccccc}
\toprule
\multirow{2}{*}{\textbf{Constraint}}
    & \multicolumn{2}{c}{\textbf{Cumulative Profit ($\times10^6$)}}
    & \multicolumn{2}{c}{\textbf{$|$Wealth Gap$|$}}
    & \multicolumn{2}{c}{\textbf{$|$Approval Rate Disparity$|$}}
    & \multicolumn{2}{c}{\textbf{Inequality Ratio $\rho$}} \\
\cmidrule(lr){2-3}\cmidrule(lr){4-5}\cmidrule(lr){6-7}\cmidrule(lr){8-9}
    & Best & Mean $\pm$ Std
    & Best & Mean $\pm$ Std
    & Best & Mean $\pm$ Std
    & Best & Mean $\pm$ Std \\
\midrule
FL & OPP & $\mathbf{1.131 \pm 0.062}$ & OUT & $\mathbf{745.0 \pm 1288.3}$ & OPP & 0.0109 $\pm$ 0.0459 & OUT & $\mathbf{0.190 \pm 0.132}$ \\
RMM & OPP & 1.105 $\pm$ 0.100 & OPP & 3177.3 $\pm$ 55.4 & OUT & $\mathbf{0.0024 \pm 0.0202}^\dagger$ & OUT & 10.198 $\pm$ 0.032 \\
SW & OPP & 1.118 $\pm$ 0.099 & OPP & 3115.9 $\pm$ 136.1 & OPP & 0.0085 $\pm$ 0.0556 & OUT & 10.294 $\pm$ 0.082 \\
\bottomrule
\end{tabular}}
\end{table}

\begin{table}[t!]
\centering
\caption{\textbf{PeRL: best fairness perspective under each fairness constraint}, per metric, at the end of deployment (mean $\pm$ std over $10$ seeds). Bold: best value in the table for that metric.}
\label{tab:app_perl_by_constraint}
\resizebox{\textwidth}{!}{%
\begin{tabular}{lcccccccc}
\toprule
\multirow{2}{*}{\textbf{Constraint}}
    & \multicolumn{2}{c}{\textbf{Cumulative Profit ($\times10^6$)}}
    & \multicolumn{2}{c}{\textbf{$|$Wealth Gap$|$}}
    & \multicolumn{2}{c}{\textbf{$|$Approval Rate Disparity$|$}}
    & \multicolumn{2}{c}{\textbf{Inequality Ratio $\rho$}} \\
\cmidrule(lr){2-3}\cmidrule(lr){4-5}\cmidrule(lr){6-7}\cmidrule(lr){8-9}
    & Best & Mean $\pm$ Std
    & Best & Mean $\pm$ Std
    & Best & Mean $\pm$ Std
    & Best & Mean $\pm$ Std \\
\midrule
FL & OPP & $\mathbf{7.443 \pm 0.410}$ & OUT & $\mathbf{71.9 \pm 298.0}$ & OPP & $\mathbf{0.0097 \pm 0.0129}$ & OUT & $\mathbf{1.059 \pm 0.094}$ \\
RMM & OPP & 6.711 $\pm$ 1.229 & OPP & 207.9 $\pm$ 1328.7 & OPP & 0.0543 $\pm$ 0.0792 & OPP & 1.229 $\pm$ 0.643 \\
SW & OPP & 6.757 $\pm$ 1.137 & OPP & 535.2 $\pm$ 1187.0 & OPP & 0.0427 $\pm$ 0.0734 & OPP & 1.776 $\pm$ 0.878 \\
\bottomrule
\end{tabular}}
\end{table}
